\documentclass[aps,twocolumn,preprintnumbers,floats]{revtex4}\def\@cite#1#2{\textsuperscript{[{#1\if@tempswa , #2\fi}]}}

\usepackage{graphicx}
\usepackage{amsmath}
\usepackage{amsfonts}
\usepackage{amssymb}
\usepackage{color}
\usepackage{subfigure}
\usepackage{epsfig}
\usepackage{morefloats}
\usepackage{multirow}
\usepackage{graphicx,booktabs}
\usepackage{mathrsfs}
\usepackage{txfonts}
\usepackage{indentfirst}
\usepackage{graphicx,booktabs}
\usepackage{pifont}
\usepackage{bbding}
\usepackage{longtable,lscape}
\usepackage[colorlinks, citecolor=blue,anchorcolor=red,menucolor=red, linkcolor=red,filecolor=red,urlcolor=blue,frenchlinks=red]{hyperref}

\newcommand{\vrho}{\mbox{\boldmath$\rho$\unboldmath}}
\newcommand{\vlab}{\mbox{\boldmath$\lambda$\unboldmath}}

\begin{document}

\title{Unified study of nucleon and $\Delta$ baryon spectra and their strong decays with chiral dynamics}

\author{Hui-Hua Zhong$^{1}$, Ming-Sheng Liu$^{2}$, Ru-Hui Ni$^{1}$, Mu-Yang Chen$^{1,5}$, Xian-Hui Zhong$^{1,5}$~\footnote {E-mail: zhongxh@hunnu.edu.cn},
Qiang Zhao$^{3,4}$~\footnote {E-mail: zhaoq@ihep.ac.cn}}

\affiliation{ 1) Department of Physics, Hunan Normal University, and Key Laboratory of Low-Dimensional Quantum Structures and Quantum Control of Ministry of Education, Changsha 410081, China }

\affiliation{ 2) College of Science, Tianjin University of Technology, Tianjin 300384, China}

\affiliation{ 3) Institute of High Energy Physics, Chinese Academy of Sciences, Beijing 100049, China}

\affiliation{ 4) University of Chinese Academy of Sciences, Beijing 100049, China}

\affiliation{ 5)  Synergetic Innovation Center for Quantum Effects and Applications (SICQEA),
Hunan Normal University, Changsha 410081, China}
%
%\date{\today}

\begin{abstract}

In this work we systematically study both the mass spectra and strong decays of the nucleon and $\Delta$ resonances
up to the $N=2$ shell within a unified quark model framework with chiral dynamics.
In this framework we achieve a good description of the strong decay properties of the well-established nucleon and $\Delta$ resonances. Meanwhile, the mass reversal between $N(1440)1/2^{+}$ as the first radial excitation state and the $1P$-wave nucleon resonances can be explained.
We show that the three-body spin-orbit potential arising from the one-gluon exchange can cause a large configuration mixing between $N(1520)3/2^-$ and $N(1700)3/2^-$, and is also responsible for the large splitting between $\Delta(1600)1/2^-$ and $\Delta(1700)3/2^-$. Some of these baryon resonances turn to weakly couple to the $N\pi$, $N\eta$, $K\Lambda$, and $K\Sigma$ channels, which may answer the question why they have not been established in these channels via the $\pi N$ and $\gamma N$ scatterings. It shows that these ``missing resonances" may have large potentials to be established in the $N\pi\pi$ final state due to their large decay rates into either the $\Delta(1232)$ or $1P$-wave nucleon resonances via the pionic decays. Further experimental search for their signals in charmonium decays at BESIII is thus strongly recommended.

\end{abstract}

%\pacs{}

\maketitle

\section{Introduction}

A better understanding of the nucleon as a bound state of quarks and gluons, as
well as the spectrum and internal structure of excited baryons remains a fundamental
challenge and goal in hadronic physics~\cite{Crede:2013kia,Klempt:2009pi}. The study of the excited nucleons can
provide us with critical insights into the nature of QCD in the confinement domain~\cite{Isgur:2000ad}.
In experiments, the excited states of nucleon and $\Delta$ (i.e., $N^*$ and $\Delta^*$) can be produced in various processes such as $\pi N$ scattering, photo- or electroproduction. In the past two decades experiments at ELSA, GRAAL,
Jlab, MAMI, and SPring-8 (see the recent reviews~\cite{Burkert:2019kxy,Burkert:2022adb,Thiel:2022xtb}) have accumulated a large data sample which have provided more clues for more nucleon resonances~\cite{Anisovich:2015gia,Anisovich:2011fc,A2:2017gwp}. With the increase of the $J/\psi$ and $\psi(3686)$ data sample accumulated at the Beijing Electron-Positron Collider (BEPC-II), an alternative approach for the nucleon resonances is to study the baryon spectroscopy via the $J/\psi$ and $\psi(3686)$ decays into baryon and anti-baryon pairs~\cite{Zou:2018lse}, where the initial isoscalar of the charmonia provides an isospin filter for the produced baryon states. In Ref.~\cite{BES:2009ufh} one new resonance $N(2040)3/2^+$ was reported by the BES-III Collaboration in $J/\psi\to \bar{p}p\pi^0$, and later two new resonances $N(2300)3/2^+$ and $N(2570)5/2^-$ were observed in $\psi(3686)\to \bar{p}p\pi^0$~\cite{BESIII:2012ssm}.

In theory, the quark model has achieved a great success in describing the baryon spectrum~\cite{Capstick:1986ter,Crede:2013kia,Klempt:2009pi,Capstick:2000qj}.
However, about the excited nucleon spectrum, there are several long-standing questions
to be answered. The first one is why the roper resonance $N(1440)1/2^+$ belonging
to the $N=2$ shell lies so much lower than the first orbitally excited nucleon
states, such as $N(1535)1/2^-$ and $N(1520)3/2^-$. This is the so-called mass reversal problem.
In the quark potential model based on one-gluon exchange (OGE)~\cite{Capstick:1986ter}, the $N(1440)1/2^+$ as a
radial excitation is predicted to be $\sim 80$ MeV above the $N(1535)1/2^-$,
rather than $\sim 100$ MeV below it. To solve this puzzle, models based
on one-boson exchange (OBE) were proposed in the literature~\cite{Glozman:1997ag,Glozman:1996wq,Glozman:1995fu,Melde:2008yr}, and it may indicate that the chiral dynamics should play a role. Questions concerning these two different ways of describing the quark potentials, i.e. OGE and OBE, were raised and initiated a lot of debates in the literature~\cite{Isgur:1999jv,Glozman:1999ms}.

Another relevant issue is the so-called ``missing resonance" problem. Namely, why many excited nucleon states in the mass region of $\sim2.0$ GeV predicted by the symmetric quark model have not been observed in experiment. Although the diquark picture was proposed as a possible solution for reducing the internal degrees of freedom, hence reducing the number of excited baryon states~\cite{Ferretti:2011zz} (see~\cite{Anselmino:1992vg} for review), the situation did not get improved even for the second orbital excitations. Nevertheless, since the quark-gluon interactions are flavor-blind, there is no obvious reason that two quarks would be bound more tightly than the third one in a baryon state. Eventually, the diquark picture should have been ruled out by the results from the Lattice QCD simulation. In Ref.~\cite{Edwards:2011jj} it is clearly shown that the SU(6)$\otimes$O(3) symmetry holds well for the first orbital excitations.

%In the diqaurk models, many states predicted in the symmetric quark models are absent. This also bring another question: Do light baryons favor the SU(6)$\times$O(3) symmetry or the qaurk-diquark structure?

To understand better the internal degrees of freedom of the excited nucleon resonances, different methods and approaches were developed and adopted in the literature, such as the quark model combining both OGE and OBE potentials~\cite{Valcarce:2005rr,Ghalenovi:2017xxv,Zhao:2007hz},
the relativistically covariant constituent quark model with instantaneous forces~\cite{Loring:2001kv,Loring:2001kx},
the Dyson-Schwinger and Faddeev equations~\cite{Eichmann:2016hgl}, lattice QCD~\cite{Burch:2006cc,Bulava:2009jb,Bulava:2010yg,Edwards:2011jj,Engel:2013ig,Fodor:2012gf,Lin:2011ti}, large $N_c$ approach ~\cite{Goity:2002pu,Goity:2003ab,Schat:2001xr,Carlson:1998vx,Matagne:2004pm,Matagne:2011fr,Matagne:2012tm},
QCD sum rules~\cite{Ioffe:1981kw,Chung:1981cc,Dosch:1988vv}, the OGE model with higher order hyperfine interactions~\cite{Chen:2009de}, and so on.

Although different approaches provide their different versions for the baryon spectra, these states can only be probed by their interactions with other particles in their productions and/or decays. Their strong decays are often investigated and compared with the experimental data, from which the structure information can be extracted. In the quark model scenario, one can find the elementary meson emission model~\cite{Koniuk:1979vy,Faiman:1968js}, the quark pair creation model~\cite{LeYaouanc:1972vsx,LeYaouanc:1973ldf}, the chiral quark model~\cite{Manohar:1983md,Zhong:2007gp}, the quark model with the $N_c$ expansion~\cite{Jayalath:2011uc}, and the covariant quark model~\cite{Melde:2005hy,Sengl:2007yq}. The status about the strong decay studies of the light baryon resonances can be found in Refs.~\cite{Capstick:2000qj,Klempt:2009pi,Burkert:2019kxy,Burkert:2022adb,Thiel:2022xtb,Eichmann:2016yit}.

%Besides the puzzles existing in the mass spectrum, in a comprehensive review of baryon spectroscopy~\cite{Klempt:2009pi}, Klempt and Richard also asked another important question related to the decay: Do we understand baryon decays, or what can be learned by studying decays? As we know, the strong decays are the main modes of the $N^*$ and $\Delta^*$. The strong decay properties are often sensitive to the inner structures of hadron excitations. Thus, one can determine their quark model classifications by the strong decays. Furthermore, the missing resonance problem is also related to the decays. If some resonances couple weakly to the $\pi N$ channel, it may be a good reason for their missing. Then, one question what we should answer is: Do all of these missing/unwell-established resonances have a small decay rate into the $\pi N$ channel indeed? Since the operators responsible for strong decays of hadron resonances are still not very clear, to deal with the strong decays, one should depend on some phenomenological models, such as the elementary meson emission model~\cite{Koniuk:1979vy,Faiman:1968js}, the quark pair creation model~\cite{LeYaouanc:1972vsx,LeYaouanc:1973ldf,Bijker:2015gyk}, the chiral quark model~\cite{Manohar:1983md,Zhong:2007gp,An:2011sb}, quark model with the $N_c$ expansion~\cite{Jayalath:2011uc}, and covariant quark model~\cite{Melde:2005hy,Sengl:2007yq}. The status about the strong decay studies of the light baryon resonances can be found in Refs.~\cite{Capstick:2000qj,Klempt:2009pi,Burkert:2019kxy,Burkert:2022adb,Thiel:2022xtb,Eichmann:2016yit}.

In this work, we attempt to provide a unified study of both the mass spectra and strong decays of
the $N^*$ and $\Delta^*$ within a constituent quark model combining the chiral dynamics, i.e.,
the chiral quark model. The chiral quark model was broadly applied to the pseudoscalar meson photoproductions~\cite{Li:1994cy,Li:1995si,Li:1997gd,Saghai:2001yd,Zhao:2002id,He:2008ty,He:2008uf,He:2010ii,Zhong:2011ht,Zhong:2011ti,Xiao:2015gra}, and $\pi N$~\cite{Zhong:2007fx,Xiao:2016dlf,Wang:2017cfp} and $\bar{K} N$ scatterings~\cite{Zhong:2008km,Xiao:2013hca,Zhong:2013oqa} in the resonance energy region. This method has also been successfully extended to
study the strong decays of the singly-heavy baryon resonances~\cite{Zhong:2007gp,Wang:2017kfr,Wang:2017hej,Liu:2012sj,Wang:2018fjm,Yao:2018jmc,Wang:2019uaj,Wang:2020gkn,Xiao:2020gjo}, strange baryon resonances~\cite{Xiao:2013xi,Liu:2019wdr,Xiao:2018pwe}, and heavy-light meson states~\cite{Zhong:2008kd,Zhong:2009sk,Zhong:2010vq,Xiao:2014ura,Ni:2021pce,li:2021hss,Ni:2023lvx}.
It clearly indicates that the light quark-meson interactions described
with chiral dynamics should play a crucial role in dynamical processes where light quarks are involved. To be more specific, it shows that in a light baryon system, besides the OGE potential for the quark-quark interactions,
there are non-negligible contributions from the OBE. In this work,
we include, in addition to the OGE potentials, the chiral potentials via the $\pi$, $\eta$
and $\sigma$ exchanges between the constituent quarks to
calculate the mass spectra of the nucleon and $\Delta$ baryons.
The parameters of the chiral potentials are
well determined by the study of the strong decays within the chiral quark model. This hybrid quark model
including both the OGE and OBE potentials has been widely applied in the study of the hadron spectra and hadron-hadron interactions in the literature, e.g., Refs.~\cite{Vijande:2004he,Brauer:1990kt,Zhang:1994pp,Yu:1995ag,Valcarce:2005em,Valcarce:2005rr,Valcarce:2008dr,
Wang:2011rga,Huang:2004sj,Yang:2017qan,Huang:2017dwn,Chen:2007qn}.

The paper is organized as follows. The framework is given in Sec.~\ref{Fram}, where
we give an introduction to the quark
model classification for the excitations of nucleons,
the potential quark model for the mass calculations, the chiral quark model for
the evaluations of the strong decays, and the model parameters.
In Sec.~\ref{RD}, the masses and strong decay
properties of $N^*$ and $\Delta^*$ within
the $N=2$ shell are presented with discussions. A summary is given in Sec.~\ref{Summary}.

\section{Framework}\label{Fram}

%\subsection{Mass Spectrum}

\subsection{Quark model classification }

In the following, we will give an introduction of the quark model
classification for light baryons up to $N=3$ shell.
The wave function of a hadron includes four parts in
the color, flavor, spin, and spatial spaces.
The color wave function $\psi^{c}$ should be a color singlet under SU(3) symmetry.

%\subsubsection{ Spin-flavor wave functions }

For a light baryon system, the flavor wave function $\phi$ can be constructed by the light $u, \ d, \ s$ quarks, which follows a flavor SU(3) symmetry. The spin wave function $\chi$ can be constructed as the eigen states of the quark spin and its projection along e.g. $z$ axis $s_z$, which follows the SU(2) symmetry. Combining together the spin and flavor symmetry, the spin-flavor wave functions follow an SU(6) symmetry and are usually denoted by $|N_6, ^{2S+1}N_3\rangle$ in the literature, where $N_6$ and $N_3$
represent the dimensions of the SU(6) and SU(3) representations, respectively, and $S$ stands
for the quantum number of the total spin of a baryon state.
The SU(6) spin-flavor wave functions are given in Table~\ref{SFfunctions}.
For more details of the flavor wave function $\phi$ and
the spin wave function $\chi$ can be found in the previous work of our group~\cite{Xiao:2013xi}.

\begin{table}[htp]
\begin{center}
\caption{\label{SFfunctions} The SU(6) spin-flavor wave functions of light baryons. $\chi$ and
$\phi$ stand for the spin and flavor wave functions, respectively. The superscripts $s$ and $a$,
$\lambda$, and $\rho$ stand for the wave functions are symmetric, antisymmetric, mixed symmetric, and mixed
antisymmetric, respectively. }
\scalebox{1.0}{
\begin{tabular}{cc|cccccccccccc}\hline\hline

$\left|N_{6},^{2S+1}N_{3}\right\rangle $  & Wave function& & $\left|N_{6},^{2S+1}N_{3}\right\rangle $ & Wave function \\
\hline
$\left|56,^{2}8\right\rangle^s $          &   $\frac{1}{\sqrt{2}}\left(\chi^\lambda \phi^\lambda+\chi^\rho\phi^\rho\right)$ &&$\left|56,^{4}10\right\rangle^s $  &   $\phi^s\chi^s$\\
$\left|70,^{2}8\right\rangle^\rho $  &   $\frac{1}{\sqrt{2}}\left(\phi^\rho\chi^\lambda +\phi^\lambda\chi^\rho\right)$ &&$\left|70,^{2}8\right\rangle^\lambda $&$\frac{1}{\sqrt{2}}\left(\phi^\rho\chi^\rho -\phi^\lambda\chi^\lambda\right)$
    \\
$\left|70,^{4}8\right\rangle^\rho $  &  $\phi^\rho\chi^s$ &&$\left|70,^{4}8\right\rangle^\lambda $  &  $\phi^\lambda\chi^s$\\
$\left|70,^{2}10\right\rangle^\rho $  &  $\phi^s\chi^\rho$&&$\left|70,^{2}10\right\rangle^\lambda $  &  $\phi^s\chi^\lambda$
    \\
$\left|70,^{2}1\right\rangle^\rho $  &  $\phi^a\chi^\lambda$&&$\left|70,^{2}1\right\rangle^\lambda $  &  $\phi^a\chi^\rho$
    \\
$\left|20,^{2}8\right\rangle^a $  & $\frac{1}{\sqrt{2}}\left(\phi^\rho\chi^\lambda-\phi^\lambda\chi^\rho\right)$
&&$\left|20,^{4}1\right\rangle^a $& $\phi^a\chi^s$    \\
\hline\hline
\end{tabular}}
\end{center}
\end{table}

%\subsubsection{ Spatial wave functions }

The spatial wave functions satisfy the O(3) symmetry
under a rotation transformation. For a baryon system containing three quarks, the spatial wave function $\Psi_{NLM_L}=[\psi_{n_{\rho}l_{\rho}m_{\rho}}(\vrho)\otimes\psi_{n_{\lambda}l_{\lambda}m_{\lambda}}(\vlab)]_{NLM_L}$
is composed of $\rho$- and $\lambda$-mode spatial wave functions.
The Jacobi coordinates $\vrho$ and $\vlab$ can be related
to the quark coordinates $\mathbf{r}_j$ ($j=1,2,3$).
The explicit expressions are given by
\begin{eqnarray}\label{Jacobiana}
\boldsymbol{\rho}&=&\frac{1}{\sqrt{2}}(\boldsymbol{r}_{1}-\boldsymbol{r}_{2}),\nonumber \\
\boldsymbol{\lambda}&=&\sqrt{\frac{2}{3}}\left(\frac{m_1\boldsymbol{r}_{1}+m_2\boldsymbol{r}_{2}}{m_1+m_2}-\boldsymbol{r}_{3}\right ),\nonumber \\
\boldsymbol{R}&=& \frac{m_1\boldsymbol{r}_{1}+m_2\boldsymbol{r}_{2}+m_3\boldsymbol{r}_{3}}{m_1+m_2+m_3}.
\end{eqnarray}
The quantum numbers $n_{\rho}, l_{\rho}$ and $ m_{\rho}$ [or $n_{\lambda}, l_{\lambda}, m_{\lambda}$]
stand for that for the radial excitation, relative orbital angular momentum and its $z$ component for the $\rho$-mode [or
$\lambda$-mode] wave function, respectively.
While $N, L$ and $M_L$ stand for the principal quantum number,
the quantum numbers of total orbital angular
momentum and its $z$ component, respectively. They are defined by
$N=2n_{\rho}+2n_{\lambda}+l_{\rho}+l_{\lambda}$, $|l_{\rho}-l_{\lambda}|\leq L \leq l_{\rho}+l_{\lambda}$ and
$M_L=m_{\rho}+m_{\lambda}$. Furthermore, the superscript $\sigma$ appearing in wave functions represents
their permutation symmetries.

For the construction of spatial wave functions, besides the method suggested in Ref.~\cite{Karl:1968aep},
one can also construct them from the permutation symmetry directly.
For a fully symmetric wave function $\varphi^{s}$, when exchanging any two quarks,
it remains unchanged, that is $\hat{P}_{ij}\varphi^{s}=\psi^{s}$, where $\hat{P}_{ij}$ is a
permutation operator with $i<j$=1,2,3. For a completely antisymmetric wave function,
one has $\hat{P}_{ij}\varphi^{a}=-\varphi^{a}$. For a mixed-symmetric wave function $\varphi^{\lambda}$
and mixed-antisymmetric wave function $\varphi^{\rho}$, when exchanging quarks 1 and 2, one has
$\hat{P}_{12}\left(\begin{array}{c}
\varphi^{\lambda}\\
\varphi^{\rho}
\end{array}\right)=\left(\begin{array}{cc}
1 & 0\\
0 & -1
\end{array}\right)\left(\begin{array}{c}
\varphi^{\lambda}\\
\varphi^{\rho}
\end{array}\right)$.
While for the permutation of quarks at positions 1 and 3, the wave function satisfies
$\hat{P}_{13}\left(\begin{array}{c}
\varphi^{\lambda}\\
\varphi^{\rho}
\end{array}\right)=U\left(\begin{array}{c}
\varphi^{\lambda}\\
\varphi^{\rho}
\end{array}\right)$,
where $U$ is a unitary matrix.
The explicit forms of the spatial wave function up to the $N=3$ shell have been given in Table~\ref{The spatial functions}.

%\subsubsection{ Total wave functions }

The total wave function of a baryon can be expressed as
\begin{equation}
\left|qqq\right\rangle _{A}=\left|color\right\rangle_{A} \otimes \left|spin,flavor,spatial\right\rangle_{S}.
\end{equation}
The color part is fully antisymmetric, thus, the spin-flavor-spatial part
$|spin,flavor,spatial\rangle$ is symmetric. Based on the requirement of the SU(6)$\otimes$O(3)
symmetry, one can obtain the configurations for the spin-flavor-spatial part.
Our results for the light-flavor baryons up to the $N=3$ shell have been given in Table~\ref{TotalWaveFunctions2}.
For convenience, the different configurations are denoted by $^{2S+1}N_{3}[N_{6}, L_N^P]~ n^{2S+1}L_{J^P}$.
\begin{table}[htp]
\begin{center}
\caption{\label{The spatial functions} The spatial wave functions of the $N\leq3$ shell $\varPsi_{NLM_{L}}^{\sigma}(\boldsymbol{\rho},\boldsymbol{\lambda})$ as the linear combination of $\psi_{n_{\rho}l_{\rho}m_{\rho}}^{\rho}  \psi_{n_{\lambda}l_{\lambda}m_{\lambda}}^{\lambda}$. In the table,
$\left\{ \psi_{02m_{\rho}}^{\rho}\psi_{01m_{\lambda}}^{\lambda}\right\}_{LM_L}$, $\left\{ \psi_{01m_{\rho}}^{\rho}\psi_{02m_{\lambda}}^{\lambda}\right\}_{LM_L}$ and $\left\{ \psi_{01m_{\rho}}^{\rho}\psi_{01m_{\lambda}}^{\lambda}\right\}_{LM_L}$
stand for coupling states with quantum numbers $L,M_L$. They are the linear combination of $\psi_{0l_{\rho}m_{\rho}}^{\rho} \psi_{0l_{\lambda}m_{\lambda}}^{\lambda}$ with the Clebsch-Gordon coefficients.}
\scalebox{1.0}{
\begin{tabular}{ccccccccccc}\hline\hline

~~~$\varPsi_{000}^{s}=$&$\psi_{000}^{\rho}\psi_{000}^{\lambda}$~~~  \\
~~~$\varPsi_{11M_L}^{\lambda}=$&$\psi_{000}^{\rho}\psi_{01M_L}^{\lambda}$~~~\\
~~~$\varPsi_{11M_L}^{\rho}=$&$\psi_{01M_L}^{\rho}\psi_{000}^{\lambda}$~~~\\
~~~$\varPsi_{200}^{s}=$&$\frac{1}{\sqrt{2}}(\psi_{100}^{\rho}\psi_{000}^{\lambda}+\psi_{000}^{\rho}\psi_{100}^{\lambda})$~~~\\
~~~$\varPsi_{200}^{\lambda}=$&$\frac{1}{\sqrt{2}}(-\psi_{100}^{\rho}\psi_{000}^{\lambda}+\psi_{000}^{\rho}\psi_{100}^{\lambda})$~~~\\
~~~$\varPsi_{200}^{\rho}=$&$\frac{1}{\sqrt{3}}(\psi_{011}^{\rho}\psi_{01-1}^{\lambda}-\psi_{010}^{\rho}\psi_{010}^{\lambda}+\psi_{01-1}^{\rho}\psi_{011}^{\lambda})$~~~\\
~~~$\varPsi_{21M_L}^{a}=$&$\frac{1}{\sqrt{2}}(\psi_{01m_{\rho}}^{\rho}\psi_{01m_{\lambda}}^{\lambda}-\psi_{01m_{\lambda}}^{\rho}\psi_{01m_{\rho}}^{\lambda})$~~~\\
~~~$\varPsi_{22M_L}^{s}=$&$\frac{1}{\sqrt{2}}(\psi_{02M_L}^{\rho}\psi_{000}^{\lambda}+\psi_{000}^{\rho}\psi_{02M_L}^{\lambda})$~~~\\
~~~$\varPsi_{22M_L}^{\lambda}=$&$\frac{1}{\sqrt{2}}(\psi_{02M_L}^{\rho}\psi_{000}^{\lambda}-\psi_{000}^{\rho}\psi_{02M_L}^{\lambda})$~~~\\
~~~$\varPsi_{22M_L}^{\rho}=$&$\left\{ \psi_{01m_{\rho}}^{\rho}\psi_{01m_{\lambda}}^{\lambda}\right\}_{2M_L} $~~~\\

~~~$\varPsi_{33M_L}^{s}=$&$-\frac{1}{2}\psi_{000}^{\rho}\psi_{03M_L}^{\lambda}+\frac{\sqrt{3}}{2}\left\{ \psi_{02m_{\rho}}^{\rho}\psi_{01m_{\lambda}}^{\lambda}\right\}_{3M_L} $~~~  \\

~~~$\varPsi_{33M_L}^{a}=$&$\frac{1}{2}\psi_{03M_L}^{\rho}\psi_{000}^{\lambda}-\frac{\sqrt{3}}{2}\left\{ \psi_{01m_{\rho}}^{\rho}\psi_{02m_{\lambda}}^{\lambda}\right\}_{3M_L}$~~~   \\

~~~$\varPsi_{33M_L}^{\lambda}=$&$\frac{\sqrt{3}}{2}\psi_{000}^{\rho}\psi_{03M_L}^{\lambda}+\frac{1}{2}\left\{ \psi_{02m_{\rho}}^{\rho}\psi_{01m_{\lambda}}^{\lambda}\right\}_{3M_L}$~~~   \\

~~~$\varPsi_{33M_L}^{\rho}=$&$\frac{\sqrt{3}}{2}\psi_{03M_L}^{\rho}\psi_{000}^{\lambda}+\frac{1}{2}\left\{ \psi_{01m_{\rho}}^{\rho}\psi_{02m_{\lambda}}^{\lambda}\right\}_{3M_L} $~~~ \\

~~~$\varPsi_{32M_L}^{\lambda}=$&$-\left\{ \psi_{02m_{\rho}}^{\rho}\psi_{01m_{\lambda}}^{\lambda}\right\}_{2M_L} $~~~\\
~~~$\varPsi_{32M_L}^{\rho}=$&$-\left\{ \psi_{01m_{\rho}}^{\rho}\psi_{02m_{\lambda}}^{\lambda}\right\}_{2M_L} $~~~\\

~~~$\varPsi_{31M_L}^{s}=$&$-\frac{\sqrt{15}}{6}\psi_{100}^{\rho}\psi_{01M_L}^{\lambda}
+\frac{1}{2}\psi_{000}^{\rho}\psi_{11M_L}^{\lambda}-\frac{\sqrt{3}}{3}\left\{ \psi_{02m_{\rho}}^{\rho}\psi_{01m_{\lambda}}^{\lambda}\right\}_{1M_L}$~~~\\

~~~$\varPsi_{31M_L}^{a}=$&$\frac{\sqrt{15}}{6}\psi_{01M_L}^{\rho}\psi_{100}^{\lambda}
-\frac{1}{2}\psi_{11M_L}^{\rho}\psi_{000}^{\lambda}+\frac{\sqrt{3}}{3}\left\{ \psi_{01m_{\rho}}^{\rho}\psi_{02m_{\lambda}}^{\lambda}\right\}_{1M_L}$~~~\\

~~~$\varPsi_{31M_L}^{\lambda}=$&$-\frac{2}{3}\psi_{100}^{\rho}\psi_{01M_L}^{\lambda}+\frac{\sqrt{5}}{3}\left\{ \psi_{02m_{\rho}}^{\rho}\psi_{01m_{\lambda}}^{\lambda}\right\}_{1M_L}$~~~\\

~~~$\varPsi_{31M_L}^{\rho}=$&$-\frac{2}{3}\psi_{01M_L}^{\rho}\psi_{100}^{\lambda}+\frac{\sqrt{5}}{3}\left\{ \psi_{01m_{\rho}}^{\rho}\psi_{02m_{\lambda}}^{\lambda}\right\}_{1M_L}$~~~\\

~~~$\varPsi_{31M_L}^{\prime\lambda}=$&$-\frac{\sqrt{5}}{6}\psi_{100}^{\rho}\psi_{01M_L}^{\lambda}
-\frac{\sqrt{3}}{2}\psi_{000}^{\rho}\psi_{11M_L}^{\lambda}-\frac{1}{3}\left\{ \psi_{02m_{\rho}}^{\rho}\psi_{01m_{\lambda}}^{\lambda}\right\}_{1M_L}$~~~\\

~~~$\varPsi_{31M_L}^{\prime\rho}=$&$-\frac{\sqrt{5}}{6}\psi_{01M_L}^{\rho}\psi_{100}^{\lambda}
-\frac{\sqrt{3}}{2}\psi_{11M_L}^{\rho}\psi_{000}^{\lambda}-\frac{1}{3}\left\{ \psi_{01m_{\rho}}^{\rho}\psi_{02m_{\lambda}}^{\lambda}\right\}_{1M_L}$~~~\\

\hline\hline
\end{tabular}}
\end{center}
\end{table}

\begin{table*}[htp]
\begin{center}
\caption{ The total wave functions of N and $\Delta$ baryons on the N$\leq$3 shell. The configurations are denoted by $^{2S+1}N_{3}[N_{6}, L_N^P]~ n^{2S+1}L_{J^P}$.
$\varPsi_{NLM_L}^\sigma$, $\chi^\sigma$ and $\phi^\sigma$ stands for the spatial, spin and flavor wave functions, respectively. The Clebsch-Gordan series for the spin and orbital angular-momentum addition $|JJ_z\rangle=\sum_{M_L+M_S=J_z}\langle L,M_L,S,M_S|JJ_z\rangle\chi(S,M_S)\varPsi_{NLM_L}$. }\label{TotalWaveFunctions2}
\setlength{\tabcolsep}{2.5mm}
\begin{tabular}{cccccccccccccccccccccccccccccc}\hline\hline
~~~& States  %$\left|N_{6},^{2S+1}N_{3},N,L,J^{P}\right\rangle n^{2S+1}L_{J^{P}}$
~~~& $I$
~~~& $L$
~~~& $S$
~~~& $J^{P}$
~~~& Wave function
~~~&                       \\

\hline
~~~& $^{2}8[56,0_{0}^+]1^{2}S_{\frac{1}{2}^{+}}$
~~~& $\frac{1}{2}$
~~~& 0
~~~& $\frac{1}{2}$
~~~& $\frac{1}{2}^{+}$
~~~& $\frac{1}{\sqrt{2}}\varPsi_{000}^{s}(\chi^{\rho}\phi^{\rho}+\chi^{\lambda}\phi^{\lambda})$\\

~~~& $^{4}8[70,1_{1}^-]$$1^{4}P_{J^{-}}$
~~~& $\frac{1}{2}$
~~~& 1
~~~& $\frac{3}{2}$
~~~& $\frac{5}{2}^{-},\frac{3}{2}^{-},\frac{1}{2}^{-}$
~~~& $\frac{1}{\sqrt{2}}(\varPsi_{11M_L}^{\lambda}\chi^{s}\phi^{\lambda}+\varPsi_{11M_L}^{\rho}\chi^{s}\phi^{\rho})$\\

~~~& $^{2}8[70,1_{1}^-]$$ 1^{2}P_{J^{-}}$
~~~& $\frac{1}{2}$
~~~& 1
~~~& $\frac{1}{2}$
~~~& $\frac{3}{2}^{-},\frac{1}{2}^{-}$
~~~& $-\frac{1}{2}\varPsi_{11M_L}^{\lambda}(\chi^{\lambda}\phi^{\lambda}-\chi^{\rho}\phi^{\rho})
      +\frac{1}{2}\varPsi_{11M_L}^{\rho}(\chi^{\rho}\phi^{\lambda}+\chi^{\lambda}\phi^{\rho})$\\

~~~& $^{2}8[56,0_{2}^+]$$2^{2}S_{\frac{1}{2}^{+}}$
~~~& $\frac{1}{2}$
~~~& 0
~~~& $\frac{1}{2}$
~~~& $\frac{1}{2}^{+}$
~~~& $\frac{1}{\sqrt{2}} \varPsi_{200}^{s}(\chi^{\rho}\phi^{\rho}+\chi^{\lambda}\phi^{\lambda})$\\

~~~& $^{4}8[70,0_{2}^+]$$2^{4}S_{\frac{3}{2}^{+}}$
~~~& $\frac{1}{2}$
~~~& 0
~~~& $\frac{3}{2}$
~~~& $\frac{3}{2}^{+}$
~~~& $\frac{1}{\sqrt{2}}(\varPsi_{200}^{\lambda}\chi^{s}\phi^{\lambda}+\varPsi_{200}^{\rho}\chi^{s}\phi^{\rho})$\\

~~~& $^{2}8[70,0^{+}_{2}]$$2^{2}S_{\frac{1}{2}^{+}}$
~~~& $\frac{1}{2}$
~~~& 0
~~~& $\frac{1}{2}$
~~~& $\frac{1}{2}^{+}$
~~~& $-\frac{1}{2}\varPsi_{200}^{\lambda}(\chi^{\lambda}\phi^{\lambda}-\chi^{\rho}\phi^{\rho})
    +\frac{1}{2}\varPsi_{200}^{\rho}(\chi^{\rho}\phi^{\lambda}+\chi^{\lambda}\phi^{\rho})$\\

~~~& $^{2}8[20,1^{+}_{2}]$$1^{2}P_{J^{+}}$
~~~& $\frac{1}{2}$
~~~& 1
~~~& $\frac{1}{2}$
~~~& $\frac{3}{2}^{+}, \frac{1}{2}^{+}$
~~~& $\frac{1}{\sqrt{2}}\varPsi_{21M_L}^{a}(\chi^{\lambda}\phi^{\rho}-\chi^{\rho}\phi^{\lambda})$\\

~~~& $^{2}8[56,2^{+}_{2}]$$1^{2}D_{J^{+}}$
~~~& $\frac{1}{2}$
~~~& 2
~~~& $\frac{1}{2}$
~~~& $\frac{5}{2}^{+}, \frac{3}{2}^{+}$
~~~& $\frac{1}{\sqrt{2}}\varPsi_{22M_L}^{s}(\chi^{\rho}\phi^{\rho}+\chi^{\lambda}\phi^{\lambda})$\\

~~~& $^{4}8[70,2^{+}_{2}]$$1^{4}D_{J^{+}}$
~~~& $\frac{1}{2}$
~~~& 2
~~~& $\frac{3}{2}$
~~~& $\frac{7}{2}^{+}, \frac{5}{2}^{+}, \frac{3}{2}^{+}, \frac{1}{2}^{+}$
~~~& $\frac{1}{\sqrt{2}}(\varPsi_{22M_L}^{\lambda}\chi^{s}\phi^{\lambda}+\varPsi_{22M_L}^{\rho}\chi^{s}\phi^{\rho})$\\

~~~& $^{2}8[70,2^{+}_{2}]$$1^{2}D_{J^{+}}$
~~~& $\frac{1}{2}$
~~~& 2
~~~& $\frac{1}{2}$
~~~& $\frac{5}{2}^{+}, \frac{3}{2}^{+}$
~~~& $-\frac{1}{2}\varPsi_{22M_L}^{\lambda}(\chi^{\lambda}\phi^{\lambda}-\chi^{\rho}\phi^{\rho})
     +\frac{1}{2}\varPsi_{22M_L}^{\rho}(\chi^{\rho}\phi^{\lambda}+\chi^{\lambda}\phi^{\rho})$\\

~~~& $^{2}8[56,1^{-}_{3}]$$2^{2}P_{J^{-}}$
~~~& $\frac{1}{2}$
~~~& 1
~~~& $\frac{1}{2}$
~~~& $\frac{3}{2}^{-}, \frac{1}{2}^{-}$
~~~& $\frac{1}{\sqrt{2}}\varPsi_{31M_L}^{s}(\chi^{\rho}\phi^{\rho}+\chi^{\lambda}\phi^{\lambda})$\\

~~~& $^{4}8[70,1^{-}_{3}]$$2^{4}P_{J^{-}}$
~~~& $\frac{1}{2}$
~~~& 1
~~~& $\frac{3}{2}$
~~~& $\frac{5}{2}^{-}, \frac{3}{2}^{-}, \frac{1}{2}^{-}$
~~~& $\frac{1}{\sqrt{2}}(\varPsi_{31M_L}^{\lambda}\chi^{s}\phi^{\lambda}+\varPsi_{31M_L}^{\rho}\chi^{s}\phi^{\rho})$\\

~~~& $^{4}8[70,1^{-}_{3}]$$2^{4}P_{J^{-\prime}}$
~~~& $\frac{1}{2}$
~~~& 1
~~~& $\frac{3}{2}$
~~~& $\frac{5}{2}^{-}, \frac{3}{2}^{-}, \frac{1}{2}^{-}$
~~~& $\frac{1}{\sqrt{2}}(\varPsi_{31M_L}^{\prime\lambda}\chi^{s}\phi^{\lambda}+\varPsi_{31M_L}^{\prime\rho}\chi^{s}\phi^{\rho})$\\

~~~& $^{2}8[70,1^{-}_{3}]$$2^{2}P_{J^{-}}$
~~~& $\frac{1}{2}$
~~~& 1
~~~& $\frac{1}{2}$
~~~& $\frac{3}{2}^{-}, \frac{1}{2}^{-}$
~~~& $-\frac{1}{2}\varPsi_{31M_L}^{\lambda}(\chi^{\lambda}\phi^{\lambda}-\chi^{\rho}\phi^{\rho})
      +\frac{1}{2}\varPsi_{31M_L}^{\rho}(\chi^{\rho}\phi^{\lambda}+\chi^{\lambda}\phi^{\rho})$\\

~~~& $^{2}8[70,1^{-}_{3}]$$2^{2}P_{J^{-\prime}}$
~~~& $\frac{1}{2}$
~~~& 1
~~~& $\frac{1}{2}$
~~~& $\frac{3}{2}^{-}, \frac{1}{2}^{-}$
~~~& $-\frac{1}{2}\varPsi_{31M_L}^{\prime\lambda}(\chi^{\lambda}\phi^{\lambda}-\chi^{\rho}\phi^{\rho})
      +\frac{1}{2}\varPsi_{31M_L}^{\prime\rho}(\chi^{\rho}\phi^{\lambda}+\chi^{\lambda}\phi^{\rho})$\\

~~~& $^{2}8[20,1^{-}_{3}]$$2^{2}P_{J^{-}}$
~~~& $\frac{1}{2}$
~~~& 1
~~~& $\frac{1}{2}$
~~~& $\frac{3}{2}^{-}, \frac{1}{2}^{-}$
~~~& $\frac{1}{\sqrt{2}}\varPsi_{31M_L}^{A}(\chi^{\lambda}\phi^{\rho}-\chi^{\rho}\phi^{\lambda})$\\

~~~& $^{4}8[70,2^{-}_{3}]$$1^{4}D_{J^{-}}$
~~~& $\frac{1}{2}$
~~~& 2
~~~& $\frac{3}{2}$
~~~& $\frac{7}{2}^{-}, \frac{5}{2}^{-}, \frac{3}{2}^{-}, \frac{1}{2}^{-}$
~~~& $\frac{1}{\sqrt{2}}(\varPsi_{32M_L}^{\lambda}\chi^{s}\phi^{\lambda}+\varPsi_{32M_L}^{\rho}\chi^{s}\phi^{\rho})$\\

~~~& $^{2}8[70,2^{-}_{3}]$$1^{2}D_{J^{-}}$
~~~& $\frac{1}{2}$
~~~& 2
~~~& $\frac{1}{2}$
~~~& $\frac{5}{2}^{-}, \frac{3}{2}^{-}$
~~~& $-\frac{1}{2}\varPsi_{32M_L}^{\lambda}(\chi^{\lambda}\phi^{\lambda}-\chi^{\rho}\phi^{\rho})
     +\frac{1}{2}\varPsi_{32M_L}^{\rho}(\chi^{\rho}\phi^{\lambda}+\chi^{\lambda}\phi^{\rho})$\\

~~~& $^{2}8[56,3^{-}_{3}]$$1^{2}F_{J^{-}}$
~~~& $\frac{1}{2}$
~~~& 3
~~~& $\frac{1}{2}$
~~~& $\frac{7}{2}^{-}, \frac{5}{2}^{-}$
~~~& $\frac{1}{\sqrt{2}}\varPsi_{33M_L}^{s}(\chi^{\rho}\phi^{\rho}+\chi_{}^{\lambda}\phi^{\lambda})$\\

~~~& $^{4}8[70,3^{-}_{3}]$$1^{4}F_{J^{-}}$
~~~& $\frac{1}{2}$
~~~& 3
~~~& $\frac{3}{2}$
~~~& $\frac{9}{2}^{-}, \frac{7}{2}^{-}, \frac{5}{2}^{-}, \frac{3}{2}^{-}$
~~~& $\frac{1}{\sqrt{2}}(\varPsi_{33M_L}^{\lambda}\chi^{s}\phi^{\lambda}+\varPsi_{33M_L}^{\rho}\chi^{s}\phi^{\rho})$\\

~~~& $^{2}8[70,3^{-}_{3}]$$1^{2}F_{J^{-}}$
~~~& $\frac{1}{2}$
~~~& 3
~~~& $\frac{1}{2}$
~~~& $\frac{7}{2}^{-}, \frac{5}{2}^{-}$
~~~& $-\frac{1}{2}\varPsi_{33M_L}^{\lambda}(\chi^{\lambda}\phi^{\lambda}-\chi^{\rho}\phi^{\rho})
     +\frac{1}{2}\varPsi_{33M_L}^{\rho}(\chi^{\rho}\phi^{\lambda}+\chi^{\lambda}\phi^{\rho})$\\

~~~& $^{2}8[20,3^{-}_{3}]$$1^{2}F_{J^{-}}$
~~~& $\frac{1}{2}$
~~~& 3
~~~& $\frac{1}{2}$
~~~& $\frac{7}{2}^{-}, \frac{5}{2}^{-}$
~~~& $\frac{1}{\sqrt{2}}\varPsi_{33M_L}^{a}(\chi^{\lambda}\phi^{\rho}-\chi^{\rho}\phi^{\lambda})$\\

~~~& $^{4}10[56,0^{+}_{0}]$$1^{4}S_{\frac{3}{2}^{+}}$
~~~& $\frac{3}{2}$
~~~& 0
~~~& $\frac{3}{2}$
~~~& $\frac{3}{2}^{+}$
~~~& $\varPsi_{000}^{s}\chi^{s} \phi^{s}$\\

~~~& $^{2}10[70,1^{-}_{1}]$$1^{2}P_{J^{-}}$
~~~& $\frac{3}{2}$
~~~& 1
~~~& $\frac{1}{2}$
~~~& $\frac{3}{2}^{-},\frac{1}{2}^{-}$
~~~& $\frac{1}{\sqrt{2}}(\varPsi_{11M_L}^{\lambda}\chi^{\lambda}+\varPsi_{11M_L}^{\rho}\chi^{\rho}) \phi^{s}$ \\

~~~& $^{4}10[56,0^{+}_{2}]$$2^{4}S_{\frac{3}{2}^{+}}$
~~~& $\frac{3}{2}$
~~~& 0
~~~& $\frac{3}{2}$
~~~& $\frac{3}{2}^{+}$
~~~& $\varPsi_{200}^{s}\chi^{s} \phi^{s}$ \\

~~~& $^{2}10[70,0^{+}_{2}]$$2^{2}S_{\frac{1}{2}^{+}}$
~~~& $\frac{3}{2}$
~~~& 0
~~~& $\frac{1}{2}$
~~~& $\frac{1}{2}^{+}$
~~~& $\frac{1}{\sqrt{2}}(\varPsi_{200}^{\lambda}\chi^{\lambda}+\varPsi_{200}^{\rho}\chi^{\rho})  \phi^{s}$ \\

~~~& $^{4}10[56,2^{+}_{2}]$$1^{4}D_{J^{+}}$
~~~& $\frac{3}{2}$
~~~& 2
~~~& $\frac{3}{2}$
~~~& $\frac{7}{2}^{+}, \frac{5}{2}^{+}, \frac{3}{2}^{+}, \frac{1}{2}^{+}$
~~~& $\varPsi_{22M_L}^{s}\chi^{s}  \phi^{s}$ \\

~~~& $^{2}10[70,2^{+}_{2}]$$1^{2}D_{J^{+}}$
~~~& $\frac{3}{2}$
~~~& 2
~~~& $\frac{1}{2}$
~~~& $\frac{5}{2}^{+}, \frac{3}{2}^{+}$
~~~&$\frac{1}{\sqrt{2}}(\varPsi_{22M_L}^{\lambda}\chi^{\lambda}+\varPsi_{22M_L}^{\rho}\chi^{\rho})  \phi^{s}$ \\

~~~& $^{4}10[56,1^{-}_{3}]$$2^{4}P_{J^{-}}$
~~~& $\frac{3}{2}$
~~~& 1
~~~& $\frac{3}{2}$
~~~& $\frac{5}{2}^{-}, \frac{3}{2}^{-}, \frac{1}{2}^{-}$
~~~& $\varPsi_{31M_L}^{s}\chi^{s}   \phi^{s}$\\

~~~& $^{2}10[70,1^{-}_{3}]$$2^{2}P_{J^{-}}$
~~~& $\frac{3}{2}$
~~~& 1
~~~& $\frac{1}{2}$
~~~& $\frac{3}{2}^{-}, \frac{1}{2}^{-}$
~~~& $\frac{1}{\sqrt{2}}(\varPsi_{31M_L}^{\lambda}\chi^{\lambda}+\varPsi_{31M_L}^{\rho}\chi^{\rho})   \phi^{s}$\\

~~~& $^{2}10[70,1^{-}_{3}]$$2^{2}P_{J^{-\prime}}$
~~~& $\frac{3}{2}$
~~~& 1
~~~& $\frac{1}{2}$
~~~& $\frac{3}{2}^{-}, \frac{1}{2}^{-}$
~~~& $\frac{1}{\sqrt{2}}(\varPsi_{31M_L}^{\prime\lambda}\chi^{\lambda}+\varPsi_{31M_L}^{\prime\rho}\chi^{\rho})   \phi^{s}$\\

~~~& $^{2}10[70,2^{-}_{3}]$$1^{2}D_{J^{-}}$
~~~& $\frac{3}{2}$
~~~& 2
~~~& $\frac{1}{2}$
~~~& $\frac{5}{2}^{-}, \frac{3}{2}^{-}$
~~~& $\frac{1}{\sqrt{2}}(\varPsi_{32M_L}^{\lambda}\chi^{\lambda}+\varPsi_{32M_L}^{\rho}\chi^{\rho})  \phi^{s}$ \\

~~~& $^{4}10[56,3^{-}_{3}]$$1^{4}F_{J^{-}}$
~~~& $\frac{3}{2}$
~~~& 3
~~~& $\frac{3}{2}$
~~~& $\frac{9}{2}^{-}, \frac{7}{2}^{-}, \frac{5}{2}^{-}, \frac{3}{2}^{-}$
~~~& $\varPsi_{33M_L}^{s}\chi^{s}  \phi^{s}$\\

~~~& $^{2}10[70,3^{-}_{3}]$$1^{2}F_{J^{-}}$
~~~& $\frac{3}{2}$
~~~& 3
~~~& $\frac{1}{2}$
~~~& $\frac{7}{2}^{-}, \frac{5}{2}^{-}$
~~~& $\frac{1}{\sqrt{2}}(\varPsi_{33M_L}^{\lambda}\chi^{\lambda}+\varPsi_{33M_L}^{\rho}\chi^{\rho})  \phi^{s}$ \\
\hline
\hline
\end{tabular}
\end{center}
\end{table*}

\subsection{Potential model}

To calculate the mass of a baryon, we adopt the following semi-relativistic Hamiltonian
\begin{eqnarray}\label{Hamiltonian}
H=\sum_{i=1}^3\sqrt{\boldsymbol{p}_i^2+m_i^2}+\sum_{i<j}V(r_{ij})+C_0,
\end{eqnarray}
where $\sqrt{\boldsymbol{p}_i^2+m_i^2}$ is the kinetic energy of the $i$-th constituent quark with mass $m_i$ and momentum $\boldsymbol{p}_i$, $V(r_{ij})$ is the effective potential between the $i$-th and $j$-th quarks with a distance $r_{ij}\equiv|\mathbf{r}_{i}-\mathbf{r}_{j}|$. $C_0$ is the zero point energy. For convenience, in the calculations one needs to express the single-partial coordinates $\boldsymbol{r}_i~(i=1,2,3)$ with the Jacobi coordinates, which has been defined by Eq.~(\ref{Jacobiana}). By combining the Jacobi coordinates, one can obtain
\begin{eqnarray}\label{momentum1}
\boldsymbol{p}_1&=&\frac{m_1}{m_{tot}}\boldsymbol{P}_{c.m.}+\frac{1}{\sqrt{2}}\boldsymbol{p}_\rho+\frac{1}{\sqrt{6}}\boldsymbol{p}_\lambda,\nonumber\\
\boldsymbol{p}_2&=&\frac{m_2}{m_{tot}}\boldsymbol{P}_{c.m.}-\frac{1}{\sqrt{2}}\boldsymbol{p}_\rho+\frac{1}{\sqrt{6}}\boldsymbol{p}_\lambda,\nonumber\\
\boldsymbol{p}_3&=&\frac{m_3}{m_{tot}}\boldsymbol{P}_{c.m.}-\sqrt{\frac{2}{3}}\boldsymbol{p}_\lambda,
\end{eqnarray}
where $\boldsymbol{p}_\rho$ and $\boldsymbol{p}_\lambda$ represent the relative momenta
corresponding to the Jacobi coordinates $\boldsymbol{\rho}$ and $\boldsymbol{\lambda}$, respectively,
$m_{tot}=m_1+m_2+m_3$, while $\boldsymbol{P}_{c.m.}$ is the momentum of the center of mass.
The contribution of the center of mass motion in the kinetic energy can be easily removed by using
the relations given in Eq.~(\ref{momentum1}).

The effective potential can be decomposed into the spin-independent
and spin-dependent parts,
\begin{equation}\label{vij}
V(r_{ij})=V^{Corn}(r_{ij})+V^{sd}(r_{ij}).
\end{equation}
The spin-independent part $V^{Corn}(r_{ij})$ is adopted the
well-known Cornell form~\cite{Eichten:1978tg}, i.e.,
\begin{equation}\label{v conf}
V^{Corn}(r_{ij})=\frac{b}{2}r_{ij}-\frac{2}{3}\frac{\alpha_{S}}{r_{ij}},
\end{equation}
which includes the linear confinement potential $V^{Coul}$, and the Coulomb-like potential $V^{Conf}$
derived from the OGE model~\cite{Capstick:1986ter,Godfrey:1985xj}. Where, the slop parameter $b$ represents the strength of the confinement potential,
$\alpha_{S}$ denotes the strong coupling constant.
While, the spin-dependent potentials in the OGE model can be decomposed into
\begin{equation}\label{vij}
V^{sd}_G(r_{ij})=V^{SS}_G(r_{ij})+V^{LS}_G(r_{ij})+V^{T}_G(r_{ij}),
\end{equation}
where $V^{SS}_G(r_{ij})$, $V^{T}_G(r_{ij})$ and $V^{LS}_G(r_{ij})$ stand for
the spin-spin, tensor, and the spin-orbit potentials, respectively.
The spin-spin and tensor potentials are given by
\begin{eqnarray}\label{v T}
V^{SS}_G(r_{ij})&=&\frac{2\alpha_{S}}{3}\left\{\frac{\pi}{2}\cdot\frac{\sigma^3_{ij}
e^{-\sigma^2_{ij}r_{ij}^2}}{\pi^{3/2}}\cdot\frac{16}{3\tilde{m}_i\tilde{m}_j}(\boldsymbol{S}_i\cdot\boldsymbol{S}_j)\right\},\\
V^{T}_G(r_{ij})
&=&\frac{2\alpha_{S}}{3}\frac{1}{\tilde{m}_i\tilde{m}_jr_{ij}^3}\Bigg\{\frac{3(\boldsymbol{S}_i\cdot \boldsymbol{r}_{ij})(\boldsymbol{S}_j\cdot \mathbf{r}_{ij})}{r_{ij}^2}-\boldsymbol{S}_i\cdot\boldsymbol{S}_j\Bigg\}.
\end{eqnarray}
The spin-orbit potential~\cite{Capstick:1986ter}
\begin{equation}\label{vLS}
V^{LS}_G(r_{ij})=V_{ij}^{so\left(\nu\right)}+V_{ij}^{so(s)},
\end{equation}
with a color-magnetic piece
\begin{eqnarray}\label{v LS}
V_{ij}^{so\left(\nu\right)}
& = & \frac{1}{r_{ij}}\frac{dV^{Coul}(r_{ij})}{dr_{ij}} \left( \frac{\boldsymbol{r}_{ij}\times \boldsymbol{p}_i\cdot \boldsymbol{S}_i}{2\tilde{m}_i^2} -  \frac{\boldsymbol{r}_{ij}\times \boldsymbol{p}_j\cdot \boldsymbol{S}_j}{2\tilde{m}_j^2} \right. \nonumber\\
&& \left. -\frac{\boldsymbol{r}_{ij}\times \boldsymbol{p}_j\cdot \boldsymbol{S}_i-\boldsymbol{r}_{ij}\times \boldsymbol{p}_i\cdot \boldsymbol{S}_j}{\tilde{m}_i \tilde{m}_j} \right),
\end{eqnarray}
and a Thomas-precession piece
\begin{eqnarray}\label{v Thomas}
V_{ij}^{so(s)}
& = & -\frac{1}{r_{ij}}\frac{dV^{Conf}(r_{ij})}{dr_{ij}} \left( \frac{\boldsymbol{r}_{ij}\times \boldsymbol{p}_i\cdot \boldsymbol{S}_i}{2\tilde{m}_i^2} -  \frac{\boldsymbol{r}_{ij}\times \boldsymbol{p}_j\cdot \boldsymbol{S}_j}{2\tilde{m}_j^2} \right).
\end{eqnarray}
In the spin-dependent potentials, the $\boldsymbol{S}_i$ and $\boldsymbol{p}_i$
are the spin and momentum operators of the $i$th quark, respectively.
It should be mentioned that in the spin-dependent potentials arising from the OGE,
the light quark mass $m_i$ has been replaced with an effective mass $\tilde{m}_i$ to include some
relativistic corrections as adopted in the literature~\cite{Liu:2013maa,Ni:2021pce}.

% $\boldsymbol{\rho}\times \boldsymbol{p_{\lambda}}$ operator can be decomposed into operators $\boldsymbol{\rho}
%\times \boldsymbol{\nabla}_{\lambda}$ and $\boldsymbol{\rho} \times \boldsymbol{\lambda}$.

For a hadron system containing light quarks, the quark-quark interactions
arising from the Goldstone boson exchanges may play important roles
due to the spontaneous breaking of chiral symmetry. To better describe the mass spectrum of light $\Delta$ and $N$ baryons,
we include the chiral potentials by the exchanges of the light pseudoscalar mesons, $\pi$ and $\eta$, and the scalar $\sigma$ meson.

In the chiral quark model, the vertex for the tree-level quark-pseudoscalar-meson interaction is described by~\cite{Manohar:1983md,Li:1994cy,Li:1997gd,Zhao:2002id,Li:1995si}
\begin{eqnarray}\label{vpa}
\mathbf{{\cal L}}_{ps}&=&\frac{\delta}{\sqrt{2}f_{\mathbb{M}}}\bar{\psi}_j\gamma_{\mu}\gamma_{5}\psi_j{\vec{I}\cdot}\partial^{\mu}\vec{\phi}_{\mathbb{M}},
\end{eqnarray}
while the vertex for the quark-$\sigma$-meson interaction is described by~\cite{Yu:1995ag}
\begin{eqnarray}\label{spb}
\mathbf{{\cal L}}_{\sigma}&=&-g_{\sigma}\bar{\psi}_j\sigma\psi_j.
\end{eqnarray}
In the above effective Lagrangians, $\psi_j$ represents the $j$th light
quark field in a hadron, $I$ is an isospin operator, $\phi_{\mathbb{M}}$ and $\sigma$
represent the pseudoscalar meson fields $(\pi,K,\eta,\eta')$ and scalar meson field $\sigma$, respectively.
$f_{\mathbb{M}}$ stands for the decay the pseudoscalar meson decay constants.
$\delta$ is a global parameter accounting for the strength of the quark-pseudoscalar-meson couplings,
while $g_{\sigma}$ is a coupling constant for the quark-$\sigma$-meson interaction.
It should be mentioned that a suppressed factor,
\begin{equation}\label{fac}
F(\boldsymbol{q}^2)=\sqrt{\frac{\Lambda^{2}}{\Lambda^{2}+\boldsymbol{q}^{2}}},
\end{equation}
is introduced to correct the interaction vertices given by Eqs.~(\ref{vpa})
and (\ref{spb}), where $\Lambda$ determines the scale at which chiral symmetry is broken,
while $\boldsymbol{q}$ is the three momentum of the meson.

By using the effective chiral Lagrangians given by Eqs.~(\ref{vpa})
and (\ref{spb}), and performing a nonrelativistic reduction,
one can obtain the chiral potentials arising from the pseudoscalar and scalar meson exchanges.
The central part of these one-meson ($\pi$-, $K$, $\eta$-, $\sigma$)-exchange potentials is given by~\cite{Vijande:2004he}
\begin{equation}\label{vch}
V^{C}_{OBE}(r_{ij})=V_{\pi}^{C}(r_{ij})+V_{K}^{C}(r_{ij})+V_{\eta}^{C}(r_{ij})+V_{\sigma}^{C}(r_{ij}),
\end{equation}
with
\begin{eqnarray}\label{vpi}
V_{\pi}^{C}(r_{ij}) &=&\frac{g_{\pi}^{2}}{4\pi}\frac{\Lambda^{2}}{\Lambda^{2}-m_{\pi}^{2}} \frac{m_{\pi}^{3}}{3m_{i}m_{j}} \left[Y\left(m_{\pi}r_{ij}\right)-\frac{\Lambda^{3}}{m_{\pi}^{3}}Y\left(\Lambda
r_{ij}\right)\right]\nonumber\\
 && \cdot(\boldsymbol{S}_i\cdot\boldsymbol{S}_j)\sum_{a=1}^{3}(\lambda_{i}^{a}\cdot\lambda_{j}^{a}),
\end{eqnarray}
\begin{eqnarray}\label{vk}
V_{K}^{C}(r_{ij}) &=&\frac{g_{K}^{2}}{4\pi}\frac{\Lambda^{2}}{\Lambda^{2}-m_{K}^{2}} \frac{m_{K}^{3}}{3m_{i}m_{j}} \left[Y\left(m_{K}r_{ij}\right)-\frac{\Lambda^{3}}{m_{K}^{3}}Y\left(\Lambda
r_{ij}\right)\right]\nonumber\\
 && \cdot(\boldsymbol{S}_i\cdot\boldsymbol{S}_j)\sum_{a=4}^{7}(\lambda_{i}^{a}\cdot\lambda_{j}^{a}),
\end{eqnarray}
\begin{eqnarray}\label{veta}
V_{\eta}^{C}(r_{ij}) &=&\frac{g_{\eta}^{2}}{4\pi}\frac{\Lambda^{2}}{\Lambda^{2}-m_{\eta}^{2}}\frac{m_{\eta}^{3}}{3m_{i}m_{j}}\left[Y\left(m_{\eta}r_{ij}\right)-\frac{\Lambda^{3}}{m_{\eta}^{3}}Y\left(\Lambda r_{ij}\right)\right]\nonumber\\
 &&\cdot(\boldsymbol{S}_i\cdot\boldsymbol{S}_j)\left[(\lambda_{i}^{8}\cdot\lambda_{j}^{8})\cos\theta_{p}-(\lambda_{i}^{0}\cdot\lambda_{j}^{0})\sin\theta_{p}\right],
\end{eqnarray}
\begin{eqnarray}\begin{aligned}\label{vsigma}
V_{\sigma}^{C}(r_{ij}) & =-\frac{g_{\sigma}^{2}}{4\pi}\frac{\Lambda^{2}}{\Lambda^{2}-m_{\sigma}^{2}}m_{\sigma}\left[Y\left(m_{\sigma}r_{ij}\right)-\frac{\Lambda}{m_{\sigma}}Y\left(\Lambda r_{ij}\right)\right].
\end{aligned}\end{eqnarray}
The non-central tensor potentials are contributed by the $\pi$, $K$ and $\eta$ exchanges, which are given by
\begin{equation}\label{vch}
V^{T}_{OBE}(r_{ij})=V_{\pi}^{T}(r_{ij})+V_{K}^{T}(r_{ij})+V_{\eta}^{T}(r_{ij}),
\end{equation}
with
\begin{eqnarray}\label{vpia}
V_{\pi}^{T}(r_{ij})  =&&\frac{g_{\pi}^{2}}{4\pi}\frac{\Lambda^{2}}{\Lambda^{2}-m_{\pi}^{2}}\frac{m_{\pi}^{3}}{3m_{i}m_{j}}\left[H\left(m_{\pi}r_{ij}\right)-\frac{\Lambda^{3}}{m_{\pi}^{3}}H\left(\Lambda
r_{ij}\right)\right]\nonumber\\
 &&\cdot S_{ij}\sum_{a=1}^{3}(\lambda_{i}^{a}\cdot\lambda_{j}^{a}),
\end{eqnarray}
\begin{eqnarray}\label{vpia}
V_{K}^{T}(r_{ij})  =&&\frac{g_{K}^{2}}{4\pi}\frac{\Lambda^{2}}{\Lambda^{2}-m_{K}^{2}}\frac{m_{K}^{3}}{3m_{i}m_{j}}\left[H\left(m_{K}r_{ij}\right)-\frac{\Lambda^{3}}{m_{K}^{3}}H\left(\Lambda
r_{ij}\right)\right]\nonumber\\
 &&\cdot S_{ij}\sum_{a=4}^{7}(\lambda_{i}^{a}\cdot\lambda_{j}^{a}),
\end{eqnarray}
and
\begin{eqnarray}\label{vetaa}
V_{\eta}^{T}(r_{ij}) =&&\frac{g_{\eta}^{2}}{4\pi}\frac{\Lambda^{2}}{\Lambda^{2}-m_{\eta}^{2}}\frac{m_{\eta}^{3}}{3m_{i}m_{j}}\left[H\left(m_{\eta}r_{ij}\right)-\frac{\Lambda^{3}}{m_{\eta}^{3}}H\left(\Lambda r_{ij}\right)\right]\nonumber \\
&&\cdot S_{ij}\left[(\lambda_{i}^{8}\cdot\lambda_{j}^{8})\cos\theta_{p}-(\lambda_{i}^{0}\cdot\lambda_{j}^{0})\sin\theta_{p}\right].
\end{eqnarray}
While the non-central spin-orbit potentials are contributed by the $\sigma$ exchange,
which are given by
\begin{eqnarray}\label{vsigmab}
V_{\sigma}^{LS}(r_{ij})  =&&-\frac{g_{\sigma}^{2}}{4\pi}\frac{\Lambda^{2}m_{\sigma}^3}{\Lambda^{2}-m_{\sigma}^{2}}
\left[G\left(m_{\sigma}r_{ij}\right)-\frac{\Lambda^3}{m_{\sigma}^3}G\left(\Lambda r_{ij}\right)\right]\nonumber \\
&& \cdot \left( \frac{\boldsymbol{r}_{ij}\times \boldsymbol{p}_i\cdot \boldsymbol{S}_i}{2m_i^2} -  \frac{\boldsymbol{r}_{ij}\times \boldsymbol{p}_j\cdot \boldsymbol{S}_j}{2m_j^2} \right).
\end{eqnarray}
In the above equations, $\lambda^a$ ($a=1-8$) stands for the SU$(3)$ Gell-Mann matrices in the flavor space,
$Y(x)=\frac{1}{x}e^{-x}$, $H(x)=(1+3/x+3/x^2)Y(x)$, $G(x)=(1/x+1/x^2)Y(x)$, $S_{ij}=3(\boldsymbol{S}_{i}\cdot \hat{\mathbf{r}}_{ij})(\boldsymbol{S}_{j}\cdot \hat{\mathbf{r}}_{ij})-\boldsymbol{S}_{i}\cdot\boldsymbol{S}_{j}$.
$m_{\pi}$, $m_{K}$, $m_{\eta}$ and $m_{\sigma}$ stand for the masses for the $\pi$, $K$, $\eta$ and $\sigma$ mesons, respectively.
In Eq.~(\ref{vetaa}), $\theta_P$ is the mixing angle for the $\eta$ meson in the SU(3) wave function basis.
In Eqs.~(\ref{vpi})-(\ref{veta}), the coupling constants $g_{\mathbb{M}}$ ($\mathbb{M}=\pi,K,\eta$) for the
pseudoscalar mesons are defined by $g_{\mathbb{M}}=\delta \frac{m_u}{f_{\mathbb{M}}}$.
It should be mentioned that in this work the isospin operator $I$ is related to the Gell-Mann matrice
$\lambda$ with $I=\lambda/\sqrt{2}$.

\subsubsection{Numerical method}

To calculate the matrix elements in coordinate space, we
follow the same method adopted in our previous work~\cite{Liu:2019wdr}. The $\rho$- and $\lambda$-mode spatial wave functions $\psi_{n_{\rho}l_{\rho}m_{\rho}}$ and $\psi_{n_{\lambda}l_{\lambda}m_{\lambda}}$ can be expressed by
\begin{equation}
\psi_{n_{\xi}l_{\xi}m_{\xi}}\left(\xi\right)=R_{n_{\xi}l_{\xi}}\left(\xi\right)Y_{l_{\xi}m_{\xi}}\left(\xi\right),
\end{equation}
where $Y_{l_{\xi}m_{\xi}}\left(\xi\right)$ ($\xi=\rho,\lambda$) is the spherical harmonic function. The radial part, $R_{n_{\xi}l_{\xi}}\left(\xi\right)$, is expanded with a series of harmonic oscillator functions:
\begin{equation}\label{spatialwavefunction}
R_{n_{\xi}l_{\xi}}\left(\xi\right)=\sum_{\ell=1}^{n}C_{\xi \ell} \Phi_{n_{\xi}l_{\xi}}(\alpha_{\xi \ell},\xi),
\end{equation}
where
\begin{equation}\label{spatialfunction}
\begin{aligned}
\Phi_{n_{\xi}l_{\xi}}(\alpha_{\xi \ell},\xi)
&=\alpha_{\xi \ell}^{\frac{3}{2}}\left[\frac{2^{l_{\xi}+2-n_{\xi}}\left(2l_{\xi}+2n_{\xi}+1\right)!!}{\sqrt{\pi}n_{\xi}!\left[(2l_{\xi}+1)!!\right]^{2}}\right]^{\frac{1}{2}}\\
&\times(\alpha_{\xi \ell}\xi)^{l}e^{-\frac{1}{2}\alpha_{\xi \ell}^{2}\xi^{2}}F\left(-n_{\xi},l_{\xi}+\frac{3}{2},\alpha_{\xi \ell}^{2}\xi^{2}\right).\\
\end{aligned}
\end{equation}
The $F\left(-n_{\xi},l_{\xi}+\frac{3}{2},\alpha_{\xi \ell}^{2}\xi^{2}\right)$ in Eq.~(\ref{spatialfunction}) is the confluent hypergeometric function, the parameter $\alpha_{\xi \ell}$ can be related to the harmonic oscillator frequency $\omega_{\xi \ell}$ with $\alpha_{\xi \ell}\equiv1/d_{\xi \ell}=\sqrt{M_{\xi}\omega_{\xi \ell}}$. The reduced masses $M_{\rho , \lambda}$ are defined by $M_{\rho}\equiv\frac{2m_{1}m_{2}}{m_{1}+m_{2}}$, $M_{\lambda}\equiv\frac{3\left(m_{1}+m_{2}\right)m_{3}}{2\left(m_{1}+m_{2}+m_{3}\right)}$.
On the other hand, the harmonic oscillator frequencies $\omega_{\xi \ell}$ can be related to the harmonic oscillator stiffness factor $K_{\ell}$ with $\omega_{\xi \ell}=\sqrt{3K_{\ell}/M_{\xi}}$.
For a baryon system containing light $u/d$ quarks, one has $d_{\rho \ell}=d_{\lambda \ell}=d_{\ell}=(3m_{u}K_{\ell})^{-1/4}$. With this relation, the spatial wave function can be simply expanded as
\begin{eqnarray}
\varPsi_{NLM_{L}}^{\sigma}\left(\boldsymbol{\rho},\boldsymbol{\lambda}\right) & = & \sum_{\ell}^{n}C_{\ell}\Psi_{NLM_{L}}^{\sigma}\left(d_{\ell},\boldsymbol{\rho},\boldsymbol{\lambda}\right).
\end{eqnarray}
Then, the Schr\"{o}dinger equation can be solved by dealing with the generalized eigenvalue problem,
\begin{equation}\label{generalized eigenvalue}
\sum_{\ell^{\prime}=1}^{n}\left(H_{\ell \ell^{\prime}}-EN_{\ell \ell^{\prime}}\right)C_{\ell^{\prime}}=0,
\end{equation}
where $H_{\ell\ell^{\prime}}\equiv\langle \zeta (d_{\ell}^{\prime})|H|\zeta(d_{\ell})\rangle $ and $N_{\ell\ell^{\prime}}\equiv\langle \zeta(d_{\ell}^{\prime})\mid \zeta (d_{\ell})\rangle $, while
the  $\zeta\left(d_{\ell}\right)$ function is given by
\begin{equation}
\zeta\left(d_{\ell}\right)=\sum_{M_{L}+M_{S}=M}\left\langle LM_{L},SM_{s}| JM_J\right\rangle \Psi_{NLM_{L}}^{\sigma}(d_{\ell},\boldsymbol{\rho},\boldsymbol{\lambda})\chi_{M_{s}}^{\sigma}\phi^{\sigma}.
\end{equation}

The variational method is applied to solve the few-body problems.
In the calculations, following the method suggested in Ref. ~\cite{Hiyama:2003cu},
the variational parameter $d_{\ell}$ is selected to form a geometric progression,
\begin{equation}\label{dn}
d_{\ell}=d_{1}a^{\ell-1}(\ell=1,\ldots,n),
\end{equation}
where $n$ represents the number of Gaussian basis functions, and $a$ is
the ratio coefficient. There are three parameters $\{d_{1}, d_{n}, n\}$ to
be determined with the variation method. It is found that when taking $d_{1} = 0.1 fm$, $d_{n} = 2 fm$, and $n = 10$,
we can obtain stable solutions for the $\Delta$ and $N$ baryons.

Finally, it should be mentioned that the spin-dependent terms can cause mixing
between the different configurations with the same $J^{P}$ quantum numbers.
Thus, in this work, we also consider the configuration mixing induced by the spin-spin, spin-orbit
and tensor potentials.

%\begin{figure*}[htbp]
% \centering \epsfxsize=16.8 cm \epsfbox{Ndecay.eps}%\vspace{-1.5 cm}
% \caption{Partial decay widths of nucleon resonances  with principal quantum number $N\leq3$ level.}\label{Ndecay}
%\end{figure*}

%\begin{figure*}[htbp]
% \centering \epsfxsize=16.8 cm \epsfbox{Deltadecay.eps} %\vspace{-1.5 cm}
% \caption{Partial decay widths of $\Delta$ baryons with principal quantum number $\leq3$ level.}\label{Deltadecay}
%\end{figure*}

\subsection{Strong decay}

In this work, we also study the OZI-allowed two-body strong decays of the baryon resonances by emitting a single pseudoscalar meson ($\pi, K, \eta,$ or $\eta' $). The strong decay amplitudes can be evaluated with the chiral quark model~\cite{Manohar:1983md}.
Within the same framework, the chiral potentials have been derived for our mass spectrum study. In this model, the low-energy quark-pseudoscalar-meson interactions are described by the effective Lagrangian given by Eq.~(\ref{vpa}).

To match the nonrelativistic wave functions of the initial and final baryon states, it is crucial to adopt the nonrelativistic form of the effective Lagrangian in the calculations. Performing a nonrelativistic reduction, at the tree level one has~\cite{Ni:2023lvx}
\begin{eqnarray}\label{HI}
H_{I}=\mathbf{{\cal H}}^{NR}+\mathbf{{\cal H}}^{RC},
\end{eqnarray}
with
\begin{eqnarray}\label{Hnr}
\mathbf{{\cal H}}^{NR}=g\sum_{j}\left(\mathcal{G}\boldsymbol{\sigma}_{j}\cdot\boldsymbol{q}
+\frac{\omega_{m}}{2\mu_{q}}\boldsymbol{\sigma}_{j}\cdot\boldsymbol{p}_{j}\right )F(\boldsymbol{q}^2)I_{j}\varphi_{m},
\end{eqnarray}
and
\begin{eqnarray}\label{Hrc}
\mathbf{{\cal H}}^{RC}  &=&-\frac{g}{32\mu_{q}^{2}}\underset{j}{\sum}\left[m_{\mathbb{P}}^2(\boldsymbol{\sigma}_{j}\cdot\boldsymbol{q})\right.\nonumber\\
&&\left.+2\boldsymbol{\sigma}_{j}\cdot(\boldsymbol{q}-2\boldsymbol{p}_{j})\times(\boldsymbol{q}\times\boldsymbol{p}_{j})\right]
F(\boldsymbol{q}^2)I_{j}\varphi_{m}.
\end{eqnarray}
In the above equations, $\boldsymbol{\sigma}_{j}$ and $\boldsymbol{p}_{j}$ are the spin operator and internal momentum operator of the $j$-th light quark within a hadron. $\varphi_m=e^{-i\boldsymbol{q}\cdot\bf{r}_j}$ is the plane wave part of the emitted light meson. The factor $g$ and $\mathcal{G}$ is defined by $g = \delta\sqrt{(E_{i}+M_{i})(E_{f}+M_{f})}/(\sqrt{2}f_{\mathbb{M}})$ and $\mathcal{G}= -\left(\frac{\omega_{m}}{E_{f}+M_{f}}+1+\frac{\omega_{m}}{2m'_{j}}\right)$, in which $(E_{i}, M_{i})$ and $(E_{f}, M_{f})$ are the energy and mass of the initial baryon and final baryon, respectively.  $\omega_m$, $\boldsymbol{q}$ and $m_{\mathbb{P}}^2$ are the energy, three momentum and mass of the final state pseudoscalar meson. The reduced mass $\mu_q$ is expressed as $1/\mu_q=1/m_j+1/m'_j$ with $m_j$ and $m'_j$ for the masses of the $j$-th quark in the initial and final baryons, respectively. $F(\boldsymbol{q}^2)$ as a factor for suppressing the unphysical contributions in the high momentum region has been given in Eq.~(\ref{fac}). The isospin operator $I_j$ is given by~\cite{Li:1997gd,Zhong:2007gp,Zhong:2008kd}
\begin{equation} I_{j}=\begin{cases}
        a^{\dagger}_j(s)a_j(u)   &$for$~K^+,\\
        a^{\dagger}_j(u)a_j(s)   &$for$~K^-,\\
        a^{\dagger}_j(s)a_j(d)   &$for$~K^0,\\
        a^{\dagger}_j(u)a_j(d)   & $for$~ \pi^-,\\
        a^{\dagger}_j(d)a_j(u)   & $for$~ \pi^+,\\
        \frac{1}{\sqrt{2}}[a^{\dagger}_j(u)a_j(u)-a^{\dagger}_j(d)a_j(d)] & $for$~ \pi^0,\\
        \frac{1}{\sqrt{2}}[a^{\dagger}_j(u)a_j(u)+a^{\dagger}_j(d)a_j(d)]\cos\phi_P,\\
        ~~~~~- a^{\dagger}_j(s)a_j(s)\sin\phi_P & $for$~ \eta ,\\
        \frac{1}{\sqrt{2}}[a^{\dagger}_j(u)a_j(u)+a^{\dagger}_j(d)a_j(d)]\cos\phi_P,\\
        ~~~~~+ a^{\dagger}_j(s)a_j(s)\sin\phi_P & $for$~ \eta' ,
       \end{cases}
\end{equation}
where $a^{\dagger}_j(u,d,s)$ and $a_j(u,d,s)$ are the creation and
annihilation operators for the $u$, $d$ and $s$ quarks,
and $\phi_P$ is the mixing angle for the $\eta$ and $\eta'$ mesons in the
flavor basis which can be related to the mixing angle $\theta_{p}$ appearing in Eq.~(\ref{vetaa}) by $\phi_{p} =\theta_{p}+54.7^{\circ}$.
In this work we take $\phi_{p}=41.2^{\circ}$ as that determined in Ref.~\cite{Zhong:2011ht}.

The chiral quark model was widely adopted to study the strong decays of
excited hadrons~\cite{Zhong:2007gp,Wang:2017kfr,Wang:2017hej,Liu:2012sj,Wang:2018fjm,Yao:2018jmc,Wang:2019uaj,Wang:2020gkn,Xiao:2020gjo,
Zhong:2008kd,Zhong:2009sk,Zhong:2010vq,Xiao:2014ura,Ni:2021pce,li:2021hss,Ni:2023lvx,Xiao:2013xi,Liu:2019wdr,Xiao:2018pwe},
and pseudoscalar meson production processes~\cite{Li:1994cy,Li:1995si,Li:1997gd,Saghai:2001yd,Zhao:2002id,He:2008ty,He:2008uf,He:2010ii,Zhong:2011ht,Zhong:2011ti,Xiao:2015gra,
Zhong:2007fx,Xiao:2016dlf,Wang:2017cfp,Zhong:2008km,Xiao:2013hca,Zhong:2013oqa}.
In these works, only the $\mathbf{{\cal H}}^{NR}$ term is kept in the calculations, while the so-called relativistic
correction term $\mathbf{{\cal H}}_{I}^{RC}$ has been overlooked. Recently, the $\mathbf{{\cal H}}_{I}^{RC}$
term was first included in the investigations of the strong decays of heavy baryon resonances and multistrangeness baryon resonances~\cite{Arifi:2021orx,Arifi:2022ntc}. It is found that the agreement with the data is significantly improved,
in particular, the decay widths of the Roper-like states are greatly increased by one order of magnitude as compared
to the results without the relativistic correction. Lately, the long-standing puzzle of the broad width for the
radially excited heavy-light mesons, such as $D_0(2550)$/$D_{s0}(2590)$ and
$D^*_1(2600)$/$D^*_{s1}(2700)$, was overcome naturally by including the $\mathbf{{\cal H}}_{I}^{RC}$ term in Ref.~\cite{Ni:2023lvx}.

Within the chiral quark model, the two-body OZI-allowed strong decay amplitude
for the $\mathcal{B} \rightarrow \mathcal{B}^{\prime} \mathbb{M}$ process can be worked out by
\begin{eqnarray}\label{Amplitudea}
\mathcal{M}\left[\mathcal{B} \rightarrow \mathcal{B}^{\prime} \mathbb{M}\right]
=\left\langle\mathcal{B}^{\prime}\left|H_{I}\right| \mathcal{B}\right\rangle,
\end{eqnarray}
where $ \mathcal{B}$ and $\mathcal{B'}$ stand for the initial and final baryon states, respectively,
and $\mathbb{M}$ is the emitting pseudoscalar meson. With derived decay amplitudes from Eq.~(\ref{Amplitudea}), the partial decay width for the $\mathcal{B} \rightarrow \mathcal{B}^{\prime} \mathbb{M}$ process can be obtained with
\begin{eqnarray}\label{TwoBodyDecay}
\Gamma=\frac{1}{8 \pi} \frac{\left|\mathbf{q}\right|}{M_{i}^{2}}\frac{1}{2J_i+1}\sum_{J_{i z} J_{f z}}\left|\mathcal{M}_{J_{i z} J_{f z}}\right|^{2},
\end{eqnarray}
where $J_i$ is the total angular momentum quantum number of the initial baryon, $J_{i z}$ and $J_{f z}$ represent the third components of the total angular momenta of the initial and final baryons, respectively.

\begin{figure}[htbp]
	\centering \epsfxsize=8cm \epsfxsize=8cm \epsfbox{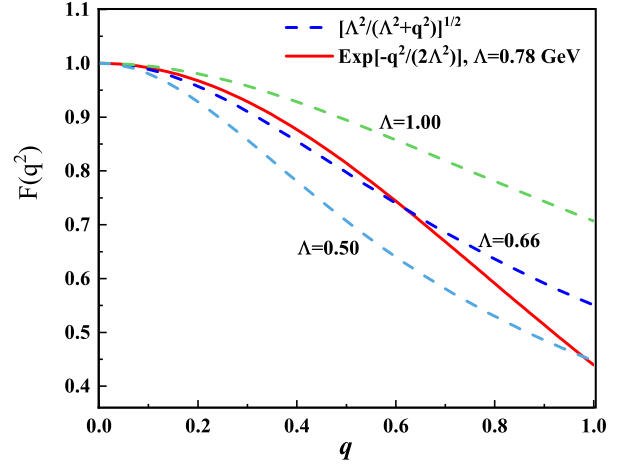} \vspace{-0.1 cm}
\caption{The shape factor in exponential form (solid line) and fractional form (solid line) as
a function of the momentum q.}\label{Formf}
\end{figure}

\subsection{Model parameters}

The model parameters adopted in this work have been presented in Table~\ref{Mass parameters}.
The decay constants for $\pi$, $K$, and $\eta$ mesons are
fixed with $f_{\pi}=93$ MeV and $f_{K/\eta}=113$ MeV.
The cut-off parameter $\Lambda$ in the factor $F(\boldsymbol{q}^2)=\sqrt{\frac{\Lambda^2}{\Lambda^2+\boldsymbol{q}^2}}$
is determined to be $\Lambda=0.66$ GeV to consist with the behavior of
the exponential form $F(\boldsymbol{q}^2)=e^{-\frac{\boldsymbol{q}^2}{2\Lambda^2}}$ with $\Lambda=0.78$ GeV (see Fig.~\ref{Formf}),
which is determined from our recent study of the heavy-light meson spectrum~\cite{Ni:2023lvx}.
The typical value of the constituent quark mass for the $u/d$ quark is
in the range of $400\pm 100$ MeV, in this work it is fixed with $m_{u/d}=420$ MeV, while the effective quark mass
in the one-gluon exchange potentials is fixed with $\tilde{m}_u=0.62$ GeV as that adopted in our previous work~\cite{Ni:2021pce}.
The dimensionless parameter $\delta$ accounting for the strength
of the quark-meson couplings is taken the same value $\delta=0.576$ as that
determined by the strong decays of light strange baryon resonances in
the previous studies of our group~\cite{Xiao:2013xi,Liu:2019wdr,Xiao:2018pwe}.
With the relation $g_{\mathbb{M}}=\delta \frac{m_u}{f_{\mathbb{M}}}$,
we obtain $g_{\pi}\simeq2.6$ and $g_{K}=g_{\eta}\simeq2.14$, which are comparable with the chiral
coupling constant $g_{ch}\simeq 2.6-2.7$ determined from the $\pi NN$ coupling in the literature, e.g. Refs. ~\cite{Yu:1995ag,Vijande:2004he,Valcarce:2008dr,Valcarce:2005rr,Wang:2011rga,Huang:2004sj,Yang:2017qan,Huang:2017dwn}.
In this work, we take $g_{\sigma}=g_{\pi}$ as an approximation.

The other four parameters \{$b$, $C_{0}$, $\alpha_S$, $\sigma$\} in the potential model are determined
by an overall description of the masses of the eight well-established states $N(938)$, $\Delta(1232)$, $\Delta(1600)3/2^+$, $N(1520)1/2^-$, $N(1535)1/2^-$, $N(1650)1/2^-$, $\Delta(1620)1/2^-$, and $ N(1675)5/2^-$ listed in the Review of Particle
Physics (RPP) by the Particle Data Group (PDG)~\cite{ParticleDataGroup:2022pth}.
The determined value of the slop parameter, $b=0.135$ $\mathrm{GeV}^{2}$, is
compatible with $b=0.10-0.14$ $\mathrm{GeV}^{2}$ adopted in the previous works of our group~\cite{li:2021hss,Li:2020xzs,
Liu:2019vtx,Liu:2019wdr,Deng:2016stx}. The parameters $\alpha_{S}=0.40$ and $\sigma=0.344$ GeV are mainly constrained by
the mass splitting between $N(938)1/2^+$ and $\Delta(1232)3/2^+$, since this splitting
is sensitive to the OGE spin-spin potentials.

It should be noted that we cannot obtain stable solutions for some states due to
the divergent behavior of $1/r^{3}$ in the OGE spin-orbit and tensor potentials, if we do not treat them as perturbative terms.
In order to overcome this problem, following the method adopted in the previous works~\cite{Deng:2016stx,Deng:2016ktl,Li:2019tbn},
we introduce a cutoff distance $r_{c}$ in the calculation.
In a small range $(0, r_c)$, we let $1/r^{3}=1/r^{3}_c$ in the spin-orbit and tensor potentials.
It is found that the mass of the nucleon excitation configuration $^{4}8[70,1_1]1^4P_{1/2^-}$
are more sensitive to the $r_{c}$ due to its relatively
larger factor $\langle\boldsymbol{L}_{ij}\cdot(\boldsymbol{S}_{i}+\boldsymbol{S}_{j})\rangle$ than the other states.
Thus, the cutoff parameter $r_c=0.23$ fm is determined by fitting the mass
of $^{4}8[70,1_1]1^4P_{1/2^-}$ obtained with the method of perturbation.
In this method, we treat the spin-orbit and tensor potentials as perturbative terms.
First, by neglecting the contributions of these terms we obtain the mass $m_0$ and
spatial wave function, then by using the wave function we further
calculate the mass correction term $\Delta m$ from the perturbative terms, finally,
we obtained the whole mass $M=m_0+\Delta m$ for our fitting.

The OBE spin-orbit and tensor potentials
is less important for our understanding the mass spectrum, thus,
they are often neglected in the literature. However,
they can affect our determining the configuration mixing between $N(1535)1/2^-$ and $N(1650)1/2^-$.
Note that the singular behavior also exists in the OBE spin-orbit and tensor potentials.
To overcome this problem, similar to the OGE case, in a small range $(0, r_c')$,
we let $1/r^{3}=1/r'^{3}_c$ and $1/r^{2}=1/r'^{2}_c$ in OBE spin-orbit and tensor potentials.
Taking $r_c'\simeq0.70$ fm, the mixing angle between $N(1535)$ and $N(1650)$
is determined to be $\theta_S\simeq 25^{\circ}$, which is consistent with that extracted from the $\pi$ and $\eta$ production processes~\cite{Zhong:2011ti,Xiao:2015gra}.
In this case, the contributions from the $1/r^{3}$ and $1/r^{2}$ terms
in the one-pion-exchange tensor potential are strongly suppressed,
the role of the potential is governed by the $1/r$ term.

\begin{table}[htp]
\begin{center}
\caption{\label{Mass parameters} Quark model parameters used in this work.}
\begin{tabular}{ccccccccccc}\hline\hline
$m_{u}$ (GeV) &$\tilde{m}_{u}$ (GeV) & $b$  $(\mathrm{GeV}^{2})$   & ${\sigma}$ (GeV) & $C_0$ (GeV)       &$r_{c}$ (fm) \\
\hline
0.42         &0.62              & 0.135                        &   0.344          & $-0.829$                  & 0.23\\
\hline\hline
 ${\alpha}_S$  &$\delta$ &$f_\pi$ (GeV)   & $f_{K/\eta}$ (GeV)   & $\Lambda$ (GeV)      & $r_{c}^{\prime}$ (fm)\\
\hline
 0.40          &0.576     &0.093           & 0.113                     & 0.66                 &  0.70               \\
\hline\hline
\end{tabular}
\end{center}
\end{table}

In the calculation of the strong decays, we adopt the numerical wave functions
for the $N^*$ and $\Delta^*$ baryons to calculate the transition amplitudes.
The details of the wave functions for the ground states $N(938)$ and $\Delta(1232)$ are crucial
for understanding the strong decay properties of the $N^*$ and $\Delta^*$ baryons
because all of these excited states should decay into $N(938)$ and/or $\Delta(1232)$.
Considering the uncertainty of the wave functions of the $N(938)$ and $\Delta(1232)$,
we adjust their size appropriately to more reasonably describe the strong decay properties
of the well-established states. For the convenience of adjustment,
the wave functions of the ground states are adopted a simple harmonic oscillator form
with an effective size parameter $\alpha$. The size parameters for $N(938)$
and $\Delta(1232)$ are determined to be $\alpha=0.49$ and $0.40$ GeV, respectively.
%We fit them with a single Gaussian form by reproducing the root-mean-square radius of the $\rho$-mode excitations,  the effective $\alpha_{\rho}=\alpha_{\lambda}$ parameter of $N(938)$ has changed from 0.75 GeV to 0.50 GeV, the value for $\Delta(1232) $ is changed from 0.54 GeV to 0.41 GeV.
The high $N^*$ and $\Delta^*$ baryons can decay into the $\Lambda$, $\Sigma$ and/or $\Sigma^{*}$.
Their wave functions are adopted the simple harmonic oscillator form as well. The harmonic oscillator size parameter $\alpha_{\rho}$ for the $\rho$-oscillator in the spatial wave function is taken as $\alpha_{\rho}=0.4$ GeV, while the parameter $\alpha_{\lambda}$ for the $\lambda$-oscillator is related to $\alpha_{\rho}$ as $\alpha_{\lambda}=\sqrt[4]{3m_{u}/(2m_{s}+m_{u})}\alpha_{\rho}$~\cite{Xiao:2013xi}.

In the calculations of the strong decays, we also need the masses of initial and final hadrons.
For the well-established states, the masses adopt the average values from the PDG~\cite{ParticleDataGroup:2022pth}.
While for the unestablished states, the masses are taken from our quark model predictions.

\begin{figure*}[htbp]
 \centering \epsfxsize=10.8 cm \epsfbox{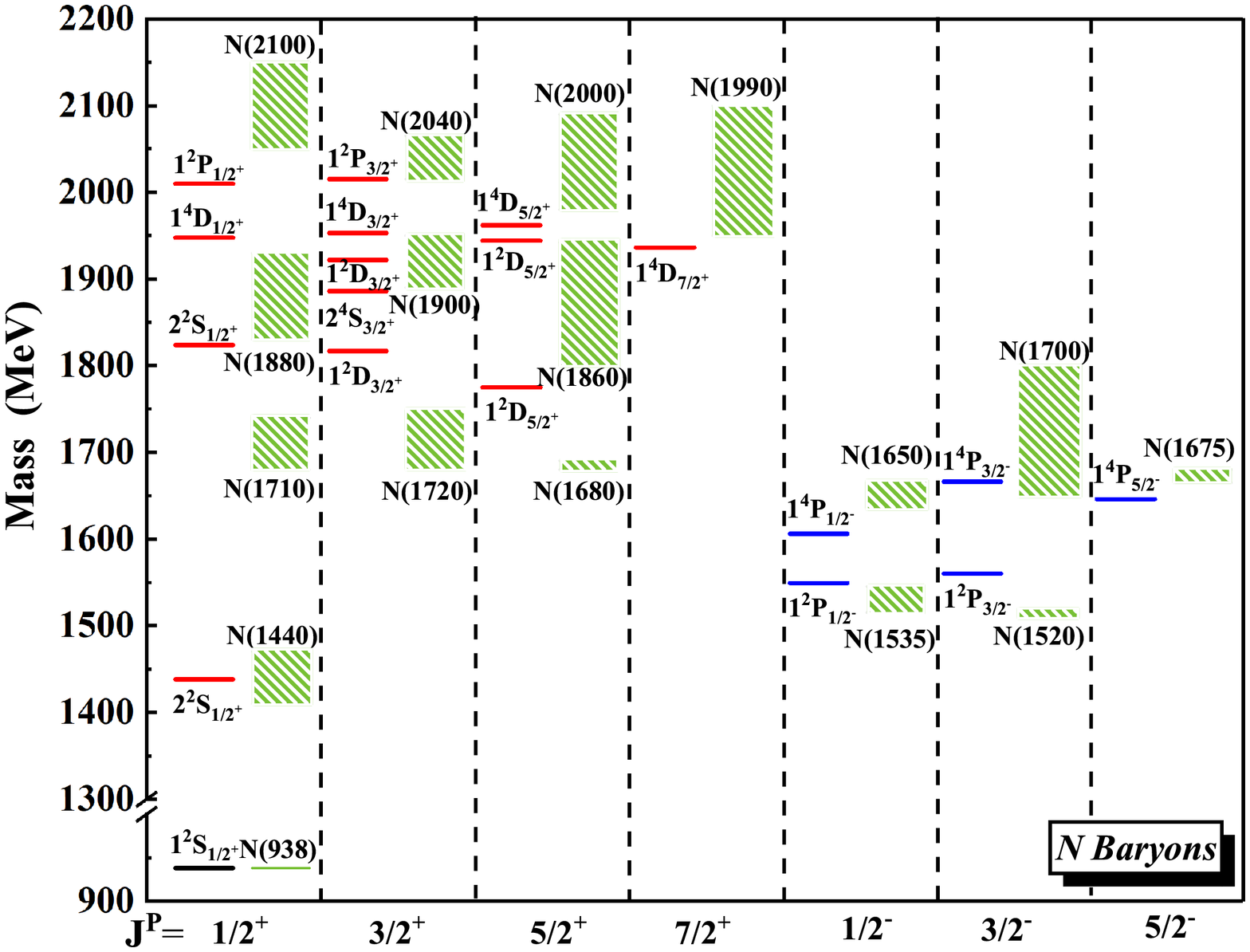}%\vspace{-1.5 cm}
 \caption{Nucleon spectrum with principal quantum number $N\leq2$. The short solid lines stand for the theoretical results,
 while the shaded boxes stand for the experimental data from the PDG~\cite{ParticleDataGroup:2022pth}.}\label{Nmass}
\end{figure*}

\begin{figure*}[htbp]
 \centering \epsfxsize=10.8 cm \epsfbox{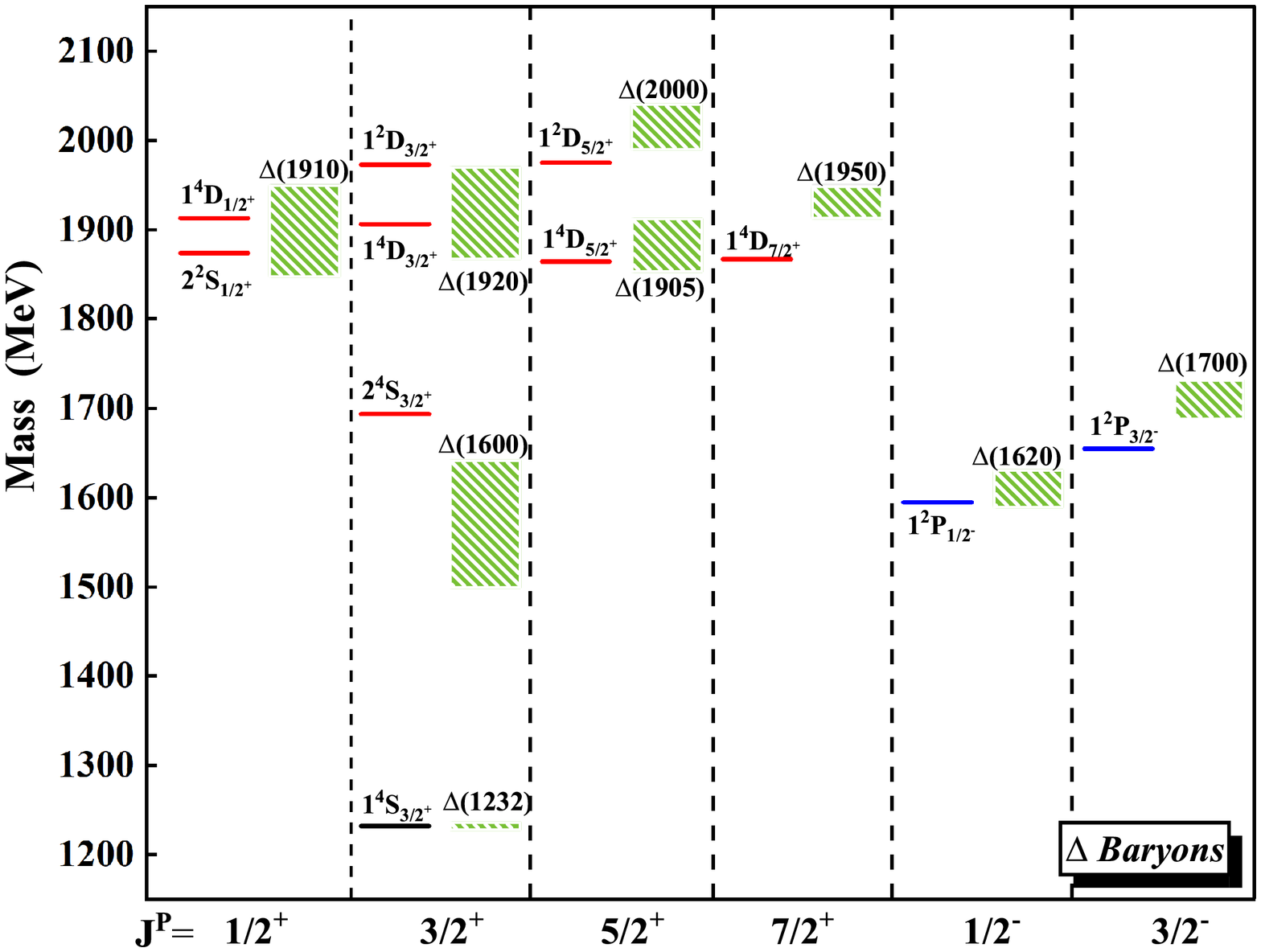}%\vspace{-1.5 cm}
 \caption{$\Delta$ spectrum with principal quantum number $N\leq2$. The short solid lines stand for the theoretical results,
 while the shaded boxes stand for the experimental data from the PDG~\cite{ParticleDataGroup:2022pth}.}\label{delmass}
\end{figure*}

\section{Results and discussion}\label{RD}

The obtained mass spectra for the nucleon and $\Delta$ baryons up to the $N=2$ shell compared with the data and some other
works are given In Tables~\ref{nucleon mass spectrum comparison}  and~\ref{Delta mass spectrum comparison}, respectively.
For clarity, the mass spectra for the nucleon and $\Delta$ baryons compared with the data are plotted in Figs.~\ref{Nmass} and \ref{delmass}, respectively. To better know about the dynamical mechanism of the strong interactions in
the nucleon and $\Delta$ baryon systems, we also analyze the contributions of each part
of the Hamiltonian to their masses. The results are given in Tables~\ref{nucleon mass spectrum contribution}
and~\ref{Delta mass spectrum contribution}. Furthermore, by combining the spectra obtained within the potential model,
the strong decay properties are studied within the chiral quark model which has been
adopted to derive the chiral potentials for the mass spectrum calculations.
The decay properties are given in the Tables~\ref{Low mass nucleon decay widths}-\ref{Delta decay widths}.
From the Tables, one can see that the mass spectra and decay properties for the well-established nucleon and $\Delta$ baryons
can be reasonably described within a unified framework.

%\begin{figure}[htbp]
%	\centering \epsfxsize=7.2cm \epsfbox{N1535width.eps}
%               \epsfxsize=7.2cm \epsfbox{N1650width.eps}\vspace{-0.1 cm}
%\caption{Decay width for the N(1535) states as a function of mixing angle $\theta$.}\label{fig:N1535width}
%\end{figure}

%\begin{figure}[htbp]
%	\centering \epsfxsize=8.5cm \epsfxsize=8.5cm \epsfbox{N1650width.eps} \vspace{-0.1 cm}
%\caption{Decay width for the N(1650) states as a function of mixing angle $\theta$.}\label{fig:N1650width}
%\end{figure}

%\begin{figure}[htbp]
%	\centering \epsfxsize=8.0cm  \epsfbox{pmixa.eps} \vspace{-0.3 cm}
%\caption{Decay width for the N(1535) and N(1650) states as a function of mixing angle $\theta$.}\label{fig:pmix width}
%\end{figure}

\subsection{N Baryons}

\subsubsection{$N(1535,1650)1/2^-$ and tensor interactions}

\begin{figure}[htbp]
	\centering \epsfxsize=7.4cm \epsfbox{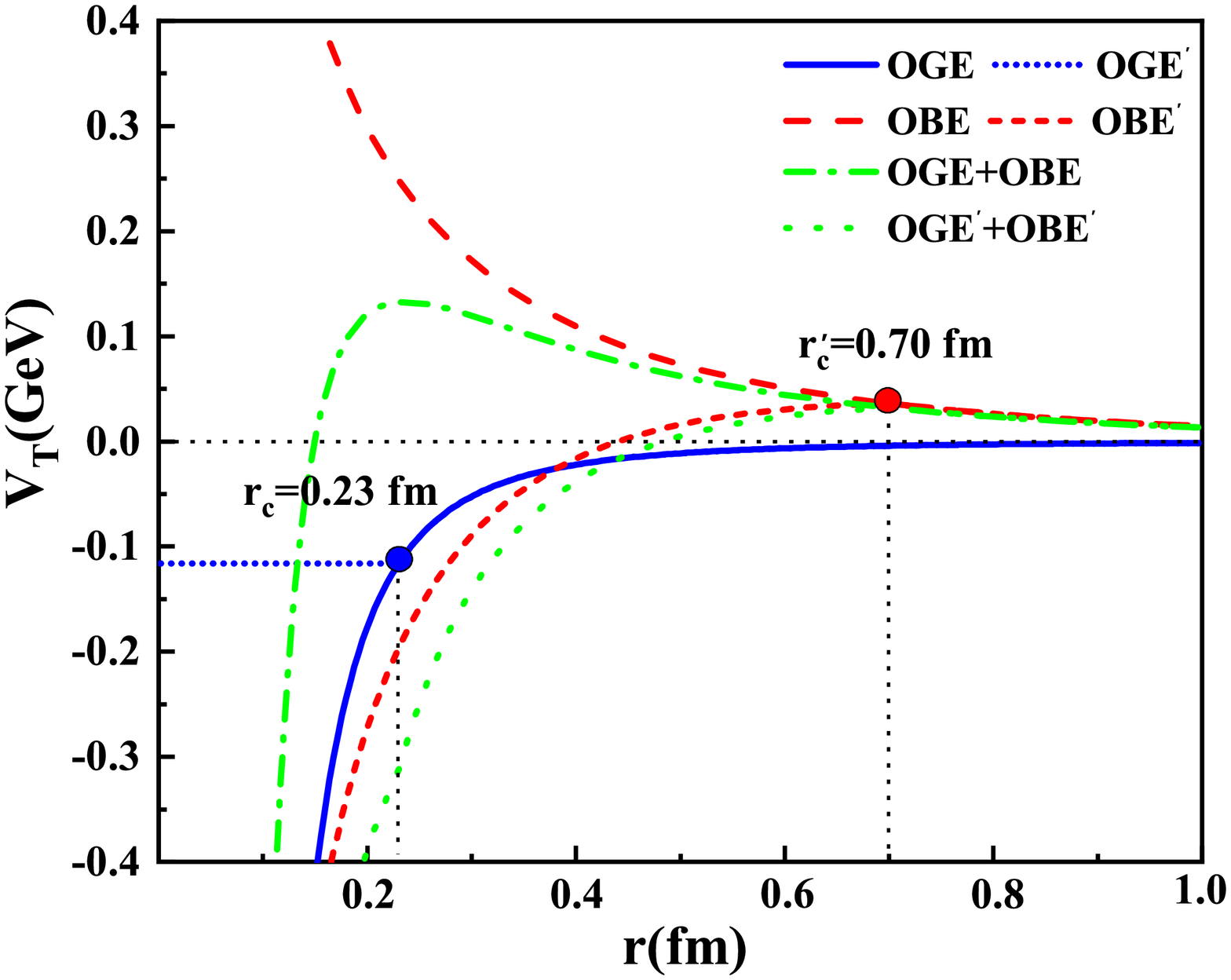} \vspace{-0.2 cm}
\caption{The tensor potentials as functions of the distance $r$ between two quarks for the $^{4}8[70,1_1^-]1^4P_{1/2^-}$
configuration, which is the dominant component of the $N(1650)1/2^-$. The solid and dashed curves represent the
potentials arising from the one-gluon and one-pion exchanges (labeled with OGE and OBE), respectively.
The short dotted and short dashed curves represent the one-gluon and one-pion exchange potentials when a truncation is carried out, they are labeled
with OGE$^{\prime}$ and OBE$^{\prime}$, respectively.
The dash-dotted curve represents the whole potential without truncations, which is labeled with OGE$+$OBE in the figure.
While the dotted curve represents the whole potential with truncations, which is labeled with OGE$^{\prime}+$OBE$^{\prime}$.}\label{tensor}
\end{figure}

\begin{figure}[htbp]
	\centering \epsfxsize=7.4cm  \epsfbox{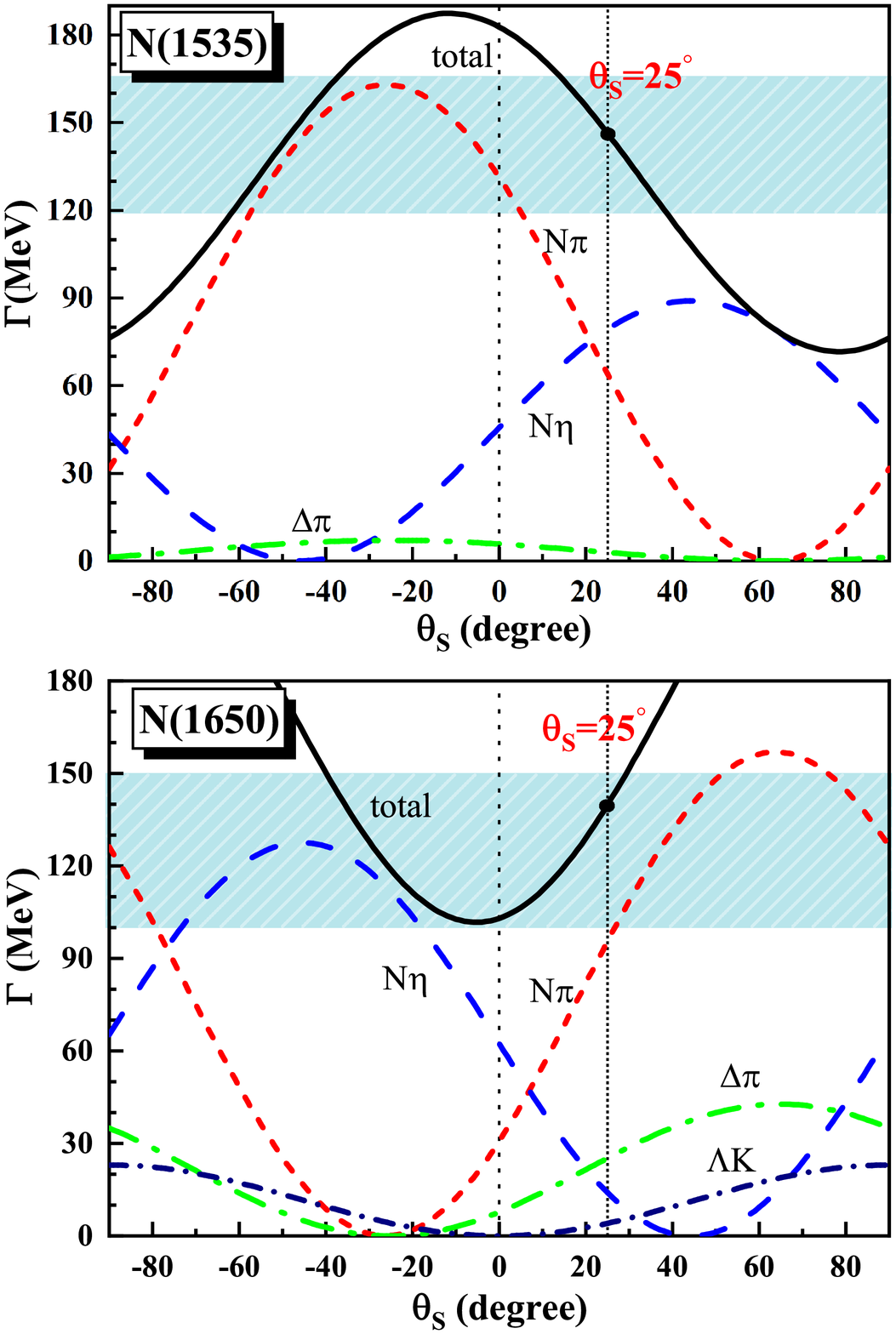} \vspace{-0.2 cm}
\caption{Partial decay widths for the $N(1535)1/2^-$ and $N(1650)1/2^-$ resonances as functions of the mixing angle $\theta_S$.}\label{fig:pmix width}
\end{figure}

In the constituent quark model, the $N(1535)1/2^{-}$ and $N(1650)1/2^{-}$ resonances are often considered
to be two mixed states via $^{2}8[70,1_1^-]1^2P_{1/2^-}$-$^{4}8[70,1_1^-]1^4P_{1/2^-}$ mixing
mainly due to the tensor interactions~\cite{Isgur:1977ef,Isgur:1978xj}.
The mixing scheme is expressed as
\begin{equation}\label{mix-s11}
\left(\begin{array}{c}N(1535)\cr N(1650)
\end{array}\right)=\left(\begin{array}{cc} \cos\theta_{S} &-\sin\theta_{S}\cr \sin\theta_{S} & \cos\theta_{S}
\end{array}\right)
\left(\begin{array}{c} ^28[70, 1_1^-] \frac{1}{2}^{-} \cr
^48[70, 1_1^-]\frac{1}{2}^{-}
\end{array}\right),
\end{equation}
where $\theta_S$ is the mixing angle. The configuration mixing
in $N(1535)1/2^{-}$ and $N(1650)1/2^{-}$ seems to be inevitable for correctly describing
the reactions $\gamma p\to \eta p$~\cite{He:2008ty,Zhong:2011ti} and $\gamma p\to \pi^0 p$~\cite{Xiao:2015gra}
in the resonances region.
By fitting the reaction data, the mixing angle is determined to be
$\theta_S \simeq 26^{\circ}$~\cite{Zhong:2011ti,Xiao:2015gra}, which is consistent
with $\theta_S \simeq 32^{\circ}$ obtained with
the OGE tensor potentials, other than the negative mixing angle determined
with the OBE tensor potentials~\cite{An:2011sb,Chen:2018gts}.
About the OGE and OBE models, there were heated
debates between Isgur and Glozman in 1999~\cite{Isgur:1999jv,Glozman:1999ms}.

%The analysis of the electromagnetic transitions of the
%nucleon resonances with a negative parity also support OGE, not OBE.

Do both the OGE and OBE potential exist in the light baryon states at the same time?
Can we explain the mixing angle $\theta_S \simeq 26^{\circ}$ determined
by the scattering data when we include both the OGE and OBE tensor potentials?
To answer these questions, we carry some analyses as follows.

In Fig.~\ref{tensor}, the tensor potentials of the OGE
and OBE for the $^{4}8[70,1_1^-]1^4P_{1/2^-}$ configuration are plotted with red
dashed and blue solid curves, respectively. It should be mentioned that
the OBE tensor potential is mainly contributed by the one pion exchange.
The total tensor potential is shown with green dash-dotted curve.
It is seen that the sign of the OGE tensor potential
is opposite to that of OBE, the two kinds of potentials
have a strong cancelation. If we adopt this potential directly,
the fairly large mixing angle $\theta_S \simeq 26^{\circ}$ cannot be explained.
Note that the singular behavior in the tensor potentials will bring unphysical
contributions in the short ranges, which should be reasonably eliminated.
Following the method adopted in Refs.~\cite{Deng:2016stx,Deng:2016ktl,Li:2019tbn},
we cut off the OGE and OBE tensor potentials at $r_c=0.23$ fm and $r_c'=0.70$ fm, respectively.
In this case, a reasonable mixing angle $\theta_S \simeq 25^{\circ}$ can be obtained.

The OGE and OBE tensor potentials after truncations are shown with blue dotted and
short dashed curves in Fig.~\ref{tensor}, respectively.
From the figure, one can find that the average depth of the OGE tensor
potential in the range of $r\leq 0.23$ fm is about $-100$ MeV.
The behavior of the OBE tensor potential after a truncation
changes notably in the short region, which is
dominated by the $1/r$ term. It is reasonable to take a
larger cutoff parameter $r_c'=0.70$ fm for the OBE tensor potential.
As we know in the short range, the quark-quark interactions are mainly due to the OGE,
while the nonperturbative interactions due to OBE should be strongly suppressed.
However, the singular behavior of the OBE potentials
notably enlarges the unphysical contributions in the short range.
Thus, we need a larger cutoff parameter $r_c'=0.70$ fm to suppress the unphysical contributions.
The whole tensor potential (OBE+OGE) after truncations is shown with a green
dotted curve in Fig.~\ref{tensor}, which is very similar
to the OGE potential. This may be an answer for the question why the OGE
tensor potential can give a mixing angle consistent with the data.
%The configuration mixing effects due to the tensor forces increase
%the mass splitting between the two $1/2^-$ states from to 37 to 93 MeV
%of $\sim 56$ MeV,  due to configuration mixing is estimated to be about 56 MeV,

With the mixing angle $\theta_S \simeq 25^{\circ}$,
the strong decay properties of the $N(1535)1/2^{-}$ and $N(1650)1/2^{-}$
are evaluated within the chiral quark model. Our results together
the experimental data and some other model predictions given in
Table~\ref{Low mass nucleon decay widths}. It is seen that our predictions
are in reasonably good agreement with the data. The anomalously
large $ N\eta$ branching ratio of the $N(1535)1/2^{-}$ and the anomalously
small $ N\eta$ branching ratio of the $N(1650)1/2^{-}$ can
be well explained. The partial width ratios between the $ N\eta$ and $N\pi $ channels
for the $N(1535)1/2^-$ and $N(1650)1/2^-$ are predicted to be
\begin{eqnarray}\label{mix-s11}
R_1^{1/2^-} &=&\frac{\Gamma[N(1535)\to  N\eta]}{\Gamma[N(1535)\to  N\pi]}\simeq 1.28,\\
R_2^{1/2^-} &=&\frac{\Gamma[N(1650)\to  N\eta]}{\Gamma[N(1650)\to  N\pi]}\simeq 0.14.
\end{eqnarray}
The predicted ratio $R_1^{1/2^-}$ for $N(1535)1/2^-$ is consistent with the ratio
$0.95\pm 0.3$ extracted from the data of $\pi$ and
$\eta$ electroproduction off the proton by the CLAS Collaboration~\cite{CLAS:2009ces}, and the new solution
$1.09^{+0.37}_{-0.28}$ extracted from the $\pi N\to \pi N,\eta N$ data by the Zagreb group~\cite{Batinic:2010zz}.
The predicted ratio $R_2^{1/2^-}=0.14$ for $N(1650)1/2^-$ is consistent with the
new solution $0.17^{+0.07}_{-0.08}$ extracted from the $\pi N\to \pi N,\eta N$ data by the Zagreb group~\cite{Batinic:2010zz}.

To see the mixing angle dependency of
the decay properties, the partial decay widths of the $N(1535)1/2^-$ and $N(1650)1/2^-$
as functions of the mixing angle $\theta_S$ are plotted in Fig.~\ref{fig:pmix width}.
It is found that without configuration mixing ($\theta_S=0$), the $N(1535)1/2^-$ dominantly
decays into the $ N\pi$ and $ N\eta$ channels with branching fractions of $\sim 70\%$ and
$\sim 25\%$, while $N(1650)1/2^-$ decays into the $ N\eta$ and $ N\pi$ channels with branching fractions of $\sim 60\%$ and
$\sim 30\%$, respectively. Taking $\theta_S=25^\circ$, the $ N\eta$ becomes the dominant decay
mode of $N(1535)1/2^{-}$ with a decay rate of $\sim 55\%$, while for $N(1650)1/2^{-}$,
the decay rate of the $ N\eta$ is suppressed to $\sim 9\%$,
the $ N\pi$ and $ \Delta\pi$ become the dominant decay modes with a
decay rate of $\sim 69\%$ and $\sim 18\%$.

As a whole, within the constituent quark model, both the mass and strong decay properties
of the $N(1535)1/2^{-}$ and $N(1650)1/2^{-}$ can be reasonably understood. Including the
OBE tensor potential together with that of OGE, one can explain
the mixing angle extracted from the $\gamma p\to \eta p,\pi^0p$ data.
To better understand the tensor potentials the more accurate
measurements of the partial width decay ratios between $ N\eta$ and $ N\pi$
for the $N(1535)1/2^{-}$ and $N(1650)1/2^{-}$ are needed in future experiments.

\begin{figure}[htbp]
	\centering \epsfxsize=7.4cm  \epsfbox{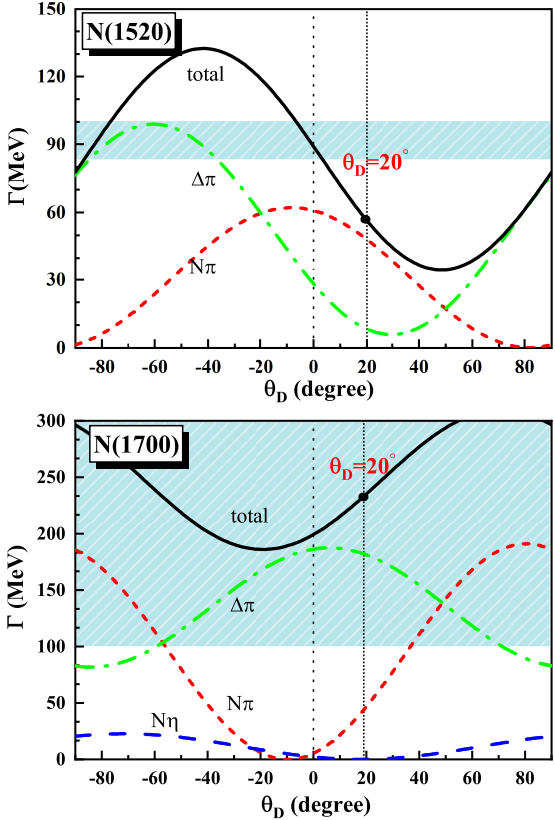} \vspace{-0.2 cm}
\caption{Partial decay widths for the $N(1520)3/2^-$ and $N(1700)3/2^-$ resonances as functions of the mixing angle $\theta_D$.}\label{fig:pmix2}
\end{figure}

\subsubsection{$N(1520,1700)3/2^{-}$ and three-body spin-orbit potentials}

In the constituent quark model, the $N(1520)3/2^{-}$ and $N(1700)3/2^{-}$ resonances
are assigned to the $^{2}8[70,1_1^-]1^4P_{3/2^-}$ and
$^{4}8[70,1_1^-]1^4P_{3/2^-}$ states, respectively.
From Table~\ref{nucleon mass spectrum comparison}, it is seen that the masses of
both $N(1520)3/2^{-}$ and $N(1700)3/2^{-}$ can be well described within various quark models.
Configuration mixing between these two states may also exist.
Taking the following mixing scheme,
\begin{equation}\label{mixd}
\left(\begin{array}{c}N(1520)\cr N(1700)
\end{array}\right)=\left(\begin{array}{cc} \cos\theta_{D} &-\sin\theta_{D}\cr \sin\theta_{D} & \cos\theta_{D}
\end{array}\right)
\left(\begin{array}{c} ^{2}8[70,1_1]\frac{3}{2}^{-} \cr
^{4}8[70,1_1]\frac{3}{2}^{-}
\end{array}\right),
\end{equation}
the mixing angle is predicted to be $\theta_{D}\simeq 8^\circ$ based on the tensor interactions
arising from the OGE~\cite{Isgur:1977ef,Isgur:1978xj}.

In the present work, it is found that the mixing angle may reach up to a fairly
large value $\theta_{D}\simeq 20^\circ$, if the three-body components of the spin-orbit interactions arising
from the OGE are included. In the following, we give a review of
the three-body components of the spin-orbit potential since it may be unfamiliar to someone. For example, considering the spin-orbit interaction between two light $u/d$ quarks, from Eq.~(\ref{momentum1}) one has
\begin{eqnarray}\label{Threebs}
V_G^{LS}(r_{12})
&  = & \frac{1}{2\sqrt{2}\tilde{m}_u^2}\left(\frac{\alpha_S}{\rho^3}-\frac{b}{2\rho}\right)
\boldsymbol{L}_{\rho}\cdot(\boldsymbol{S}_1+\boldsymbol{S}_2)\nonumber\\
&&-\frac{1}{4\sqrt{6}\tilde{m}_u^2}\left(\frac{2\alpha_S}{\sqrt{3}\rho^3}+\frac{b}{\rho}\right)\boldsymbol{\rho}\times \boldsymbol{p}_{\lambda}\cdot(\boldsymbol{S}_1-\boldsymbol{S}_2).
\end{eqnarray}
In the above equation, the first term is the two-body spin-orbit potential which is widely adopted in the literature,
while the second term proportional to $\boldsymbol{\rho}\times \boldsymbol{p}_{\lambda}$ just corresponds to the three-body spin-orbit potential~\cite{Isgur:1978xj,Capstick:1986ter}.
Unfortunately, this term was often neglected in the previous study of the baryon spectrum.
In our calculations, it is found that the contributions from the OGE two-body spin-orbit potentials nearly cancel out those
from the OGE tensor potentials, thus, the configuration mixing between the two low-lying $J^P=3/2^-$ states
is mainly caused by the three-body spin-orbit potentials.

The sizeable configuration mixing between $N(1520)3/2^{-}$ and $N(1700)3/2^{-}$ seems to be needed for reasonably describing
the reactions $\gamma p\to \eta p$ in the resonances region~\cite{Zhong:2011ti}.
By fitting the reaction data, the mixing angle is determined to be
$\theta_D \simeq 21^{\circ}$~\cite{Zhong:2011ti}, which is consistent
with $\theta_D \simeq 20^{\circ}$ obtained with the three-body OGE spin-orbit potentials in the present work.

With the mixing angle $\theta_D \simeq 20^{\circ}$,
the strong decay properties of the $N(1520)3/2^{-}$ and $N(1700)3/2^{-}$
are evaluated within the chiral quark model. Our results together with
the experimental data and some other model predictions are given in
Table~\ref{Low mass nucleon decay widths}. It is seen that our predictions
are in good agreement with the data. The predicted width
of $N(1520)3/2^{-}$, $\Gamma\simeq57$ MeV, is much narrower (about a factor of 3)
smaller than $\Gamma\simeq180$ MeV predicted for $N(1700)3/2^{-}$.
The $N\pi$ and $\Delta\pi$ are the two important decay channels of both $N(1520)3/2^{-}$ and $N(1700)3/2^{-}$.
The partial width ratios between the $N\pi$ and $\Delta\pi$ for these two states are predicted to be
\begin{eqnarray}\label{mix-d13}
R_1^{3/2^-} &=&\frac{\Gamma[N(1520)\to N\pi]}{\Gamma[N(1520)\to \Delta\pi ]}\simeq 5.7,\\
R_2^{3/2^-} &=&\frac{\Gamma[N(1700)\to N\pi]}{\Gamma[N(1700)\to \Delta\pi ]}\simeq 0.26,
\end{eqnarray}
which are close to the upper limit of the estimated ratios $\sim1.6-3.0$, $\sim0.08-0.30$, respectively from the PDG~\cite{ParticleDataGroup:2022pth}.
In should be mentioned that these ratios predicted from various works are very different,
which can seen from Table~\ref{Low mass nucleon decay widths}.
From this table, it is also found that the ratio $R_2^{3/2^-}$ is sensitive to the mixing angle. Thus,
a precise measurement of $R_2^{3/2^-}$ is very useful for not only determining the mixing angle, but also testing
the various predictions in the literature.

As a whole, within the constituent quark model, both the mass and strong decay properties
of the $N(1520)3/2^{-}$ and $N(1700)3/2^{-}$ can be reasonably understood.
The three-body spin-orbit potential arising from the OGE can cause a fairly large mixing angle
$\theta_D\simeq 20^\circ$ for these two resonance states.
To test the large mixing angle due to the three-body spin-orbit potentials, an accurate
measurement of the partial width decay ratio between $N\pi$ and $\Delta\pi$
for the $N(1700)3/2^{-}$ may be helpful.

\begin{figure}[htbp]
	\centering \epsfxsize=7.5cm \epsfbox{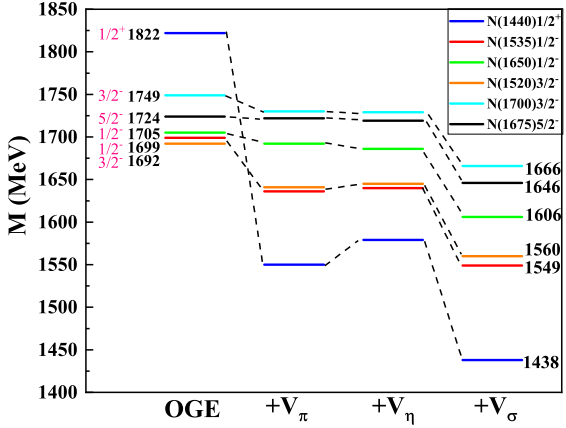} \vspace{-0cm}
\caption{Illustration of the dynamic mechanism for the
mass reversal between $N(1440)1/2^{+}$ and the $1P$-wave nucleon states. In the figure, four scenarios are shown.
From left to right, OGE stands for the case with the OGE potentials only,
while $+V_{\pi}$, $+V_{\eta}$, and $+V_{\sigma}$ stand for the cases that the $\pi$-, $\eta$-, and $\sigma$-exchange
potentials are further included one by one.}\label{fig:NV_GBE}
\end{figure}

\subsubsection{$N(1440)1/2^+$ and mass reversal problem}

In the quark model, the Roper resonance $N(1440)1/2^{+}$ is usually assigned as the first radially excited state $^{2}8[56, 0_2^+]2^2S_{1/2^+}$. The measured mass of $N(1440)1/2^{+}$ is about 90 MeV smaller than the first orbitally excited state $N(1535)1/2^{-}$.
However, in the common quark potential model based on OGE,
the predicted mass of the $N(1440)1/2^{+}$ is about $80$ MeV larger than the $N(1535)1/2^{-}$.
This is the well-known ``mass reversal problem''.

In the present work, it is found that the so-called ``mass reversal problem" can be naturally overcome by including the OBE interactions
together with the OGE potentials. It should be emphasized that
the OBE interactions adopted for the mass spectrum calculations are
described within the same framework, i.e. the chiral quark model, for dealing with
the strong decays. The parameters in the OBE potentials have been well determined
by the strong decay properties of the baryon resonances.
The dynamic mechanism of mass reversal is explained as follows.

From Fig.~\ref{fig:NV_GBE}, it is clearly seen that the one-pion-exchange (OPE) potential
is responsible for the mass reversal between the $1P$ states with negative parity
and the Roper resonance $N(1440)1/2^{+}$. Our conclusion is consistent
with that based on the Goldstone-boson exchange interaction models~\cite{Glozman:1995fu,Melde:2008yr,Ghalenovi:2017xxv},
and the lattice calculation of valence QCD~\cite{Liu:1998um,Liu:2014jua,Liu:2016rwa}.
In Table~\ref{nucleon mass spectrum contribution}, the contributions of
each part of the Hamiltonian to the baryon mass are given. One can see that the center
part of the OPE potential decreases the mass of $N(1440)1/2^{+}$ with a fairly
large value of $\sim500$ MeV, however, for the $1P$ states, the mass reduction due to
OPE interactions is no more than 70 MeV. This leads to a mass reversal between the $N(1440)1/2^{+}$ and the $1P$ states.
On the other hand, one can find that in the $1P$ states there is an
obvious gap of about 100 MeV between the $|70, ^{2}8\rangle$ (dominant)
states $N(1535)1/2^{-}$ and $N(1520)3/2^{-}$ and the $|70, ^{4}8\rangle$ (dominant)
states $N(1650)1/2^{-}$, $N(1700)3/2^{-}$, and $N(1675)5/2^{-}$ due to the center part of the OPE interactions.
Furthermore, $\eta$ and $\sigma$ exchanged potentials also have some contributions.
From Fig.~\ref{fig:NV_GBE} and Table~\ref{nucleon mass spectrum contribution},
one can see that the center part of the $\eta$-exchanged potential increases
the mass of $N(1440)1/2^{+}$ with a sizeable value of $\sim30$ MeV,
while its contributions to the $1P$ states are tiny.
The center part of the $\sigma$-exchanged potential causes an overall
shift for $N(1440)1/2^{+}$ and the $1P$ states to the low mass region with a value of
about $60-150$ MeV.

\begin{table}
\begin{center}
\tabcolsep=0.08cm
\caption{\label{1440 matrix element} The spin-isospin factors of the matrix elements contributed
by the chiral potentials for the $2S$ nucleon states. }
\begin{tabular}{c|ccc|ccc|ccc}\hline\hline
                                          &\multicolumn{3}{|c|}{$V_{\pi}^C$}  &\multicolumn{3}{|c|}{$V_{\eta}^C$}&\multicolumn{3}{|c}{$V_{\sigma}^C$}\\
                                          &$\langle \hat{o} \rangle_{1,2} $&$\langle \hat{o} \rangle _{3,4}$&$\langle \hat{o} \rangle _{5}$
                                          &$\langle \hat{o} \rangle_{1,2} $&$\langle \hat{o} \rangle _{3,4}$&$\langle \hat{o} \rangle _{5}$
                                          &$\langle \hat{o} \rangle_{1,2} $&$\langle \hat{o} \rangle _{3,4}$&$\langle \hat{o} \rangle _{5}$\\
                                          \hline
$|56, ^{2}8, \rangle\frac{1}{2}^{+}$  &5/2 &5/2  &0      &-1/2 &-1/2&0     &1/2 &1/2  &0  \\
$|70, ^{2}8, \rangle\frac{1}{2}^{+}$  &5/4 &-5/4 &-3/2   &1/4  &-1/4&-1/2  &1/4 &-1/4 &1/2\\
$|70, ^{4}8, \rangle\frac{3}{2}^{+}$  &1/4 &-1/4 &-3/2   &-1/4 &1/4 &1/2   &1/4 &-1/4 &1/2\\
\hline\hline
\end{tabular}
\end{center}
\end{table}

To better understand the large contribution of the OPE interactions to the mass of $^{2}8[56, 0_2^+]2^2S_{1/2^{+}}$
($N(1440)1/2^{+}$), as an example we further carry out an analysis of the matrix elements of the OBE potentials for
the $2S$ states $^{2}8[56, 0_2^+]2^2S_{1/2^{+}}$,
$^{2}8[70, 0_2^+]2^2S_{1/2^{+}}$, and $^{4}8[70, 0_2^+]2^2S_{3/2^{+}}$.
The matrix elements of the center parts of OBE potentials can be decomposed
into the product of the spatial integral part and the spin-isospin factor contributed by
the $(\boldsymbol{S}_i\cdot\boldsymbol{S}_j)(\lambda_{i}^{a}\cdot\lambda_{j}^{a})$ operator in
the spin-flavor space. For the $2S$ states, in the spatial integral part there are five terms
\begin{eqnarray}\label{abc}
\langle o\rangle_1 & \equiv & \langle\psi^{\rho}_{100}(d_{\ell})\psi^{\lambda}_{000}(d_{\ell})|V^C_{\pi/\eta/\sigma}
|\psi^{\rho}_{100}(d_{\ell'})\psi^{\lambda}_{000}(d_{\ell'})\rangle, \nonumber\\
\langle o\rangle_2 &\equiv & \langle\psi^{\lambda}_{100}(d_{\ell})\psi^{\rho}_{000}(d_{\ell})|V^C_{\pi/\eta/\sigma}
|\psi^{\lambda}_{100}(d_{\ell'})\psi^{\rho}_{000}(d_{\ell'})\rangle,\nonumber\\
\langle o\rangle_3 &\equiv & \langle\psi^{\lambda}_{100}(d_{\ell})\psi^{\rho}_{000}(d_{\ell})|V^C_{\pi/\eta/\sigma}
|\psi^{\rho}_{100}(d_{\ell'})\psi^{\lambda}_{000}(d_{\ell'})\rangle,\nonumber\\
\langle o\rangle_4 &\equiv & \langle\psi^{\rho}_{100}(d_{\ell})\psi^{\lambda}_{000}(d_{\ell})|V^C_{\pi/\eta/\sigma}
|\psi^{\lambda}_{100}(d_{\ell'})\psi^{\rho}_{000}(d_{\ell'})\rangle,\nonumber\\
\langle o\rangle_5 &\equiv & \langle\psi^{\lambda}_{01m}(d_{\ell})\psi^{\rho}_{01m'}(d_{\ell})|V^C_{\pi/\eta/\sigma}
|\psi^{\rho}_{01m''}(d_{\ell'})\psi^{\lambda}_{01m'''}(d_{\ell'})\rangle.\nonumber
\end{eqnarray}
The spin-isospin factors corresponding to these integral terms are given in Table~\ref{1440 matrix element}.
From the table, it is seen that large contribution of the OPE potential to the $N(1440)1/2^{+}$
is due to the two aspects: (i) the spin-isospin factors of $\pi$ exchange are much larger than those for the other two states;
(ii) the different integral terms contributed to the matrix element are constructive due to the same sign
of the spin-isospin factors, however, for the other two $2S$ states the matrix element contributed
by $\langle o\rangle_{1,2}$ has a large cancellation with the $\langle o\rangle_{3,4}$ term
due to the opposite signs of the spin-isospin factors.

It should be mentioned that in Ref.~\cite{Zhao:2007hz},
the authors also studied the nucleon spectrum within a similar potential model
including both the OGE and OBE potentials. Unfortunately, they did not obtain a
successful explanation of the low-mass nature of $N(1440)1/2^+$. The reason is because
in their calculations, the cross-terms $\langle o\rangle_{3,4}$ for $N(1440)1/2^{+}$ were neglected as an approximation. These terms are found important here and can account for the mysterious mass reversal problem between the $2S$ and $1P$ states.

The strong decay properties of the $N(1440)1/2^{+}$ resonance
can be well understood with the chiral dynamics as well.
Within the chiral quark model, the decay width of $N(1440)1/2^{+}$ is predicted to be
\begin{eqnarray}\label{N(1440) decay width}
\Gamma\simeq 389~\mathrm{MeV},
\end{eqnarray}
which is consistent with the PGD value $\Gamma_{exp}=350\pm 100$ MeV~\cite{ParticleDataGroup:2022pth}.
The decays are governed by the $N\pi$ channel with a branching fraction of $\sim 90\%$,
which is slightly larger than the observation $55-75\%$~\cite{ParticleDataGroup:2022pth}. Furthermore, there is a
sizeable decay rate into the $\Delta\pi$ channel, the branching fraction is predicted
to be $\sim 7\%$, which is comparable with the experimental data $6-27\%$~\cite{ParticleDataGroup:2022pth}.
It should be mentioned that the relativistic correction term
$\mathbf{{\cal H}}^{RC}$ plays a crucial role in
the $N(1440)\to N\pi$ decay process. If only with the nonrelativistic term $\mathbf{{\cal H}}^{NR}$, the partial width of
$\Gamma[N(1440)\to N\pi]$ will be notably (a factor of $\sim 2$) underestimated.
In the recent studies, it is found that the $\mathbf{{\cal H}}^{RC}$ is crucial not only for describing the strong decays of the
Roper(-like) baryon states~\cite{Arifi:2022ntc,Arifi:2021orx}, but also for describing the strong decays of
the first radially excited heavy-light meson states~\cite{Ni:2023lvx}.

As a whole with the chiral dynamics, one can not only explain the
mass reversal between $N(1440)1/2^{+}$ and the $1P$-wave nucleon states, but also understand their
strong decay properties (see Table~\ref{Low mass nucleon decay widths}). The center part of the one-pion-exchange interaction
is responsible for the mass reversal.
%There still exist some puzzles about
%the mass order for the $1P$-wave nucleon states. For example, in theory the mass
%of $N(1535)1/2^{-}$ is slightly (about 10 MeV) smaller than $N(1520)3/2^{-}$,
%which should be further confirmed in experiments.

\subsubsection{Positive parity nucleon resonances around 1.7 GeV}

In the mass range of 1.7 GeV, there are three positive parity nucleon resonances
$N(1710)1/2^+$, $N(1720)3/2^+$, and $N(1680)5/2^+$ listed in the RPP~\cite{ParticleDataGroup:2022pth}. According to the observed mass and parity,
they may be candidates of some low-lying states in the $N=2$ shell predicted in the quark model.

Commonly, the four-star resonance $N(1710)1/2^+$ is assigned
to the $2S$-wave state $^{2}8[70, 0_2^+]2^2S_{1/2^{+}}$
classified in the quark model. This state represents a new feature:
its wave function contains components of both orbital and radial excitations, while in the orbital excitation component the two
angular momenta $l_{\lambda}$ and $l_{\rho}$ are both one and couple to zero.
With this assignment, our predicted mass
\begin{eqnarray}\label{N(1710) mass}
M\simeq1824~\mathrm{MeV},
\end{eqnarray}
is about 80 MeV larger than the upper
limit of the observed value $M_{exp}=1710\pm 30$ MeV~\cite{ParticleDataGroup:2022pth}. From Table~\ref{nucleon mass spectrum comparison}, it is seen that the mass of $N(1710)1/2^+$ is also overestimated by $\sim50-90$ MeV in most of the quark models, e.g., Refs.~\cite{Capstick:1986ter,Chen:2009de,Melde:2008yr,Ghalenovi:2017xxv,Ferretti:2011zz}.
On the other hand, the strong decay properties are also studied,
our results have been given in Table~\ref{Low mass nucleon decay widths}. Our predicted width
\begin{eqnarray}\label{N(1710) width}
\Gamma\simeq168~\mathrm{MeV},
\end{eqnarray}
is comparable with the width $\Gamma_{exp}=140\pm 60$ MeV extracted from the reaction data~\cite{ParticleDataGroup:2022pth}.
From Table~\ref{Low mass nucleon decay widths}, one can see that the $^{2}8[70, 0_2^+]2^2S_{1/2^{+}}$ state mainly decays into
the $N(1440)\pi$, $N(1535)\pi$, $N\pi$, and $\Delta\pi$ channels with branching fractions
$\sim 55\%$, $\sim 31\%$, $6\%$, and $5\%$, respectively. Our predicted branching fractions
for the $N(1535)\pi$, $N\pi$, and $\Delta\pi$ channels are consistent with
the observations~\cite{ParticleDataGroup:2022pth}. The large decay rate of $N(1710)1/2^+\to N\pi\pi$ ($14-48\%$)
observed in experiments may be mainly contributed by the $N(1440)\pi$ and $N(1535)\pi$
channels via cascade decays. Our predicted branching fraction for the $K\Lambda$ channel is a tiny
value $\sim 0.3\%$, which is consistent with the result $1.8\pm 1.5\%$ extracted
by using an updated multichannel energy-dependent partial-wave analysis method~\cite{Hunt:2018wqz}.
However, our predicted branching fraction for
the $\eta N$ channel, $\sim 3\%$, is notably smaller than the experimental data $10-50\%$~\cite{ParticleDataGroup:2022pth}.

The four-star resonance $N(1720)3/2^+$ is usually assigned
to the $1D$-wave state $^{2}8[56, 2_2^+]1^2D_{3/2^{+}}$
classified in the quark model. With this assignment, our predicted mass
\begin{eqnarray}\label{N(1720) mass}
M\simeq1817~\mathrm{MeV},
\end{eqnarray}
is about 60 MeV larger than the upper limit of the observed mass $M_{exp}=1715\pm 35$ MeV.
The mass of $N(1720)3/2^+$ is often overestimated in the literature, e.g.,
Refs.~\cite{Capstick:1986ter,Ferretti:2011zz}. Furthermore,
our predicted width
\begin{eqnarray}\label{N(1720) decay width}
\Gamma\simeq136~\mathrm{MeV},
\end{eqnarray}
is in the measured range of $\sim150-400$ MeV~\cite{ParticleDataGroup:2022pth}.
The details of the strong decay properties have been given in
Table~\ref{Low mass nucleon decay widths}. According to our predictions,
the $^{2}8[56, 2_2^+]1^2D_{3/2^{+}}$ mainly decays into
the $N(1440)\pi$, $N(1520)\pi$, and $N\pi$ channels with branching fractions
$\sim 65\%$, $\sim 13\%$, and $15\%$, respectively. While the decay rates into the $\Delta\pi$, $N\eta$,
and $K\Lambda$ channels are on the order of $1\%$. Our predicted branching
fractions for the $N\pi$, $N\eta$, and $K\Lambda$ channels are in
agreement with the observations~\cite{ParticleDataGroup:2022pth}. However, the observed branching fractions,
$47-89\%$ and $<2\%$, for both the $\Delta\pi$ and $N(1440)\pi$ channels~\cite{ParticleDataGroup:2022pth}
are inconsistent with our predictions. The large decay rate of $N(1720)3/2^+\to N\pi\pi$ ($>50\%$)
observed in experiments may be mainly contributed by the $N(1440)\pi$
channel via cascade decays.

The four-star resonance $N(1680)5/2^+$ listed in RPP might correspond
to the low-mass mixed state via the $^{2}8[56, 2_2^+]1^2D_{5/2^{+}}$-$^{2}8[70, 2_2^+]1^2D_{5/2^{+}}$ mixing (see Table~\ref{nucleon mass spectrum contribution}).
Their mixing is mainly induced by the spin-spin interactions.
Our predicted mass
\begin{eqnarray}\label{N(1680a)}
M\simeq1775~\mathrm{MeV},
\end{eqnarray}
is slightly ($\sim80$ MeV) larger than
the PDG average data $M_{exp}=1685\pm5$ MeV~\cite{ParticleDataGroup:2022pth}. When neglecting the
contributions from the $N\rho$, $N\sigma$, and $N\pi\pi$ channels,
our predicted width
\begin{eqnarray}\label{N(1680b)}
\Gamma\simeq46~\mathrm{MeV},
\end{eqnarray}
is about one half of the observed one $\Gamma_{exp}\simeq115-130$ MeV~\cite{ParticleDataGroup:2022pth}.
The details of the strong decay properties have been given in
Table~\ref{Low mass nucleon decay widths}. According to our predictions,
the $N\pi$ should be the dominant channel of the $N(1680)5/2^+$,
while its decay rates in the other channels, such as $\Delta\pi$, $N\eta$, and $K\Lambda$
are tiny. These predicted decay properties are consistent with the observations~\cite{ParticleDataGroup:2022pth}.

As a whole, assigning the $N(1710)1/2^+$, $N(1720)3/2^+$,
and $N(1680)5/2^+$ resonances to the low-lying nucleon excitations in the $N=2$ shell,
only a very rough description of their properties can be obtained in theory.
Our predicted masses are systematically larger than the center values of PDG~\cite{ParticleDataGroup:2022pth} by $\sim80$ MeV.
Some other dynamic mechanisms, such as coupled-channel effects, may be responsible for their low mass nature.
The $N(1710)1/2^+$ and $N(1720)3/2^+$ resonances may be better established
in the three-body final state $N\pi\pi$ via the cascade decays, since they
may dominantly decay into the $N(1440)\pi$ channel.

%It should be emphasized that there are still large uncertainties in the
%extracted resonance parameters for the $N(1710)1/2^+$ and $N(1720)3/2^+$ resonances, since it was never clearly
%seen in any reactions.

\subsubsection{Positive parity nucleon resonances around 1.9 GeV}

In the higher mass region around 1.9 GeV, there
are several positive parity nucleon resonances, such as $N(1860)5/2^+$, $N(1880)1/2^+$,
$N(1900)3/2^+$, and $N(1990)7/2^+$, listed in the RPP~\cite{ParticleDataGroup:2022pth}.
These states may be candidates of the nucleon excitations in the $N=2$ shell
predicted within the constituent quark models (see Table~\ref{nucleon mass spectrum comparison}).

The four-star resonance $N(1900)3/2^+$ is usually assigned to
the second orbital excitation of nucleon $^{2}8[70, 2_2^+]1^2D_{3/2^{+}}$.
With this assignment, our predicted mass and width
\begin{eqnarray}\label{N(1900)}
M\simeq1922~\mathrm{MeV},~~\Gamma\simeq222~\mathrm{MeV},
\end{eqnarray}
are consistent with the PDG values $M_{exp}\simeq1920\pm30$ MeV and
$\Gamma_{exp}\simeq210\pm110$ MeV, respectively~\cite{ParticleDataGroup:2022pth}. The details of our predicted decay properties have been given
Table~\ref{positive parity nucleon decay widths}. It is found that
the $N(1900)3/2^+$ as the $^{2}8[70, 2_2^+]1^2D_{3/2^{+}}$ assignment should
dominantly decays into the $N(1520)\pi$, $N(1440)\pi$ and $\Delta\pi$ with branching fractions
$\sim46\%$, $\sim28\%$ and $\sim11\%$, respectively. Large decay rates into $\Delta\pi$
and $N(1520)\pi$ have been observed in experiments, however,
the extracted branching fractions for the $\Delta\pi$ and $N(1520)\pi$ channels, $\sim30-70\%$ and $\sim7-23\%$~\cite{CBELSATAPS:2015kka}
are notably different from our predictions. The large decay rate of
$N(1900)3/2^+ \rightarrow N\pi\pi$ $ (> 56\%)$ observed in experiments~\cite{ParticleDataGroup:2022pth} may be mainly contributed by the $N(1440)\pi$ and $N(1520)\pi$
channels via cascade decays. Our predicted branching fractions for the $N\pi$ and $N\eta$ channels,
$\sim4.3\%$ and $\sim1.5\%$, are consistent with the values extracted from the reaction data~\cite{Hunt:2018wqz,CBELSATAPS:2015kka}.
It should be mentioned that the decay rates of $N(1900)3/2^+$ into
the $K\Lambda$ and $K\Sigma$ channels are predicted to be in the order of
$\mathcal{O}(10^{-3})$, which is about $1-2$ orders of magnitude
smaller than the PDG values~\cite{ParticleDataGroup:2022pth}.

The three-star resonance $N(1880)1/2^+$ may be assigned to
the excited nucleon state $^{4}8[70, 2_2^+]1^4D_{1/2^{+}}$ classified
in the quark model. According to our predictions, the $^{4}8[70, 2_2^+]1^4D_{1/2^{+}}$
slightly mixes with the $^{2}8[20, 1_2^+]1^2P_{1/2^{+}}$
(see Table~\ref{nucleon mass spectrum contribution}).
%With this assignment, both the mass and broad width of $N(1880)1/2^+$ can be well understood.
The mass and width are predicted to be
\begin{eqnarray}\label{N(1880)}
M\simeq1948~\mathrm{MeV},~~\Gamma\simeq636~\mathrm{MeV},
\end{eqnarray}
which are in good agreement with the determinations $M_{exp}\simeq1967\pm20$ MeV and
$\Gamma_{exp}\simeq500\pm70$ MeV from an updated multichannel energy-dependent
partial-wave analysis of $\pi N$ scattering in Ref.~\cite{Hunt:2018wqz}.
It should be mentioned that our predicted mass is about 60 MeV larger than
the PDG averaged mass, $\sim1880$ MeV~\cite{ParticleDataGroup:2022pth}, while our predicted width
is much broader than the PDG value $\sim200-400$ MeV~\cite{ParticleDataGroup:2022pth}.
From Table~\ref{positive parity nucleon decay widths}, one find that
the $N(1880)1/2^+$, as the $^{4}8[70, 2_2^+]1^4D_{1/2^{+}}$
dominant state, mainly decays into the $\Delta\pi$, $\Delta(1620)\pi$,
and $N(1535)\pi$ channels with branching fractions $\sim46\%$,
$\sim33\%$, and $\sim16\%$, respectively, while the branching
fractions for the $N\pi$, $N\eta$, and $K\Sigma$ channels are
predicted to be in the range of $\sim0.5-1.0\%$. Most of the
predicted branching fractions are in the PDG ranges.

The two-star resonance $N(1860)5/2^+$ may be experimental evidence for
the mixed state dominated by the $^{2}8[70, 2_2^+]1^2D_{5/2^{+}}$
(see Table~\ref{nucleon mass spectrum contribution}).
For this state, the mass and width are predicted to be
\begin{eqnarray}\label{N(1860a)}
M\simeq1944~\mathrm{MeV}, ~\Gamma\simeq70~\mathrm{MeV},
\end{eqnarray}
which are comparable with the mass $M_{exp}\simeq1882\pm10$ MeV and
$\Gamma_{exp}\simeq95\pm20$ MeV determined by the partial wave analysis of
the $\pi N$ scattering data~\cite{Hohler:1979yr}.
Our predicted width is also close to the lower limit of the width $\Gamma_{exp}\simeq122\pm 41$ MeV
extracted by using the Laurent+Pietarinen method from the $\pi N\to \pi N$ reaction~\cite{Svarc:2014zja}.
It should be mentioned that there exist large uncertainties in the
extracted masses and widths from different PWA groups. A very broad width
$\Gamma_{exp}\simeq376\pm 58$ MeV was extracted by the PWA group of Kent State University~\cite{Hunt:2018wqz}.
According to our predicted decay properties given in Table~\ref{positive parity nucleon decay widths},
the $N(1860)5/2^+$ as the $^{2}8[70, 2_2^+]1^2D_{5/2^{+}}$
assignment should dominantly decay into the $\Delta\pi$ channel with a
branching fraction $\sim58\%$, furthermore, there are sizeable decay rates into the
$N\pi$, $N\eta$, $N(1520)\pi$, $N(1675)\pi$ and $N(1680)\pi$ with a
comparable branching fraction $\sim1-12\%$.
Our predicted branching fractions for $N(1860)5/2^+\to \Delta\pi,N\pi,N\eta$ ($\sim58\%$, $\sim1\%$, and $\sim3\%$) are
consistent with the PDG values~\cite{ParticleDataGroup:2022pth}.

The two-star resonance $N(2000)5/2^+$ may be experimental evidence for the $^{4}8[70, 2_2^+]1^4D_{5/2^{+}}$ state.
Assigning the $N(2000)5/2^+$ to this state, the pole
mass $M_{exp}\simeq1900$ MeV and width $\Gamma_{exp}\simeq123$ MeV
extracted from the pion- and photon-induced reactions with the coupled-channel PWA method
~\cite{Shklyar:2012js} are consistent with our
predicted mass and width
\begin{eqnarray}\label{N(2000)}
M\simeq1962~\mathrm{MeV},~~\Gamma\simeq143~\mathrm{MeV},
\end{eqnarray}
respectively. It should be mentioned that a larger mass, $2030\pm 30$ MeV, and a
broader width, $380\pm 60$ MeV, were extracted by the CBELSA/TAPS Collaboration
from the $\gamma p\to p\pi^0\pi^0$ reaction~\cite{CBELSATAPS:2015kka}.
According to our predicted decay properties
given in Table~\ref{positive parity nucleon decay widths},
the $N(2000)5/2^+$ as the $^{4}8[70, 2_2^+]1^4D_{5/2^{+}}$
assignment should dominantly decay into the $\Delta\pi$ and $N(1675)\pi$ channels with
branching fractions $\sim31\%$ and $\sim62\%$, respectively.
The decay rates into the $N\pi$ and $N\eta$ are in the order of magnitude $\sim1\%$.
The branching fractions for $N(2000)5/2^+\to \Delta\pi,N\eta,N\pi$ predicted in theory
are comparable with the PDG values~\cite{ParticleDataGroup:2022pth}. The large decay rate ($35-90\%$)
of $N(2000)5/2^+\to N\pi\pi$ may be contributed by the cascade decays
$N(2000)5/2^+\to \Delta\pi,N(1675)\pi\to N\pi\pi$.

The two-star resonance $N(1990)7/2^+$ may correspond to the second
orbital excitation of nucleon $^{4}8[70, 2_2^+]1^4D_{7/2^{+}}$,
which is the only nucleon state with $J^P=7/2^+$ in the $N=2$ shell classified in the quark model.
This $J^P=7/2^+$ state should be a fairly narrow state. Its mass and width are
predicted to be
\begin{eqnarray}\label{N(1990)}
M\simeq1936~\mathrm{MeV},~~\Gamma\simeq63~\mathrm{MeV},
\end{eqnarray}
respectively.
As a two-star resonance rated by the PDG, the mass and
width of $N(1990)7/2^+$ are still not well constrained by experiments~\cite{ParticleDataGroup:2022pth}.
Recently, the pole position of the $N(1990)7/2^+$ determined with the J\"{u}lich-Bonn
dynamical coupled-channel model~\cite{Ronchen:2022hqk} changes notably compared
with the PDG values~\cite{ParticleDataGroup:2022pth}. The newly determined mass $M_{exp}=1861\pm 9$ MeV is
much smaller than the PDG value $2020\pm80$ MeV~\cite{ParticleDataGroup:2022pth},
while the newly determined width $\Gamma_{exp}=72\pm 5$ MeV is much more narrow than the PDG value $300\pm 100$ MeV~\cite{ParticleDataGroup:2022pth}.
Considering the $N(1990)7/2^+$ as the $^{4}8[70, 2_2^+]1^4D_{7/2^{+}}$ assignment,
both the mass and the narrow width determined in Ref.~\cite{Ronchen:2022hqk} are
reasonably consistent with our predictions.
According to our predicted strong decay properties given in
Table~\ref{positive parity nucleon decay widths}, the $N(1990)7/2^+$ should dominantly decay into
$\Delta\pi$ channel with a branching fraction $\sim 50\%$. Furthermore, the branching fractions
for the $N(1990)7/2^+\to N\pi, N\eta$ decays are estimated
to be $\sim 10\%$ and $\sim 6\%$, respectively, which are slightly larger than
the PDG values $2-6\%$ and $<3\%$~\cite{ParticleDataGroup:2022pth}.

As a whole, the $N(1880)1/2^+$, $N(1900)3/2^+$, $N(1860)5/2^+$, $N(2000)5/2^+$, and $N(1990)7/2^+$
may favor the assignments of excited nucleon states in the $N=2$ shell. However, in this shell
four quark model states $^{4}8[70, 0_2^+]2^4S_{3/2^{+}}$,
$^{4}8[70, 2_2^+]1^4D_{3/2^{+}}$, $^{2}8[20, 1_2^+]1^2P_{1/2^{+}}$
and $^{2}8[20, 1_2^+]1^2P_{3/2^{+}}$ are still missing.
These missing states may have a mass of $M\sim1.9$ GeV and a relatively narrow width
of $\Gamma\sim$10s-100 MeV. Their decay properties have been given in
Table~\ref{positive parity nucleon decay widths}.
These missing nucleon states very weakly couple to the $N\pi$, $N\eta$,
$K\Lambda$, and $K\Sigma$ channels, which can naturally explain
why they have not been observed in the $N\pi$, $N\eta$,
$K\Lambda$, and $K\Sigma$ final states via the $\pi N$
and $\gamma N$ scattering. It is interesting to find that
these missing nucleon states couple strongly to the $\Delta\pi$ channel,
they may have large potentials to be established in the $N\pi\pi$ final state by using the
charmonium cascade decays, such as $J/\psi$ and $\psi(2S)\to \bar{p}(\Delta\pi)_{N^*}\to \bar{p}(p\pi)_{\Delta}\pi$, at BESIII.

\subsection{$\Delta$ baryons}

\subsubsection{$\boldsymbol{\Delta(1232)}$}
%\subsection{1P states}

The $\Delta(1232)3/2^+$ resonance is the ground state
in the $\Delta$ baryon spectrum. Its mass can be well described in theory, however,
its partial decay width of $\Gamma[\Delta(1232) \to N\pi]$ is notably underestimated by a factor of $\sim2$ in various models (see Table~\ref{Delta decay widths}). This puzzle may arise from the phase-space factor what we adopt. By adopting an effective
phase space factor (EPSF) as suggested in Ref.~\cite{Kokoski:1985is}, the predicted width can be consistent with the data~\cite{Bijker:2015gyk}.
Similar puzzle also exists in the meson sector. For example, it is found that with the standard relativistic phase space
the predicted partial width of $\Gamma[\phi(1020)\to K\bar{K}]=2.5$ MeV in a previous work of our group is notably smaller than the measured
value $3.5$ MeV, with the EPSF this discrepancy is overcome~\cite{Li:2020xzs}.

\subsubsection{$\Delta(1620)1/2^-$ and $\Delta(1700)3/2^-$}

Commonly, the $\Delta(1620)1/2^-$ and $\Delta(1700)3/2^-$ resonances
are assigned to the two $1P$-wave states $^{2}10 [70, 1_1^-]1^2P_{1/2^-}$ and
$^{2}10 [70, 1_1^-]1^2P_{3/2^-}$, respectively. From Table~\ref{Delta mass spectrum comparison},
it is found that the masses of these two states
were predicted to be degenerate in many works, e.g. Refs.~\cite{Chen:2009de,Melde:2008yr,Ghalenovi:2017xxv,Ferretti:2011zz}. However, the mass splitting between $\Delta(1620)1/2^-$ and $\Delta(1700)3/2^-$ is a
fairly large value $\Delta M\simeq80$ MeV.

The three-body spin-orbit potentials arising from the OGE may be responsible for
the mass splitting between $\Delta(1620)1/2^-$ and $\Delta(1700)3/2^-$.
It is found that the masses of the two $1P$-wave states with
$J^P=1/2^-$ and $J^P=3/2^-$ are exactly degenerated when only the two-body spin-orbit potentials (i.e., the first term of Eq.~(\ref{Threebs})) are included. In Ref.~\cite{Capstick:1986ter}, Capstick and Isgur realized the importance of the three-body spin-orbit potentials
which are proportional to $\rho\times \boldsymbol{p}_\lambda$ (i.e., the second term of Eq.~(\ref{Threebs})).
Including the three-body spin-orbit potentials, the masses of
the two $1P$-wave states $J^P=1/2^-$ and $J^P=3/2^-$ are predicted to be
\begin{eqnarray}\label{N(16001/2) decay width}
M\simeq 1595,~1655~\mathrm{MeV},
\end{eqnarray}
respectively. Our predicted masses together with the mass splitting, $\Delta M\simeq60$ MeV, are consistent with
the experimental data~\cite{ParticleDataGroup:2022pth}. Similarly, the three-body spin-orbit potentials should cause
a notable mass splitting between the two $1P$-wave $\Omega$ states
$\Omega(1P)1/2^-$ and $\Omega(1P)3/2^-$, which can be further test in future experiments.

Furthermore, our predicted decay properties have been given in Table~\ref{Delta decay widths}.
It is seen that the $\Delta(1620)1/2^-$ dominantly decays into the $\Delta\pi$ channel,
while the decay rate into the $N\pi$ channel is also sizeable. Only including the contributions from the $N\pi$
and $\Delta\pi$ channels, our predicted width of $\Delta(1620)1/2^-$
\begin{eqnarray}\label{N(16001/2) decay width}
\Gamma\simeq 46~\mathrm{MeV},
\end{eqnarray}
is comparable with those predictions with the
$^3P_0$ model in Ref.~\cite{Bijker:2015gyk}, however, is a factor of $\sim3$ smaller than the PDG averaged data $\Gamma_{exp}=130\pm 20$ MeV~\cite{ParticleDataGroup:2022pth}.
This disparity may be notably reduced if the contributions from
the $N\rho$ channel are included. For the $\Delta(1620)1/2^-$, the partial width ratio between the $N\pi$
and $\Delta\pi$ channels is predicted to be
\begin{eqnarray}\label{N(16001/2) decayb}
\frac{\Gamma[\Delta(1620)1/2^-\to \Delta\pi]}{\Gamma[\Delta(1620)1/2^-\to N\pi]}\simeq 1.9,
\end{eqnarray}
which is consistent with the ratio $\sim2.0$ extracted from the reaction data
by the KSU group~\cite{Hunt:2018wqz} and CBELSA/TAPS Collaboration~\cite{CBELSATAPS:2015kka}.

The $\Delta(1700)3/2^-$ resonance mainly decays into the $\Delta\pi$ and $N\pi$
channels. Its width is predicted to be
\begin{eqnarray}\label{N(16001/2) decay width}
\Gamma\simeq 138~\mathrm{MeV},
\end{eqnarray}
which is consistent with the value $\Gamma_{exp}=119\pm70$ MeV extracted from $\pi N$
data by the Pitt-ANL group~\cite{Vrana:1999nt}, however, is much narrower
than the PDG value $\Gamma_{exp}=300\pm80$ MeV~\cite{ParticleDataGroup:2022pth} and the predictions within the
$^3P_0$ model in Ref.~\cite{Bijker:2015gyk} (see Table~\ref{Delta decay widths}).
The partial width ratio between the $N\pi$
and $\Delta\pi$ channels is predicted to be
\begin{eqnarray}\label{N(17003/2) decayb}
\frac{\Gamma[\Delta(1700)3/2^-\to \Delta\pi]}{\Gamma[\Delta(1700)3/2^-\to N\pi]}\simeq 4.2,
\end{eqnarray}
which is consistent with the value extracted by the KSU group in their recent work~\cite{Hunt:2018wqz}.

As a whole, the masses and widths of the $\Delta(1620)1/2^-$
and $\Delta(1700)3/2^-$ can be reasonably understood by assigning them to $1P$-wave $\Delta$ baryons.
The three-body spin-orbit potentials arising from the OGE, which are often neglected in the literature,
may be responsible for the splitting between the two $1P$-wave states.

\subsubsection{$\Delta(1600)3/2^+$}

The $\Delta(1600)3/2^+$ is often assigned to the first radially excited $\Delta$ resonance
with $J^P=3/2^+$, $^{4}10 [56, 0_2^+]2^4S_{1/2^+}$.
The mass of this state is predicted to be
\begin{eqnarray}\label{N(1686) decay width}
M\simeq 1694~\mathrm{MeV},
\end{eqnarray}
which is close to the upper limit of the PDG value $M_{exp}=1570\pm70$ MeV~\cite{ParticleDataGroup:2022pth}.

Within our chiral quark model, taking the mass in the range of $1570\pm70$ MeV, the width of the $\Delta(1600)3/2^+$ is
estimated to be
\begin{eqnarray}\label{N(1600) decay width}
\Gamma\simeq 101^{+58}_{-44}~\mathrm{MeV},
\end{eqnarray}
which is consistent with the recently extracted data $\Gamma_{exp}\simeq 136$ MeV
from a PWA by including $K\Sigma$ photoproduction~\cite{Ronchen:2022hqk} within uncertainties, however,
it is about a factor of $2$ narrower than the PDG averaged width $\Gamma_{exp}=250\pm50$ MeV~\cite{ParticleDataGroup:2022pth}.
The $\Delta(1600)3/2^+$ should mainly decay into the $N\pi$ and $\Delta\pi$ channels.
The partial width of the $\Delta\pi$ channel strongly depend on the mass of $\Delta(1600)3/2^+$ what we adopt.
In the mass range of $M=1570\pm70$ MeV, the partial width ratio between the $N\pi$ and $\Delta\pi$ channels is predicted to be
\begin{eqnarray}\label{N(16003/2) decayb}
\frac{\Gamma[\Delta(1600)3/2^+\to \Delta\pi]}{\Gamma[\Delta(1600)3/2^+\to N\pi]}\simeq 2-6,
\end{eqnarray}
which is consistent with the PDG estimated range~\cite{ParticleDataGroup:2022pth}. The precise measurement of this ratio
may be helpful to accurately determine the mass of $\Delta(1600)3/2^+$ from the scattering data.

There may also exist a mass reversal problem between the radially excited state
$\Delta(1600)3/2^+$ and the orbitally excited state $\Delta(1620)1/2^-$.
Some PWA groups extracted a relatively small mass, $\sim1500$ MeV, for the $\Delta(1600)3/2^+$,
which is about 100 MeV smaller than that extracted for the $\Delta(1620)1/2^-$~\cite{CBELSATAPS:2015kka,Anisovich:2011fc}.
However, in most studies in theory the radially excited $\Delta$ state
should have a larger mass than $\Delta(1620)1/2^-$ (see Table~\ref{Delta mass spectrum comparison}).
In the present work, it is predicted that the $\Delta(1600)3/2^+$ as the radially excited state
$^{4}10 [56, 0_2^+]2^4S_{1/2^+}$ should lie $\sim80$ MeV above $\Delta(1620)1/2^-$.
It should be mentioned that our result is consistent with that extracted by the KSU group in their recent work~\cite{Hunt:2018wqz}.

As a whole, the still exist large uncertainties in the determined mass and width
for the $\Delta(1620)1/2^-$ resonance. To better understand the properties
of the $\Delta(1620)1/2^-$, more accurate measurements of the resonance
parameters and the partial width ratio between the $N\pi$ and $\Delta\pi$ channels are needed.

\subsubsection{Positive parity $\Delta$ resonances around 1.9 GeV}

In the higher mass range around 1.9 GeV, there
are several positive parity resonances $\Delta(1910)1/2^+$, $\Delta(1920)3/2^+$,
$\Delta(1905)5/2^+$, $\Delta(2000)5/2^+$, and $\Delta(1950)7/2^+$ listed in the RPP~\cite{ParticleDataGroup:2022pth}.
These states just lie in the mass range of the $1D$-wave $\Delta$ states
predicted within the constituent quark models (see Table~\ref{Delta mass spectrum comparison}).

The four-star resonance $\Delta(1910)1/2^+$ has mass and width are $M_{exp}=1900\pm 50$ MeV and $\Gamma_{exp}=300\pm 100$ MeV,
respectively~\cite{ParticleDataGroup:2022pth}. It may be a candidate of
the $^{4}10 [56, 2_2^+]1^4D_{1/2^+}$ state classified in the quark model.
With this assignment, both our predicted mass and width
\begin{eqnarray}\label{delta(1910)}
M\simeq1913~\mathrm{MeV},~~\Gamma\simeq369~\mathrm{MeV},
\end{eqnarray}
are consistent with the observations~\cite{ParticleDataGroup:2022pth}. The branching fractions
for the $N\pi$ and $N(1440)\pi$ channels are predicted to be
$\sim5\%$ and $\sim45\%$, which are close to the observed ranges $10-30\%$ and
$3-45\%$, respectively~\cite{ParticleDataGroup:2022pth}. However, our predicted branching fraction, $\sim 1\%$,
for the $\Delta\pi$ channel is too tiny to be comparable with the observed value
$34-66\%$~\cite{ParticleDataGroup:2022pth}. Furthermore, according to our predictions (see Table~\ref{Delta decay widths}),
the $\Delta(1910)1/2^+$ may have large decay rates into the
$N(1650)\pi$ and $\Delta(1620)\pi$ channels with branching fractions
$\sim16\%$ and $\sim31\%$, respectively.
These branching fractions are waiting to be test in future experiments.

The four-star resonance $\Delta(1950)7/2^+$ listed in RPP~\cite{ParticleDataGroup:2022pth} should be assigned
to the quark model state $^{4}10 [56, 2_2^+]1^4D_{7/2^+}$.
With this assignment, both our predicted mass and width
\begin{eqnarray}\label{delta(1910)}
M\simeq1867~\mathrm{MeV},~~\Gamma\simeq106~\mathrm{MeV},
\end{eqnarray}
are consistent with the pole mass $M_{exp}\simeq1875$ MeV
and width $\Gamma_{exp}\simeq 166$ MeV determined by the $\pi N$ and $\gamma N$ reaction dada with the J\"{u}lich-Bonn
dynamical coupled-channel approach~\cite{Ronchen:2022hqk}.
However, our predicted width is about a factor of $2$ narrower than the measured width
$\Gamma_{exp}=285\pm 50$ MeV and the predicted values
within the $^3P_0$ model~\cite{Bijker:2015gyk}.
The strong decay properties have been given in Table~\ref{Delta decay widths}. It is seen
that the $\Delta(1950)7/2^+$ as the $^{4}10 [56, 2_2^+]1^4D_{7/2^+}$ state
may mainly decays into the $N\pi$ and $\Delta\pi$ channels with branching
fractions $\sim48\%$ and $\sim35\%$, respectively.
Our predicted branching fraction for the $N\pi$ channel is in the
range of $35-45\%$ estimated by the PDG~\cite{ParticleDataGroup:2022pth}. The large decay rate
into the $N\pi\pi$ channel, $57\pm20\%$, determined from the
$\gamma N\to \pi^+\pi^-p$ reaction~\cite{CLAS:2018drk} may be mainly contributed by
the $\Delta\pi$ channel via a cascade decay.

The four-star resonance $\Delta(1905)5/2^+$
and two-star resonance $\Delta(2000)5/2^+$ listed in the RPP~\cite{ParticleDataGroup:2022pth} may relate
to the $^{4}10 [56, 2_2^+]1^4D_{5/2^+}$
and $^{2}10 [70, 2_2^+]1^2D_{5/2^+}$ states.
Our study shows that these two quark model
configurations highly mix with each other with a mixing angle
$\theta_{1D}\simeq 38^\circ$ mainly due to the spin-dependent interactions,
which is consistent with the prediction in Ref.~\cite{Capstick:1986ter}.
For the low-mass mixed state, the mass and width are predicted to be
\begin{eqnarray}\label{delta(1905)}
M\simeq1864~\mathrm{MeV},~~\Gamma\simeq56~\mathrm{MeV}.
\end{eqnarray}
It dominantly decays into the
$\Delta \pi$ and $N\pi$ channel with branching fractions $\sim 62\%$ and $\sim 23\%$, respectively.
While for the high-mass mixed state, the mass and width are predicted to be
\begin{eqnarray}\label{delta(1966)}
M\simeq1975~\mathrm{MeV},~~\Gamma\simeq127~\mathrm{MeV}.
\end{eqnarray}
This high-mass state mainly decays into the $\Delta \pi$ and $N(1675) \pi$ channels with branching fractions $\sim 14\%$
and $\sim 55\%$, respectively.

Assigning the $\Delta(1905)5/2^+$ to
the high mixed state, the measured mass $M_{exp}\simeq1880$ MeV~\cite{ParticleDataGroup:2022pth} is
consistent with the theoretical prediction, moreover, the measured branching fractions
of the $\Delta \pi$ and $N\pi$ channels, $>48\%$ and $\sim 9-15\%$~\cite{ParticleDataGroup:2022pth},
are also comparable with our predictions. However, the average width
$\Gamma_{exp}\simeq 300$ MeV from the PDG~\cite{ParticleDataGroup:2022pth} is much broader than our prediction.
It should pointed out that there are large differences
in the resonance parameters of $\Delta(1905)5/2^+$ extracted from various PWA groups.
To well determined the resonance parameters of $\Delta(1905)5/2^+$,
more measurements are needed in future experiments.

The $\Delta(2000)5/2^+$ as a two-star resonance is still not well established in experiments.
Its mass and width determined from different PWA groups scatter in fairly large ranges
$M_{exp}\sim 1700-2300$ MeV and $\Gamma_{exp}\sim 100-600$ MeV, respectively~\cite{ParticleDataGroup:2022pth}.
According to the poor observations of $\Delta(2000)5/2^+$, it is difficult to determine whether
it corresponds to the high mass mixed state or not.
The high mixed $J^P=5/2^+$ state couples weakly to the $N\pi$ channel, which might be its missing reason
in the observations. To establish
the $J^P=5/2^+$ states, the observation of the
$N\pi\pi$ final state via the cascade decay $\Delta^*\to \Delta\pi\to N\pi\pi$ may be helpful.

For the three-star resonance $\Delta(1920)3/2^+$, the extracted mass
and width from the reaction data are $M_{exp}=1920\pm 50$ MeV and
$\Gamma_{exp}=300\pm 60$ MeV, respectively~\cite{ParticleDataGroup:2022pth}.
The $\Delta(1920)3/2^+$ may be a good candidate of the $^{4}10 [56, 2_2^+]1^4D_{3/2^+}$ state
classified in the quark model. With this assignment, our predicted mass and width
\begin{eqnarray}\label{delta(1920)}
M\simeq1906~\mathrm{MeV},~~\Gamma\simeq156~\mathrm{MeV},
\end{eqnarray}
are comparable with the observations.
According to our predicted decay properties given in Table~\ref{Delta decay widths},
it is seen that the $^{4}10 [56, 2_2^+]1^4D_{3/2^+}$ dominantly decays into
the $N(1440)\pi$ channel with a branching fraction of $\sim 27\%$, which is
in the range of measurements, $4-86\%$~\cite{ParticleDataGroup:2022pth}. Furthermore, the $\Delta\pi$, $N(1520)\pi$, and $N(1680)\pi$
channels play an important role in the decays, they have a comparable branching fraction of $\sim 24\% $, $\sim31\%$, $\sim9\%$.
The large decay rate of the $\Delta(1920)3/2^+\to N\pi\pi$ observed in experiments
may be contributed by the $\Delta\pi$, $N(1520)\pi$, and $N(1680)\pi$ channels.
Our predicted sizeable branching fraction $4\%$ for the $\Delta(1920)3/2^+\to N\pi$ is also
in the measured range of $5-20\%$~\cite{ParticleDataGroup:2022pth}. Finally, it should be mentioned that there are large discrepancies
in both the mass and width of $\Delta(1920)3/2^+$ extracted by various PWA groups~\cite{ParticleDataGroup:2022pth}.

The other $J^P=3/2^+$ state $^{2}10 [70, 2_2^+]1^2D_{3/2^+}$
is still missing. The mass and width are predicted to be
\begin{eqnarray}\label{delta(1920)}
M\simeq1973~\mathrm{MeV},~~\Gamma\simeq259~\mathrm{MeV},
\end{eqnarray}
respectively. More details about its decay
properties have been given in Table~\ref{Delta decay widths}. This state dominantly
decays into the $\Delta(1700)\pi$, $\Delta\pi$, and $N(1520)\pi$ with branching fractions
$\sim49\%$, $\sim9\%$, and $\sim22\%$, respectively.
However, the decay rate into the $N\pi$ is very tiny ($\sim 2\%$),
which can naturally explain why the $^{2}10 [70, 2_2^+]1^2D_{3/2^+}$ state has not been established in the
$\pi N/\gamma N\to \pi N$ reactions. This missing state is likely to be established in the $N\pi\pi$ final
state via the cascade decays $\Delta^*\to \Delta\pi/ N(1520)\pi\to N\pi\pi$.

Besides the $1D$-wave states, in the $N=2$ shell there is a $2S$-wave state
$^{2}10 [70, 0_2^+]2^2S_{1/2^+}$. This state is still missing.
The mass and width of this state is predicted to be
\begin{eqnarray}\label{delta(1883)}
M\simeq1874~\mathrm{MeV},~~\Gamma\simeq332~\mathrm{MeV},
\end{eqnarray}
respectively. More details about the mass and decay properties have
been given in Tables~\ref{Delta mass spectrum contribution} and~\ref{Delta decay widths}.
Our predicted mass is consistent with the predictions in the literature~\cite{Capstick:1986ter,Chen:2009de,Ghalenovi:2017xxv,Ferretti:2011zz,
Loring:2001kx}. This state mainly decays into the $N(1535)\pi$,
$N(1650)\pi$, and $N(1520)\pi$ with branching fractions $\sim81\%$, $\sim9\%$, and $\sim6\%$, respectively.
The decay rate into $N\pi$ is rather small, $\sim 1\%$. Thus, this state is hard to be
established by using the conventional $\pi N/\gamma N\to \pi N$ reactions.
This missing state is most likely to be established in the $N\pi\pi$ final
state via the cascade decay $\Delta^*\to N(1535)\pi\to N\pi\pi$.

As a whole, the masses and decay properties of $\Delta(1910)1/2^+$, $\Delta(1920)3/2^+$,
and $\Delta(1905)7/2^+$ can be reasonably understood within the quark model.
However, for the $\Delta(1950)5/2^+$ resonance, there is a discrepancy between
our predicted width and the observations. The existence
of the $\Delta(2000)5/2^+$ should be further confirmed in experiments.
Around the mass range of 1.9 GeV, three states, i.e., the $2S$-wave state
$^{2}10 [70, 0_2^+]2^2S_{1/2^+}$, the $1D$-wave state
$^{2}10 [70, 0_2^+]1^2D_{3/2^+}$, and the
high-mass mixed $1D$-wave state with $J^P=5/2^+$, are still missing due to
their very weak couplings to the $N\pi$ channel. The observations
of the three-body final state $N\pi\pi$ via the
cascade decays might be very helpful to not only establish the missing resonances,
but also better understand the nature of the resonances observed
in the previous experiments.

\section{summary}\label{Summary}

In this work, we systematically study both the mass spectrum and strong decay of the $N^*$ and $\Delta^*$
within the quark model framework by combing the chiral dynamics.
A reasonable description of the masses and strong decay properties
for the well-established states has been achieved. Some key points are emphasized as follows.

The chiral dynamics play an important role in the light baryon spectra.
With the chiral dynamics, the mass reversal between $N(1440)1/2^{+}$ and the $1P$-wave
nucleon resonances together with their strong decay properties can be naturally understood.
The center part of the one-pion-exchange interaction is responsible for the mass reversal.

The mixing between the $N(1535)1/2^-$ and $N(1650)1/2^-$ resonances is mainly caused
by the OGE tensor potentials in the short distance range. The singular behavior
of the $1/r^3$ terms in the OBE tensor potentials (mainly contributed by
the one-pion exchange) enlarges the unphysical contributions
in the short distance region, which cannot be effectively suppressed by
the spatial wave functions and form factors.
These unphysical contributions should be reasonably removed to reproduce the correct mixing angle.

The OGE three-body spin-orbit potentials, which are often neglected in the literature, play an important role in the baryon spectrum.
They not only cause a large configuration mixing between $N(1520)3/2^-$ and $N(1700)3/2^-$,
but also are responsible for the $\Delta(1600)1/2^-$-$\Delta(1700)3/2^-$ splitting.

The $N(1880)1/2^+$, $N(1900)3/2^+$, $N(1860)5/2^+$, $N(2000)5/2^+$, $N(1990)7/2^+$,
$\Delta(1910)1/2^+$, $\Delta(1920)3/2^+$, and $\Delta(1905)7/2^+$ listed in the RPP
can be assigned to the excitations in the $N=2$ shell. For the $\Delta(1950)5/2^+$,
there is a discrepancy between our predicted width and the observations. The existence
of the $\Delta(2000)5/2^+$ should be further confirmed in future experiments.

In the $N=2$ shell, four iso-spin 1/2 states and three
iso-spin 3/2 states are still missing. These missing states may have a mass of $M\sim1.9$ GeV and a width
of $\Gamma\sim$10s-200 MeV. They very weakly couple to the $N\pi$, $N\eta$,
$K\Lambda$, and $K\Sigma$ channels, which can naturally explain
why they have not been observed in these channels via the $\pi N$
and $\gamma N$ scatterings.

Most of the nucleon and $\Delta$ baryons in $N=2$ shell have large
decay rates into the $\Delta(1232)$, or the $1P$-wave
excitations of nucleons. The missing resonances in this shell may have large potentials to be
established in the $N\pi\pi$ final state by using the charmonium cascade decays at BESIII.

The low mass nature of the low-lying positive parity states $N(1710)1/2^+$, $N(1720)3/2^+$,
and $N(1680)5/2^+$ cannot be well understood in theory.
Our predicted masses are systematically larger than the center values of PDG by $\sim80$ MeV.
Some other dynamic mechanisms, such as coupled-channel effects, may be responsible for their low mass nature.

For lack of space, our results and discussions for the nucleon and $\Delta$ baryons
in $N=3$ shell will be given in another work.

%%%%%%%%%%%%%%%%%%%%%%%%%%%%%%%%%%%%%%%%%%%%%%%%%%%%%%%%%%%%%%%
\begin{table*}[htp]
\begin{center}
\caption{ Mass (MeV) spectrum of the nucleon baryon compared with the experimental data from the PDG~(labeled with Exp)~\cite{ParticleDataGroup:2022pth}, and the results obtained from the OGE model~\cite{Capstick:1986ter},
the OGE model with higher order hyperfine interactions (labeled with OGEh)~\cite{Chen:2009de},
the quark model combining both OGE and OBE potentials (labeled with Hyb)~\cite{Ghalenovi:2017xxv},
the quark-diquark model~\cite{Ferretti:2011zz}, the relativistically covariant
constituent quark model with instantaneous forces(labeled with Inst)~\cite{Loring:2001kx},
the Dyson-Schwinger and Faddeev equations~\cite{Eichmann:2016hgl}, and the large $1/N_c$ expansion approach (labeled with $1/N_c$) ~\cite{Goity:2002pu,Goity:2003ab,Goity:2008zz,Schat:2001xr,Carlson:1998vx,Matagne:2004pm,Matagne:2011fr,Matagne:2012tm}. If a resonance as a mixed state dominated by a certain quark model configuration, it is labeled with an underline.}\label{nucleon mass spectrum comparison}
\begin{tabular}{cccccccccccccccccccccccc}\hline\hline
&State
&~~~Ours~~~
&OGE~\cite{Capstick:1986ter}
&~OGEh~\cite{Chen:2009de}
&~OBE~\cite{Melde:2008yr}
&~Hyb~\cite{Ghalenovi:2017xxv}
&~Diq~\cite{Ferretti:2011zz}
&~Inst~\cite{Loring:2001kx}
&~$1/N_c$~\cite{Goity:2002pu,Goity:2003ab,Goity:2008zz,Schat:2001xr,Carlson:1998vx,Matagne:2004pm,Matagne:2011fr,Matagne:2012tm}
&~Exp~\cite{ParticleDataGroup:2022pth}
\\
\hline

&$^{2}8[56,0^{+}_{0}]$$1^{2}S_{\frac{1}{2}^{+}}$             &938   &960   &939  &939  &939   &939   &939   &     &938         \\  %1
&$^{2}8[56,0^{+}_{2}]$$2^{2}S_{\frac{1}{2}^{+}}$             &1438  &1540  &1462 &1459 &1445  &1513  &1518  &1450 &$1410-1470$ \\  %7
&$^{2}8[70,0^{+}_{2}]$$2^{2}S_{\frac{1}{2}^{+}}$             &1824  &1770  &1748 &1776 &1833  &1768  &1729  &1712 &$1680-1740$ \\  %9
&$\underline{^{4}8[70,2^{+}_{2}]1^{4}D_{\frac{1}{2}^{+}}}$   &1948  &1880  &1887 &     &      &1893  &1950  &     &$1830-1930$ \\  %15
&$\underline{^{2}8[20,1^{+}_{2}]1^{2}P_{\frac{1}{2}^{+}}}$   &2010  &1975  &2060 &     &      &      &1996  &1983 &$2050-2150$ \\ %49

&$^{2}8[56,2^{+}_{2}]$$1^{2}D_{\frac{3}{2}^{+}}$             &1817  &1795  &1734 &     &1690  &1768  &1688  &1674 &$1680-1750$ \\%11
&$^{4}8[70,0^{+}_{2}]$$2^{4}S_{\frac{3}{2}^{+}}$             &1886  &1870  &1857 &     &      &1808  &1809  &1885 &
\multirow{4}{*}{$\begin{cases} 1890-1950\\ 2015-2065\end{cases}$}\\%8
&$^{2}8[70,2^{+}_{2}]$$1^{2}D_{\frac{3}{2}^{+}}$             &1922  &1910  &1975 &     &      &      &1936  &     &     \\%17
&$\underline{^{4}8[70,2^{+}_{2}]1^{4}D_{\frac{3}{2}^{+}}}$   &1953  &1950  &1952 &     &      &      &1969  &     &     \\%14
&$\underline{^{2}8[20,1^{+}_{2}]1^{2}P_{\frac{3}{2}^{+}}}$   &2015  &2030  &2058 &     &      &      &      &     &     \\%48

&$\underline{^{2}8[56,2^{+}_{2}]1^{2}D_{\frac{5}{2}^{+}}}$   &1775  &1770  &1738 &     &1690 &1808  &1723  &1689 &$1680-1690$  \\%10
&$\underline{^{2}8[70,2^{+}_{2}]1^{2}D_{\frac{5}{2}^{+}}}$   &1944  &1980  &1953 &     &     &      &1934  &     &$1800-1980$  \\%16
&$^{4}8[70,2^{+}_{2}]$$1^{4}D_{\frac{5}{2}^{+}}$             &1962  &1995  &2007 &     &     &      &1959  &1850 &$1942-2090$  \\%13

&$^{4}8[70,2^{+}_{2}]$$1^{4}D_{\frac{7}{2}^{+}}$             &1936  &2000  &1943 &     &     &      &1989  &1872 &$1950-2100$  \\%12

&$\underline{^{2}8[70,1^{-}_{1}]1^{2}P_{\frac{1}{2}^{-}}}$   &1549  &1460  &1497 &1519 &1568 &1527  &1435  &1541 &$1515-1545$  \\%6
&$\underline{^{4}8[70,1^{-}_{1}]1^{4}P_{\frac{1}{2}^{-}}}$   &1606  &1535  &1650 &1647 &1658 &1671  &1660  &1660 &$1635-1665$  \\%4

&$\underline{^{2}8[70,1^{-}_{1}]1^{2}P_{\frac{3}{2}^{-}}}$   &1560  &1495  &1548 &1519 &1568 &1527  &1476  &1532 &$1510-1520$  \\%5
&$\underline{^{4}8[70,1^{-}_{1}]1^{4}P_{\frac{3}{2}^{-}}}$   &1666  &1625  &1731 &1647 &1658 &1671  &1606  &1699 &$1650-1800$  \\%3

&$^{4}8[70,1^{-}_{1}]$$1^{4}P_{\frac{5}{2}^{-}}$             &1646  &1630  &1655 &1647 &1658 &1671  &1655  &1671 &$1665-1680$  \\%2

\hline\hline
\end{tabular}
\end{center}
\end{table*} 
\begin{table*}[htp]
\begin{center}
\caption{ Mass (MeV) spectrum of the $\Delta$ baryon compared with the experimental data from the PDG~(labeled with Exp)~\cite{ParticleDataGroup:2022pth}, and the results obtained from the OGE model~\cite{Capstick:1986ter},
the OGE model with higher order hyperfine interactions (labeled with OGEh)~\cite{Chen:2009de},
the quark model combining both OGE and OBE potentials (labeled with Hyb)~\cite{Ghalenovi:2017xxv},
the quark-diquark model~\cite{Ferretti:2011zz}, the relativistically covariant
constituent quark model with instantaneous forces (labeled with Inst)~\cite{Loring:2001kx},
the Dyson-Schwinger and Faddeev equations~\cite{Eichmann:2016hgl}, and the large $1/N_c$ expansion approach (labeled with $1/N_c$) ~\cite{Goity:2002pu,Goity:2003ab,Goity:2008zz,Schat:2001xr,Carlson:1998vx,Matagne:2004pm,Matagne:2011fr,Matagne:2012tm}. If a resonance as a mixed state dominated by a certain quark model configuration, it is labeled with an underline. }\label{Delta mass spectrum comparison}
\begin{tabular}{cccccccccccccccccccccccccc}\hline\hline
&State
&~~~Ours~~~
&OGE~\cite{Capstick:1986ter}
&~OGEh~\cite{Chen:2009de}
&~OBE~\cite{Melde:2008yr}
&~Hyb.~\cite{Ghalenovi:2017xxv}
&~Diq~\cite{Ferretti:2011zz}
&~Inst.~\cite{Loring:2001kx}
&~$1/N_c$~\cite{Goity:2002pu,Goity:2003ab,Goity:2008zz,Schat:2001xr,Carlson:1998vx,Matagne:2004pm,Matagne:2011fr,Matagne:2012tm}
&~Exp.~\cite{ParticleDataGroup:2022pth}
\\
\hline
& $^{4}10[56,0^{+}_{0}]$$1^{4}S_{\frac{3}{2}^{+}}$               & 1232  &1230 &1231 &1240 &1232  &1233 &1260 &      & $1230-1234$  \\%1
& $^{4}10[56,0^{+}_{2}]$$2^{4}S_{\frac{3}{2}^{+}}$               & 1694  &1795 &1790 &1718 &1659  &1602 &1810 &1625  & $1500-1640$  \\%5
& $\underline{^{2}10[70,0^{+}_{2}]2^{2}S_{\frac{1}{2}^{+}}}$     & 1874  &1835 &1867 &     &1874  &1858 &1866 &1746  & $1660-1782$  \\%4
& $\underline{^{4}10[56,2^{+}_{2}]1^{4}D_{\frac{1}{2}^{+}}}$     & 1913  &1875 &1900 &     &      &1952 &1906 &1897  & $1850-1950$  \\%8
& $\underline{^{4}10[56,2^{+}_{2}]1^{4}D_{\frac{3}{2}^{+}}}$     & 1906  &1915 &1904 &     &2090  &1952 &1871 &1906  & $1870-1970$  \\%9
& $\underline{^{2}10[70,2^{+}_{2}]1^{2}D_{\frac{3}{2}^{+}}}$     & 1973  &1985 &1972 &     &      &     &1950 &      & $1855-1910$  \\%6
& $\underline{^{4}10[56,2^{+}_{2}]1^{4}D_{\frac{5}{2}^{+}}}$     & 1864  &1910 &1931 &     &1874  &1952 &1897 &1921  &              \\%10
& $\underline{^{2}10[70,2^{+}_{2}]1^{2}D_{\frac{5}{2}^{+}}}$     & 1975  &1990 &1967 &     &      &     &1985 &1756  & $1991-2039$  \\%7
& $^{4}10[56,2^{+}_{2}]$$1^{4}D_{\frac{7}{2}^{+}}$               & 1867  &1940 &1902 &     &1874  &1952 &1956 &1942  & $ 1915-1950$ \\%11
& $^{2}10[70,1^{-}_{1}]$$1^{2}P_{\frac{1}{2}^{-}}$               & 1595  &1555 &1668 &1642 &1667  &1554 &1654 &1645  & $1590-1630$  \\%2
& $^{2}10[70,1^{-}_{1}]$$1^{2}P_{\frac{3}{2}^{-}}$               & 1655  &1620 &1668 &1642 &1667  &1554 &1628 &1720  & $1690-1730$  \\%3

\hline\hline
\end{tabular}
\end{center}
\end{table*}

%%%%%%%%%%%%%%%%%%%%%%%%%%%%%%%%%%%%%%%%%%%%%%%%%%%%%%%%%%%%%%%%%%%%%%

\begin{table*}[htp]
\begin{center}
\caption{ Predicted masses of nucleon resonances with principal quantum number $N \leq 2$ and the average contributions of each part of the Hamiltonian (in MeV). $ T$ stands for the contribution of the kinetic energy term. $ V^{Conf}$ and $ V^{Coul}$ stand for the contributions from the linear confinement and Coulomb-like potentials, respectively. $V_{G}^{SS}$, $V_{G}^{LS}$ and $V_{G}^{T}$ denote the contributions of the spin-spin, spin-orbit, and tensor terms, respectively, in a single gluon exchange potential. $ V_{\pi}^C$, $ V_{\eta}^C$, and $ V_{\sigma}^C$ represent the contributions of the central potentials of one-boson exchanges, while $ V_{\pi}^{T}$, $ V_{\eta}^{T}$, $ V_{\sigma}^{LS}$ denote the corresponding contributions of the tensor and spin-orbit terms, respectively. The $\Delta m_{mix}$ in the last column represents the mass shift due to configuration mixing. The zero-point energy for each of them is $-829$ MeV. }\label{nucleon mass spectrum contribution}
\tabcolsep=0.16cm
%\renewcommand\arraystretch{1.00}
%\scalebox{0.8}{
\begin{tabular}{cccccccccccccccccccccccc}\hline\hline
State
&Mixing matrix
&Mass
&$ T$
&$ V^{Conf}$
&$ V^{Coul}$
&$ V^{SS}_{G}$
&$ V^{LS}_{G}$
&$ V^{T}_{G}$
&$ V_{\pi}^C$
&$ V_{\eta}^C$
&$ V_{\sigma}^C$
&$ V_{\pi}^{T}$
&$ V_{\eta}^{T}$
&$ V_{\sigma}^{LS}$
&$\Delta m_{mix}$
\\
\hline
 $^{2}8[56,0^{+}_{0}]$$1^{2}S_{\frac{1}{2}^{+}}$ & & 938
&2716 	&400 	&-597 	&-20 	&0 	&0 	&-571 	&38 	&-198 	&0 	&0 	&0 	&0\\  %1

 $^{2}8[56,0^{+}_{2}]$$2^{2}S_{\frac{1}{2}^{+}}$ & & 1438
&2747 	&656 	&-508 	&-14 	&0 	&0 	&-499 	&34 	&-149 	&0 	&0 	&0 	&0\\  %7

 $^{2}8[70,0^{+}_{2}]$$2^{2}S_{\frac{1}{2}^{+}}$ & & 1824
&2148 	&884 	&-246 	&-7 	&0 	&0 	&-62 	&5 	&-68 	&0 	&0 	&0 	&0\\  %9

 $^{4}8[70,2^{+}_{2}]$$1^{4}D_{\frac{1}{2}^{+}}$ &
\multirow{2}{*}{$\left(\begin{array}{cc} 0.96 & 0.29 \\ 0.29 & -0.96 \\ \end{array}\right)$}& 1948
&2209 	&882 	&-231 	&6 	&-14 	&-13 	&7 	&-3 	&-63 	&-5 	&-1 	&12 	&-9\\%15

 $^{2}8[20,1^{+}_{2}]$$1^{2}P_{\frac{1}{2}^{+}}$ & & 2010
&2071 	&983 	&-193 	&-4 	&-2 &-1 &18 	&0 	&-45 	&0 	&0 	&4 	&9\\ %49

 $^{2}8[56,2^{+}_{2}]$$1^{2}D_{\frac{3}{2}^{+}}$ & & 1817
&2280 	&823 	&-263 	&-8 	&1 	&0 	&-111 	&5 	&-85 	&0 	&0 	&4 	&0\\%11

 $^{2}8[70,2^{+}_{2}]$$1^{2}D_{\frac{3}{2}^{+}}$ & & 1922
&2137 	&918 	&-219 	&-6 	&0 	&0 	&-29 	&2 	&-57 	&4 	&0 	&0 	&0\\%17

 $^{4}8[70,0^{+}_{2}]$$2^{4}S_{\frac{3}{2}^{+}}$ & & 1886
&2076 	&933 	&-232 	&7 	&0 	&0 	&-3 	&-4 	&-62 	&0 	&0 	&0 	&0\\%8

 $^{4}8[70,2^{+}_{2}]$$1^{4}D_{\frac{3}{2}^{+}}$ &
\multirow{2}{*}{$\left(\begin{array}{cc} 0.88 & 0.47 \\ 0.47 & -0.88 \\ \end{array}\right)$}   & 1953
&2092 	&954 	&-208 	&4 	&2 	&0 	&7 	&-2 	&-51 	&0 	&0 	&6 	&-21\\%14

 $^{2}8[20,1^{+}_{2}]$$1^{2}P_{\frac{3}{2}^{+}}$ & & 2015
&2049 	&995 	&-191 	&-2 &1 	&0 	&14 &0 	&-44 	&0 	&0 	&1 	&21\\%48

 $^{2}8[56,2^{+}_{2}]$$1^{2}D_{\frac{5}{2}^{+}}$ &
\multirow{2}{*}{$\left(\begin{array}{cc} 0.90 & -0.43 \\ -0.43 & -0.90 \\ \end{array}\right)$}   & 1775
&2194 	&870 	&-243 	&-7 	&-3 &0 	&-84 	&4 	&-73 	&0 	&0 	&-3 &-51  \\%10	

 $^{2}8[70,2^{+}_{2}]$$1^{2}D_{\frac{5}{2}^{+}}$ & & 1944	
&2116 	&930 	&-218 	&-6 &-2 &0 	&-40 	&2 	&-57 	&0 	&0 	&-3 &51\\%16

 $^{4}8[70,2^{+}_{2}]$$1^{4}D_{\frac{5}{2}^{+}}$ & & 1962
&2044 	&984 	&-202 	&5 	&1 	&6 	&2 	&-2 	&-49 	&-2 	&1 	&1 	&0\\%13

 $^{4}8[70,2^{+}_{2}]$$1^{4}D_{\frac{7}{2}^{+}}$ & & 1936
&2038 	&988 	&-201 	&5 	&-10 	&-2 	&2 	&-2 	&-48 	&1 	&-1 	&-7 	&0\\%12

 $^{2}8[70,1^{-}_{1}]$$1^{2}P_{\frac{1}{2}^{-}}$ &
\multirow{2}{*}{$\left(\begin{array}{cc} 0.90 & -0.43\\ 0.43 & 0.90 \\\end{array}\right)$} &1549
&2179 	&708 	&-300 	&-6 	&-21 &-5 &-69 	&4 	&-94 	&-5 	&0 	&4 	&-17\\%6

 $^{4}8[70,1^{-}_{1}]$$1^{4}P_{\frac{1}{2}^{-}}$  & & 1606
&2253 	&679 	&-318 	&8 	&-46 	&-20 	&-15 	&-6 	&-102 	&-23 	&1 	&7 	&17\\%4

 $^{2}8[70,1^{-}_{1}]$$1^{2}P_{\frac{3}{2}^{-}}$ &
\multirow{2}{*}{$\left(\begin{array}{cc} 0.94 & -0.34\\ 0.34 & 0.94 \\ \end{array}\right)$} & 1560
&2075 	&759 	&-275 	&-7 	&4 	&1 	&-66 	&4 	&-83 	&0 	&0 	&-1 	&-22\\%5

 $^{4}8[70,1^{-}_{1}]$$1^{4}P_{\frac{3}{2}^{-}}$ & & 1666
&1998 	&807 	&-256 	&6 	&-3 	&9 	&-11 	&-3 	&-74 	&-3 	&1 	&2 	&22\\  %3

 $^{4}8[70,1^{-}_{1}]$$1^{4}P_{\frac{5}{2}^{-}}$ & & 1646
&1983 	&816 	&-253 	&8 	&6 	&-3 	&-3 	&-4 	&-73 	&1 	&0 	&-4 	&0\\  %2

\hline\hline
\end{tabular}
\end{center}
\end{table*} 
\begin{table*}[htp]
\begin{center}
\caption{ Predicted masses of $\Delta$ resonances with principal quantum number $N \leq 2$ and the average contributions of each part of the Hamiltonian (in MeV). The average contributions of each part of the $\Delta$ resonances. The caption is the same as that of Table~\ref{nucleon mass spectrum contribution}.}\label{Delta mass spectrum contribution}
\tabcolsep=0.16cm
\begin{tabular}{cccccccccccccccccccccccccc}\hline\hline
\tabcolsep=0.1cm
&State%$n^{2S+1}L_{J^P} |N_6,^{2S+1}N_3,N,L,J^P\rangle$
&Mixing matrix
&Mass
&$T$
&$V^{Conf}$
&$V^{Coul}$
&$V^{SS}_{G}$
&$V^{LS}_{G}$
&$V^{T}_{G}$
&$V^C_{\pi}$
&$V^C_{\eta}$
&$V^C_{\sigma}$
&$V_{\pi}^{T}$
&$V_{\eta}^{T}$
&$V_{\sigma}^{LS}$
&$\Delta m_{mix}$ \\
\hline
& $^{4}10[56,0^{+}_{0}]$$1^{4}S_{\frac{3}{2}^{+}}$   &   & 1232
&2066 	&569 	&-391 	&15 	&0 	   &0 	&-52 	&-15 	&-132 	&0 	&0 	&0 	&0\\%1

& $^{4}10[56,0^{+}_{2}]$$2^{4}S_{\frac{3}{2}^{+}}$   &   & 1694
&2208 	&816 	&-344 	&11 	&0 	   &0 	&-48 	&-15 	&-105 	&0 	&0 	&0 	&0\\%5

& $^{2}10[70,0^{+}_{2}]$$2^{2}S_{\frac{1}{2}^{+}}$   &
\multirow{2}{*}{$\left(\begin{array}{cccccc}   0.97 &	0.25    \\   0.25 &	-0.97 \\ \end{array}\right)$} & 1874
&2090 	&923 	&-234 	&-3 	&1 	   &0 	&-4 	&-63 	&-3 	&0 	&0 	&1 	&-4\\%4

& $^{4}10[56,2^{+}_{2}]$$1^{4}D_{\frac{1}{2}^{+}}$   &   & 1913
&2165 	&886 	&-242 	&7 	    &14    &-7 	&-17 	&-67 	&-4 	&-5 &-2 &10 &4\\%8

& $^{2}10[70,2^{+}_{2}]$$1^{2}D_{\frac{3}{2}^{+}}$   &
\multirow{2}{*}{$\left(\begin{array}{cccccc}   0.92 &	0.38    \\   -0.38 &	0.92 \\ \end{array}\right)$} & 1973
&2083 	&953 	&-212 	&-3 	&14    &0 	&0 	    &-53 	&-1 	&0 	&0 	&4 	&17\\%6

& $^{4}10[56,2^{+}_{2}]$$1^{4}D_{\frac{3}{2}^{+}}$   &   & 1906
&2124 	&914 	&-232 	&5 	    &13 	&0 	&-14 	&-62 	&-3 	&0 	&0 	&6 	&-17\\%9

& $^{2}10[70,2^{+}_{2}]$$1^{2}D_{\frac{5}{2}^{+}}$   &
\multirow{2}{*}{$\left(\begin{array}{cccccc}   0.78 &	0.63    \\   -0.63 &	0.78  \\ \end{array}\right)$} & 1975
&2080 	&950 	&-215 	&0 	    &-6 	&2 	&-4 	&-55 	&-2 	&2 	&0 	&-1 	&53\\%7

& $^{4}10[56,2^{+}_{2}]$$1^{4}D_{\frac{5}{2}^{+}}$   &   & 1864
&2086 	&942 	&-220 	&2 	    &-3 	&3 	&-8 	&-57 	&-2 	&3 	&1 	&0 	&-53\\%10

& $^{4}10[56,2^{+}_{2}]$$1^{4}D_{\frac{7}{2}^{+}}$   &   & 1867
&2080 	&938 	&-224 	&6 	    &-18 	&-2 &-15 	 &-3 	&-59 	&-2 &-1 &-6 &0\\%11

& $^{2}10[70,1^{-}_{1}]$$1^{2}P_{\frac{1}{2}^{-}}$   &   & 1595
&2061 	&767 	&-272 	&-4 	&-38   &0 	&-2 	&-3 	&-82 	&0 	&0 	&-4 &0\\%2

& $^{2}10[70,1^{-}_{1}]$$1^{2}P_{\frac{3}{2}^{-}}$   &   & 1655
&2000 	&805 	&-257 	&-3 	&17    &0 	&-3 	&-2 	&-75 	&0 	&0 	&2 	&0\\%3

\hline\hline
\end{tabular}
\end{center}
\end{table*}

%%%%%%%%%%%%%%%%%%%%%%%%%%%%%%%%%%%%%%%%%%%%%%%%%%%%%%%%%%%%%%%%%%%%%%

\begin{table*}[htp]
\begin{center}
\caption{ Partial decay widths (MeV) of nucleon resonances compared with the data and the other theoretical results. For the $\Sigma K$ decay channel, the calculated partial widths of the nucleon resonances are less than 1 MeV, therefore they are not listed in the table. The unopened decay channels for a resonance are labeled by $\cdot\cdot\cdot$. If a resonance as a mixed state dominated
by a certain quark model configuration, it is labeled with an underline.}
\label{Low mass nucleon decay widths}
\setlength{\tabcolsep}{2mm}
\begin{tabular}{lcccccccccccccc}\hline\hline
State&$N  \pi$
&$\Delta \pi$
&$N \eta $
&$\varLambda K $
&$N(1440)\pi$
&$N(1520)\pi$
&$N(1535)\pi$
&Sum
\\
\hline
%% Npi   Deltapi Neta    LambdaK SigmaK  Sigma*K Neta'   1440    1520    1535    1650    1680    1620    1700
%% 1076  1370    1486    1610    1687    1877    1896    1538    1658    1673    1788    1818    1758    1838

%$|70, ^{2}8, 1, 1,\frac{1}{2}^{-}\rangle 1^{2}P_{\frac{1}{2}^{-}}$   &1517   %6
$N(1535) \frac{1}{2}^{-}$
&$40\sim91$      &$1\sim7$       &$38\sim96$    & & & &  &$125\sim175$ &PDG~\cite{ParticleDataGroup:2022pth}\\
$\underline{^{2}8[70,1^{-}_{1}]}$
&$62.3^{+0.2}_{-0.4}$ &$2.8^{+0.9}_{-0.8}$  &$79.5^{+9.6}_{-12.9}$  &$\cdot\cdot\cdot$  &$\cdot\cdot\cdot$ &$\cdot\cdot\cdot$ &$\cdot\cdot\cdot$ &$144.6^{+10.7}_{-14.1}$ &Ours\\
$1530^{+15}_{-15}$
&63        &16     &75         & & & & &154 &   $U(7)$~\cite{Bijker:2015gyk}\\
****
&84        &6      &50         & & & & &140        & hQM~\cite{Bijker:2015gyk}\\
&$51\pm21$ &       &$121\pm15$ & & & & &$172\pm36$ & CQM~\cite{An:2011sb}\\
&57        &0.9    &73         & & & & &130.9      & $1/{N_c}$~\cite{Jayalath:2011uc}\\
[1ex]
\cline{2-10}

%$|70, ^{4}8, 1, 1,\frac{1}{2}^{-}\rangle 1^{4}P_{\frac{1}{2}^{-}}$    &1616    %4
$N(1650) \frac{1}{2}^{-}$
&$50\sim105$  &$6\sim27$ 	&$15\sim53$   &$5\sim23 $    &$6\sim39$ &  &  &$100\sim150$    &PDG~\cite{ParticleDataGroup:2022pth}\\
$\underline{^{4}8[70,1^{-}_{1}]}$
&$96.5^{-1.5}_{+1.2}$ &$25.5^{+4.7}_{-4.1}$ &$13.2^{+0.0}_{-0.1} $  &$4.3^{+0.5}_{-0.7}  $ &$0.9^{-0.1}_{+0.1}$&$\cdot\cdot\cdot$  &$\cdot\cdot\cdot$  &$140.4^{+3.6}_{-3.6}$ &Ours\\
$1650^{+15}_{-15}$
&41         &18     &72        &           & & & &131 &$U(7)$~\cite{Bijker:2015gyk}\\\
****
&51         &4      &29        &           & & & &84      &hQM~\cite{Bijker:2015gyk}\\
&$81\pm22$  &       &$28\pm22$   &$9\pm6$    & & & &$118\pm50$ &CQM~\cite{An:2011sb}\\
&133        &5.1    &12.5      &$11.5$     & & & &162.1  &$1/{N_c}$~\cite{Jayalath:2011uc}\\
[1ex]
\cline{2-10}

%$|70, ^{2}8, 1, 1,\frac{3}{2}^{-}\rangle 1^{2}P_{\frac{3}{2}^{-}}$   &1539   %5
$N(1520) \frac{3}{2}^{-}$
&$55\sim78$  &$22\sim41$   &$0.07\sim0.1 $  & & & & &$100\sim120$   &PDG~\cite{ParticleDataGroup:2022pth}\\
$\underline{^{2}8[70,1^{-}_{1}]}$
&$48.2^{+1.7}_{-1.6}$ &$8.4^{+0.7}_{-0.6}$  &$0.1^{+0.1}_{-0.1}$ &$\cdot\cdot\cdot$  &$\cdot\cdot\cdot$ &$\cdot\cdot\cdot$ &$\cdot\cdot\cdot$ &$56.7^{+2.5}_{-2.3}$  &Ours\\
$1515^{+5}_{-5}$
&134       &207  &0.0     & & & & &341 &$U(7)$~\cite{Bijker:2015gyk}\\
****
&111       &206  &0.0     & & & & &317        &hQM~\cite{Bijker:2015gyk}\\
&$66\pm7$  &     &0.19    & & & & &$66.2\pm7$ &CQM~\cite{An:2011sb}\\
&72        &19   &0.26    & & & & &91.3       &$1/{N_c}$~\cite{Jayalath:2011uc}\\ [1ex]
\cline{2-10}

%$|70, ^{4}8, 1, 1,\frac{3}{2}^{-}\rangle 1^{4}P_{\frac{3}{2}^{-}}$    &1639  %3
$N(1700) \frac{3}{2}^{-}$
&$7\sim51$   &$55\sim255$   &$1\sim6$   &$1\sim6 $    &$3\sim33$ &$<12$ 	& &$100\sim300$      &PDG~\cite{ParticleDataGroup:2022pth}\\
$\underline{^{4}8[70,1^{-}_{1}]}$
&$46.4^{+19.1}_{-13.8}$  &$181.2^{+4.0}_{-15.8}$ &$0.0$	&$0.4^{+1.1}_{-0.4}$ &$0.5^{+1.6}_{-0.4}$ &$3.9^{+15.0}_{-3.9}$ &$2.2^{+10.7}_{-1.9}$ &$234.6^{+51.5}_{-36.2}$  &Ours\\
$1720^{+80}_{-70}$
&9          &561 &3          &               &  & & &573 &$U(7)$~\cite{Bijker:2015gyk}\\
***
&$13\pm10$          &   & $0.5\pm0.5$         & $0.1\pm0.1$              &  & & &   &CQM~\cite{An:2011sb}\\
&12         &297.3&$\leq0.15$  &$\leq0.03$   &0.0 & & &$309.3$    &$1/{N_c}$~\cite{Jayalath:2011uc}\\
[1ex]
\cline{2-10}

%$|70, ^{4}8, 1, 1,\frac{5}{2}^{-}\rangle 1^{4}P_{\frac{5}{2}^{-}}$    &1616   %2
$N(1675) \frac{5}{2}^{-}$
&$49\sim67$ &$30\sim59$    &$<2$     &$<0.06$   & & & &$130\sim160$     &PDG~\cite{ParticleDataGroup:2022pth}\\
$^{4}8[70,1^{-}_{1}]$
&$25.3^{+0.6}_{-0.6}$  &$59.2^{+3.1}_{-3.0}$ &$7.6^{+0.5}_{-0.5}$  &$0.0$  &0.0  &0.0 &0.0  &$92.1^{+4.2}_{-4.1}$   &Ours\\
$1675^{+5}_{-5}$
&47      &108        &11         &             & & & &166 &$U(7)$~\cite{Bijker:2015gyk}\\
****
&41      &85         &9          &             & & & &135 &hQM~\cite{Bijker:2015gyk}\\
&51      &75         &6.3        &$\leq0.1$    & & & &132.3 &$1/{N_c}$~\cite{Jayalath:2011uc}\\%1/Nc
[1ex]
\cline{2-10}

%$|70, ^{2}8, 1, 1,\frac{3}{2}^{-}\rangle 1^{2}P_{\frac{3}{2}^{-}}$   &1539   %7
$N(1440) \frac{1}{2}^{+}$
&$138\sim338$    &$15\sim122$       &  &  & & & &$250\sim450$ &PDG~\cite{ParticleDataGroup:2022pth}\\
$^{2}8[56,0^{+}_{2}]$
&$363.1^{+49.6}_{-48.7}$ &$25.8^{+24.1}_{-16.2}$  &$\cdot\cdot\cdot$ 	&$\cdot\cdot\cdot$  &$\cdot\cdot\cdot$ &$\cdot\cdot\cdot$ &$\cdot\cdot\cdot$ &$388.9^{+73.7}_{-64.9}$ &Ours\\
$1440^{+30}_{-30}$
&85  &13  & & & & & &98 &$U(7)$~\cite{Bijker:2015gyk}\\
****
&105 &12  & & & & & &117 &hQM~\cite{Bijker:2015gyk}\\
[1ex]
\cline{2-10}

%$|70, ^{2}8, 2, 0,\frac{1}{2}^{+}\rangle 2^{2}S_{\frac{1}{2}^{+}}$   &1780    %9
$N(1710) \frac{1}{2}^{+}$
&$4\sim40$  &$2\sim18$   &$8\sim100$    &$4\sim50$ & & &$7\sim42$    &$80\sim200$ &PDG~\cite{ParticleDataGroup:2022pth}\\
$^{2}8[70,0^{+}_{2}]$
&$10.2^{-1.2}_{+1.0}$ &$8.2^{-0.3}_{-0.0}$   &$4.2^{+0.1}_{-0.2}$  &$0.5^{+0.1}_{-0.1}$  & $92.7^{+36.9}_{-32.1}$
&$0.3^{+0.7}_{-0.2}$  &$51.7^{+39.2}_{-32.5}$ &$167.8^{+75.5}_{-64.1}$  &Ours\\
$1710^{+30}_{-30}$
&5  &56 &9  &3    &   & & &73  &$U(7)$~\cite{Bijker:2015gyk}\\
****
&18 &70 &12 &14.1 &   & & &114.1  &hQM~\cite{Bijker:2015gyk}\\
[1ex]
\cline{2-10}

%$|56, ^{2}8, 2, 2,\frac{3}{2}^{+}\rangle 1^{2}D_{\frac{3}{2}^{+}}$   &1777     %11
$N(1720) \frac{3}{2}^{+}$
&$12\sim56$  &$71\sim356$ &$2\sim20$   &$6\sim76$   &$<8$ 	&$2\sim20$ & &$150\sim400$  &PDG~\cite{ParticleDataGroup:2022pth}\\
$^{2}8[56,2^{+}_{2}]$
&$21.0^{+2.1}_{-2.2}$ 	 &$6.7^{+2.8}_{-2.1}$ 	   &$0.9^{+0.2}_{-0.2}$ 	 &$1.3^{+0.5}_{-0.4}$ &$88.7^{+30.2}_{-27.7}$ 	 &$17.7^{+8.8}_{-7.7}$ 	 &$0.0$  	 &$136.3^{+44.6}_{-40.3}$      &Ours\\
$1720^{+30}_{-40}$
&111 &36 &7    &14 &  & & &168 &$U(7)$~\cite{Bijker:2015gyk}\\
****
&141 &30 &8    &12 &  & & &191 &hQM~\cite{Bijker:2015gyk}\\
[1ex]
\cline{2-10}

%$|56, ^{2}8, 2, 2,\frac{5}{2}^{+}\rangle 1^{2}D_{\frac{5}{2}^{+}}$   &1739  %10
$N(1680) \frac{5}{2}^{+}$
&$69\sim91$ &$13\sim30$ &$<1$   &        &           &     & 	&$115\sim130$ &PDG~\cite{ParticleDataGroup:2022pth}\\
$\underline{^{2}8[56,2^{+}_{2}]}$
&$43.8^{+1.5}_{-1.4}$ 	 &$1.7^{+0.1}_{-0.1}$  &$0.5^{+0.0}_{-0.0}$ &$0.1^{+0.0}_{-0.0}$   &$0.1^{+0.0}_{-0.0}$ &0.0 &$0.0$  &$46.2^{+1.6}_{-1.5}$ &Ours\\
$1685^{+5}_{-5}$
&121  &100 &1    &     & & & &222  & $U(7)$~\cite{Bijker:2015gyk}\\
****
&91   &92  &0    &     & & & &183  & hQM~\cite{Bijker:2015gyk}\\
[1ex]
\cline{2-10}

\hline\hline
\end{tabular}
\end{center}
\end{table*}

\begin{table*}[htp]
\begin{center}
%\begin{adjustwidth}{-1cm}{-1cm} %调节表格页边距 % 负值表示增加页边距
\caption{ Partial decay widths (MeV) of nucleon resonances compared with the data and the other theoretical results. The calculated partial widths for the $N\eta^{\prime}$ and $\Sigma^{*}K$ channels are usually small, less than 2 MeV, thus, they are not presented in the table. For the $N^{*}\pi/\Delta^{*} \pi$ column, the partial widths for the $N(1520)\pi$, $N(1535)\pi$, $N(1650)\pi$, $N(1675)\pi$, $N(1680)\pi$, and $\Delta(1620)\pi$ and $\Delta(1700)\pi$ channels are given, to save space, the less important decay channels
with a tiny partial width (less than 5 MeV) are not listed in the table. If a resonance as a mixed state dominated
by a certain quark model configuration, it is labeled with an underline.  }
\label{positive parity nucleon decay widths}
\setlength{\tabcolsep}{1mm} %调节表格列间距
\begin{tabular}{lcccccccccccccc}\hline\hline
State&$N  \pi$
&$\Delta \pi$
&$N \eta $
&$\varLambda K / \Sigma K $
&$N(1440)\pi$
&\multicolumn{3}{c}{$N^{*}\pi/\Delta^{*} \pi$}
&Sum
\\
\hline
%$|70, ^{2}8, 2, 2,\frac{3}{2}^{+}\rangle 1^{2}D_{\frac{3}{2}^{+}}$    &1892  %17
& & & & & &$N(1520)\pi$ &$N(1680)\pi$ &$\Delta(1700)\pi$  &\\
%\cline{7-9}

$N(1900) \frac{3}{2}^{+}$
&$1\sim64$ &$30\sim224$ &$2\sim45$  &$2\sim64/3\sim22$  &  &$7\sim74$ &  &   & $100\sim320$   &PDG~\cite{ParticleDataGroup:2022pth}\\
$^{2}8[70,2^{+}_{2}]$
&$9.6^{+0.2}_{-0.3}$ 	&$25.2^{+5.8}_{-5.0}$     &$3.3^{+0.1}_{-0.2}$ 	 &$2.0^{+0.1}_{-0.1}/0.2^{+0.0}_{-0.0}$  	 &$61.7^{+4.0}_{-4.9}$   &$102.4^{+5.9}_{-7.6}$     &$5.5^{+3.5}_{-2.6}$   &$11.6^{+6.1}_{-5.0}$ 	  &$221.5^{+25.7}_{-25.7}$  &Ours\\
$1920^{+30}_{-30}$
&11  &63 &12   &$13/1$ & & & & &100  &$U(7)$~\cite{Bijker:2015gyk}\\
****
&15  &70 &12   &$13/1$ & & & & &111 &hQM~\cite{Bijker:2015gyk}\\
[1ex]
\cline{2-11}

%$|70, ^{4}8, 2, 2,\frac{1}{2}^{+}\rangle 1^{4}D_{\frac{1}{2}^{+}}$   &1934   %15
& & & & & &$N(1520)\pi$ &$N(1535)\pi$  &$\Delta(1620)\pi$ \\
%\cline{7-9}
$N(1880) \frac{1}{2}^{+}$
&$6\sim124$    &$10\sim168$    &$2\sim220$     &$2\sim12/20\sim96$  &   & &$8\sim48$    &   &$200\sim400$    &PDG~\cite{ParticleDataGroup:2022pth}\\
$\underline{^{4}8[70,2^{+}_{2}]}$
&$2.4^{+0.1}_{-0.1}$  &$293.7^{+47.3}_{-47.6}$  &$3.2^{+0.1}_{-0.2}$  &$0.0/3.0^{+0.3}_{-0.5}$   &$19.7^{+1.5}_{-2.3}$ &$3.5^{+2.3}_{-1.6}$	 &$99.8^{+19.3}_{-20.8}$   &$210.4^{+50.2}_{-56.5}$ 	   &$635.7^{+121.1}_{-129.6}$   &Ours\\
$1948^{+50}_{-50}$\\
***\\
[1ex]
\cline{2-11}

%$|20, ^{2}8, 2, 1,\frac{1}{2}^{+}\rangle 1^{2}P_{\frac{1}{2}^{+}}$   &1961  %49
& & & & &  &$N(1535)\pi$ &$N(1650)\pi$ &$\Delta(1620)\pi$  \\
%\cline{7-9}
$N(2100) \frac{1}{2}^{+}$
&$16\sim102$ &$12\sim45$ &$10\sim144$ &$<3/\ldots$   &       &$52\sim109$  &  & &$200\sim320$   &PDG~\cite{ParticleDataGroup:2022pth}\\
$\underline{^{2}8[20,1^{+}_{2}]}$
&$0.2^{+0.0}_{-0.0}$  &$32.1^{+4.1}_{-4.3} $ &$0.3^{+0.0}_{-0.0}$ &$0.0/0.3^{+0.0}_{-0.0}$ &$2.0^{+0.0}_{-0.1} $ &$46.4^{+5.4}_{-6.6}$ &$41.4^{+10.6}_{-11.0}$  &$68.3^{-11.7}_{-13.7}$  &$191.0^{+8.4}_{-35.7}$  &Ours\\
$2010^{+50}_{-50}$\\
***\\
[1ex]
\cline{2-11}

%$|70, ^{2}8, 2, 2,\frac{5}{2}^{+}\rangle 1^{2}D_{\frac{5}{2}^{+}}$    &1879   %16
& & & & &  &$N(1520)\pi$  &$N(1675)\pi$ &$N(1680)\pi$\\
%\cline{7-9}
$N(1860) \frac{5}{2}^{+}$
&$13\sim87$    &$64\sim234$   &$0\sim26$  &$<0.17/\ldots$  	    	    &         &	    &       &    &$318\sim434$         &PDG~\cite{ParticleDataGroup:2022pth}\\
$\underline{^{2}8[70,2^{+}_{2}]}$
&$0.8^{+0.1}_{-0.1}$	    &$40.6^{+11.6}_{-9.9}$ 	   &$2.2^{+0.8}_{-0.6}$ 	 &$0.3^{+0.2}_{-0.1}/0.4^{+0.3}_{-0.2}$ 	  &$0.2^{+0.2}_{-0.1}$ 	  	 &$12.6^{+7.8}_{-5.5}$  	 &$4.2^{+2.0}_{-1.8}$    &$8.6^{+9.0}_{-5.5}$ 	 &$69.9^{+32.0}_{-23.8}$      &Ours\\
$1944^{+50}_{-50}$\\
**\\
[1ex]
\cline{2-11}

%$|70, ^{4}8, 2, 2,\frac{5}{2}^{+}\rangle 1^{4}D_{\frac{5}{2}^{+}}$    &1929    %13
& & & & &  &$N(1520)\pi$ &$N(1650)\pi$  &$N(1675)\pi$\\
%\cline{7-9}
$N(2000) \frac{5}{2}^{+}$
&$20\sim45$ &$100\sim356$  &$<17.8$   &	    &	    &       &       &       &$335\sim445$        &PDG~\cite{ParticleDataGroup:2022pth}\\
$^{4}8[70,2^{+}_{2}]$
&$1.6^{+0.3}_{-0.3}$ 	   &$44.7^{+13.0}_{-11.0}$ 	  &$1.0^{+0.3}_{-0.3}$ 	  &$0.0/0.5^{+0.3}_{-0.2}$ 	 &$0.3^{+0.2}_{-0.1}$ 	 &$3.7^{+2.2}_{-1.6}$ 	  &$2.3^{+2.8}_{-1.5}$ 	 &$89.3^{+32.0}_{-32.5}$  &$143.4^{+51.1}_{-47.5}$	 &Ours\\
$1962^{+50}_{-50}$\\
**         \\
[1ex]
\cline{2-11}

%$|70, ^{4}8, 2, 2,\frac{7}{2}^{+}\rangle 1^{4}D_{\frac{7}{2}^{+}}$   &1874    %12
& & & & & &$N(1520)\pi$ &$N(1675)\pi$  &$N(1680)\pi$\\
%\cline{7-9}
$N(1990) \frac{7}{2}^{+}$
&$4\sim24$    &  &$<12$    &$12\sim24/\ldots$     	   	     &	        &       &       &   &$200\sim400$    &PDG~\cite{ParticleDataGroup:2022pth}\\
$^{4}8[70,2^{+}_{2}]$
&$6.3^{+1.4}_{-1.2}$ &$31.7^{+12.4}_{-9.8}$ 	&$3.7^{+1.3}_{-1.1}$ 	  &$0.0/1.6^{+1.2}_{-0.8}$ 	  &$1.0^{+0.7}_{-0.5}$ &$15.0^{+10.3}_{-7.0}$   	 &$2.5^{+4.9}_{-2.0}$ 	 &$1.1^{+1.2}_{-0.7}$  &$62.9^{+33.4}_{-23.1}$ &Ours\\
$1936^{+50}_{-50}$\\
**\\
[1ex]
\cline{2-11}

%$|70, ^{4}8, 2, 0,\frac{3}{2}^{+}\rangle 2^{4}S_{\frac{3}{2}^{+}}$   &1849   %8
& & & & & &$N(1520)\pi$ &$N(1535)\pi$  &$\Delta(1700)\pi$\\
%\cline{7-9}
$N(1886) \frac{3}{2}^{+}$
 &$0.3^{-0.2}_{+0.2}$ 	   &$3.4^{-2.3}_{+2.7}$   &$1.3^{-0.4}_{+0.4}$ 	    &$0.0/0.4^{-0.2}_{+0.1}$ &$36.4^{+8.5}_{-8.4}$ 	&$69.9^{+17.2}_{-17.7}$	 &$3.7^{+3.5}_{-2.1}$  &$48.6^{+68.8}_{-34.6}$   &$164.0^{+94.9}_{-59.4}$	 &Ours\\
$^{4}8[70,0^{+}_{2}]$\\
$1886^{+50}_{-50}$\\
[1ex]
\cline{2-11}

%$|70, ^{4}8, 2, 2,\frac{3}{2}^{+}\rangle 1^{4}D_{\frac{3}{2}^{+}}$    &1954  %14
& & & & & &$N(1520)\pi$ &$N(1535)\pi$  &$\Delta(1700)\pi$\\
%\cline{7-9}
$N(1953) \frac{3}{2}^{+}$
&$0.9^{+0.0}_{-0.0}$ 	&$17.5^{+4.7}_{-4.1}$ 	  &$1.3^{+0.0}_{-0.1}$ 	&$0.0/1.3^{+0.1}_{-0.2}$	 &$6.0^{+0.4}_{-0.6}$ 	 &$7.1^{+1.5}_{-1.5}$  &$6.7^{+4.7}_{-3.2}$    	 &$9.3^{+4.7}_{-4.4}$ 	 &$50.1^{+16.1}_{-14.1}$	 &Ours\\
$^{4}8[70,2^{+}_{2}]$\\
$1953^{+50}_{-50}$\\
[1ex]
\cline{2-11}

%$|20, ^{2}8, 2, 1,\frac{3}{2}^{+}\rangle 1^{2}P_{\frac{3}{2}^{+}}$   &1971   %48
& & & & & &$N(1520)\pi$ &$\Delta(1620)\pi$  &$\Delta(1700)\pi$\\
%\cline{7-9}
$N(2015) \frac{3}{2}^{+}$
&$0.3^{+0.0}_{-0.0}$ 	&$7.2^{+1.3}_{-1.1}$ 	&$0.4^{+0.0}_{-0.0}$ 	&$0.0/0.4^{+0.0}_{-0.0}$  	&$1.8^{+0.0}_{-0.1}$	    &$5.2^{-0.3}_{-0.0}$ &$13.1^{+9.8}_{-6.5}$	 &$63.2^{+25.5}_{-22.8}$	 &$91.6^{+36.3}_{-30.5}$ &Ours\\
$^{2}8[20,1^{+}_{2}]$\\
$2015^{+50}_{-50}$\\
[1ex]
\cline{2-11}

\hline\hline
\end{tabular}
%\end{adjustwidth}
\end{center}
\end{table*} 
\begin{table*}[htp]
\begin{center}
\caption{Partial decay widths (MeV) of $\Delta$ resonances compared with the data and the other theoretical results.
For the $N^{*}\pi/\Delta^{*} \pi$ column, the partial widths for the $N(1520)\pi$, $N(1535)\pi$, $N(1650)\pi$, $N(1675)\pi$, $N(1680)\pi$, and $\Delta(1620)\pi$, $\Delta(1700)\pi$ channels are given, to save space, the less important decay channels
with a tiny partial width are not listed in the table. }
\label{Delta decay widths}
\setlength{\tabcolsep}{2mm} %调节表格列间距
\begin{tabular}{lccccccccccccccccc}\hline\hline
State &$N  \pi$
&$\Delta \pi$
&$\Sigma K$
&$\Delta\eta$
&$N(1440)\pi$
&\multicolumn{3}{c}{$N^{*}\pi/\Delta^{*} \pi$}
&Sum\\
\hline

%% Npi   Deltapi  SigmaK  Deltaeta  Sigma*K Delteta' 1440    1520    1535    1650    1680    1620    1700
%% 1076  1370     1687    1780      1877    2190     1538    1658    1673    1788    1818    1758    1838

%$|56, ^{4}10, 0, 0,\frac{3}{2}^{+}\rangle 1^{4}S_{\frac{3}{2}^{+}}$  &1232  %1
$\Delta(1232) \frac{3}{2}^{+}$
&$113\sim119$ & & 	& 	&	& 	& 	&  &$114\sim120$  &PDG~\cite{ParticleDataGroup:2022pth}\\
$^{4}10[56,0^{+}_{0}]$
&$49.8^{+1.3}_{-1.3}$ &$\cdot\cdot\cdot$ 	&$\cdot\cdot\cdot$ 	&$\cdot\cdot\cdot$ &$\cdot\cdot\cdot$ 	&\multicolumn{3}{c}{$\cdot\cdot\cdot$} &$49.8^{+1.3}_{-1.3}$ &Ours\\
$1232^{+2}_{-2}$
&71   & & & & &  & & &71 &$U(7)$~\cite{Bijker:2015gyk}\\
****
&63   & & & & &  & & &63 & hQM~\cite{Bijker:2015gyk}\\ [1.2ex]
\cline{2-11}

%$|56, ^{4}10, 2, 0,\frac{3}{2}^{+}\rangle 2^{4}S_{\frac{3}{2}^{+}}$  &1570   1654   %5
$\Delta(1600) \frac{3}{2}^{+}$
&$16\sim72$   &$116\sim246$   & &  &$34\sim81$  &  & 	& &$200\sim300$   &PDG~\cite{ParticleDataGroup:2022pth}\\
$^{2}10[70,2^{+}_{2}]$
&$21.6^{+0.0}_{-1.6}$ 	   &$79.3^{+53.5}_{-43.3}$	    &$\cdot\cdot\cdot$ &$\cdot\cdot\cdot$	 &$0.1^{+4.0}_{+0.5}$ 	  &\multicolumn{3}{c}{$\cdot\cdot\cdot$}	 &$101.0^{+57.5}_{-44.4}$ 	 &Ours\\
$1570^{+70}_{-70}$\\
**** \\ [1ex]
\cline{2-11}

%$ ^{2}10, 1, 1,\frac{1}{2}^{$-$}\rangle 1^{2}P_{\frac{1}{2}^{$-$}}$  &1610  1625  %2
$\Delta(1620) \frac{1}{2}^{-}$
&$28\sim53$  &$48\sim108$     & &	 &$<14$    & & & &$110\sim150$   &PDG~\cite{ParticleDataGroup:2022pth}\\
$^{2}10[70,1^{-}_{1}]$
&$16.1^{-0.2}_{+0.2}$ 	   &$29.9^{+8.8}_{-7.3}$ 	     &$\cdot\cdot\cdot$ 	&$\cdot\cdot\cdot$ &$0.0$   &\multicolumn{3}{c}{$\cdot\cdot\cdot$}	 &$46.0^{+8.6}_{-7.1}$  &Ours\\
$1610^{+20}_{-20}$
&5 &76 &  & & & &  & &81 &$U(7)$~\cite{Bijker:2015gyk}\\
****
&9 &59 &  & & & &  & &68 &hQM~\cite{Bijker:2015gyk}\\
[1ex]
\cline{2-11}

%$|70, ^{2}10, 1, 1,\frac{3}{2}^{$-$}\rangle 1^{2}P_{\frac{3}{2}^{$-$}}$  &1710   1625   %3
& & & & & &$N(1520)\pi$ &$N(1535)\pi $  &$N(1650)\pi $  &\\
%\cline{7-9}
$\Delta(1700) \frac{3}{2}^{-}$
&$22\sim76$   &$20\sim266$   &      &   & &$2\sim19$  &$1\sim6$ &  &$220\sim380 $    &PDG~\cite{ParticleDataGroup:2022pth}\\
$^{2}10[70,1^{-}_{1}]$
&$25.2^{+2.4}_{-2.3}$	   &$107.1^{+9.3}_{-8.7}$  &$0.0^{+0.1}_{-0.0}$ &$\cdot\cdot\cdot$	 &$0.2^{+0.1}_{-0.1}$ 	 &$5.7^{+4.1}_{-3.0}$ 	 &$0.2^{+0.2}_{-0.2}$ 	   &$\cdot\cdot\cdot$   &$138.4^{+16.2}_{-14.3}$   &Ours\\
$1710^{+20}_{-20}$
&46 &311 &  & & & & & &357 & $U(7)$~\cite{Bijker:2015gyk}\\
****
&40 &333 &  & & & & & &373 & hQM~\cite{Bijker:2015gyk}\\
[1ex]
\cline{2-11}

%$|56, ^{4}10, 2, 2,\frac{1}{2}^{+}\rangle 1^{4}D_{\frac{1}{2}^{+}}$  &1900    1882   %8
& & & & &  &$N(1535)\pi $  &$N(1650)\pi $ &$\Delta(1620)\pi $ &\\
%\cline{7-9}
$\Delta(1910) \frac{1}{2}^{+}$
&$20\sim120$ &$68\sim264$ &$8\sim56$ &$10\sim52$    &$6\sim180$ &   &  &   	 &$200\sim400$ 	 &PDG~\cite{ParticleDataGroup:2022pth}\\
$\underline{^{4}10[56,2^{+}_{2}]}$
&$17.1^{+1.2}_{-1.5}$ 	  &$3.7^{-0.6}_{+0.3}$ 	    &$4.9^{+1.1}_{-1.3}$  &$1.3^{+0.4}_{-0.6}$	 &$164.6^{+22.5}_{-28.2}$  &$4.3^{+0.2}_{-0.6}$ 	 &$57.5^{+32.6}_{-28.0}$ 	 &$115.8^{+39.6}_{-41.0}$  &$369.2^{+97.0}_{-100.9}$  &Ours\\
$1900^{+50}_{-50}$
&26 &32 &38 &4  & & & & &100 &$U(7)$~\cite{Bijker:2015gyk}\\
****
&49 &34 &38 &4  & & & & &125 &hQM~\cite{Bijker:2015gyk}\\
[1ex]
\cline{2-11}

%$|56, ^{4}10, 2, 2,\frac{5}{2}^{+}\rangle 1^{4}D_{\frac{5}{2}^{+}}$  &1880         1860                 %10
& & & & &  &$N(1675)\pi $  &$N(1680)\pi $ &$\Delta(1620)\pi $  &\\
%\cline{7-9}
$\Delta(1905) \frac{5}{2}^{+}$
&$24\sim60$ &$>130$   &    &$5\sim24$  & 	    & 	        &$14\sim60$   &  &$270\sim400$ 	&PDG~\cite{ParticleDataGroup:2022pth}\\
$\underline{^{4}10[56,2^{+}_{2}]}$
&$12.9^{+1.9}_{-1.4}$ &$34.8^{+9.0}_{-6.4}$ &$0.4^{+0.3}_{-0.2}$  &$0.4^{+0.4}_{-0.2}$ &$0.2^{-0.1}_{+0.0}$ &$2.5^{+1.3}_{-0.8}$  &$2.5^{+3.1}_{-1.6}$ &$2.6^{+2.4}_{-1.3}$  &$56.3^{+18.3}_{-11.9}$  &Ours\\
$1880^{+30}_{-25}$
&31 &188 &1 &19 & & & & &239   &$U(7)$~\cite{Bijker:2015gyk}\\
****
&26 &182 &0 &15 & & & & &223  &hQM~\cite{Bijker:2015gyk}\\
[1ex]
\cline{2-11}

%$|56, ^{4}10, 2, 2,\frac{7}{2}^{+}\rangle 1^{4}D_{\frac{7}{2}^{+}}$  &1930    1821   %11
& & & & &  &$N(1520)\pi $  &$N(1675)\pi $ &$N(1680)\pi $ &\\
%\cline{7-9}
$\Delta(1950) \frac{7}{2}^{+}$
&$82\sim151$  &$2\sim30$   &$1\sim2$	&$<2$   	& 	    & 	    &	    &$7\sim30$   &$235\sim335$  &PDG~\cite{ParticleDataGroup:2022pth}\\
$^{4}10[56,2^{+}_{2}]$
&$50.8^{+4.4}_{-3.1}$ 	  &$36.7^{+5.5}_{-3.8}$ 	&$2.8^{+0.8}_{-0.5}$ &$0.8^{+0.4}_{-0.2}$ 	  &$4.7^{+1.2}_{-0.8}$ 	 &$3.2^{+0.8}_{-0.5}$		 &$0.7^{+0.5}_{-0.2}$ 	 &$6.3^{+2.5}_{-1.6}$ 	 &$106.0^{+16.1}_{-10.7}$ &Ours\\
$1930^{+20}_{-15}$
&172 &92 &5 &1 & & & & &270 & $U(7)$~\cite{Bijker:2015gyk}\\
****
&146 &70 &3 &1 & & & & &220 & hQM~\cite{Bijker:2015gyk}\\
[1ex]
\cline{2-11}

%$|56, ^{4}10, 2, 2,\frac{3}{2}^{+}\rangle 1^{4}D_{\frac{3}{2}^{+}}$  &1920      1885    %9
& & & & &  &$N(1520)\pi $  &$N(1680)\pi $ &$\Delta(1700)\pi $ &\\
%\cline{7-9}
$\Delta(1920) \frac{3}{2}^{+}$
&$12\sim72$   &$>110$   &$5\sim22$  &$12\sim61$  &$10\sim310$ &$<18$ 	&   & 	&$240\sim360$   &PDG~\cite{ParticleDataGroup:2022pth}\\
$\underline{^{4}10[56,2^{+}_{2}]}$
&$6.0^{+0.2}_{-0.3}$ 	   &$38.0^{+13.9}_{-11.2}$ 	 &$1.9^{+0.3}_{-0.4}$  &$1.8^{+1.4}_{-1.0}$		 &$41.6^{+4.2}_{-5.9}$ 	  &$48.7^{+10.0}_{-10.6}$ &$14.4^{+17.3}_{-10.3}$ 	 &$3.4^{+2.2}_{-2.1}$ 		 &$155.8^{+49.5}_{-41.8}$   &Ours\\
$1920^{+50}_{-50}$
&7  &132 &23 &22 & & & & &184 &  $U(7)$~\cite{Bijker:2015gyk}\\
***
&17 &137 &22 &20 & & & & &196 & hQM~\cite{Bijker:2015gyk}\\
[1ex]
\cline{2-11}

%$|70, ^{2}10, 2, 2,\frac{5}{2}^{+}\rangle 1^{2}D_{\frac{5}{2}^{+}}$  &1921     1906     %7
& & & & &  &$N(1535)\pi $  &$N(1675)\pi $ &$N(1680)\pi $ &\\
%\cline{7-9}
$\Delta(1975) \frac{5}{2}^{+}$
&$0.2^{+0.0}_{-0.0}$  &$17.4^{+0.0}_{-0.9}$  & $0.0$  & $5.6^{+0.9}_{-1.4}$   &$3.5^{+0.8}_{-0.7}$  &$13.8^{+8.5}_{-6.0}$  &$70.7^{+21.3}_{-23.0}$   &$16.3^{+12.9}_{-8.9}$ &$127.5^{+44.4}_{-40.9}$ &Ours\\
$\underline{^{2}10[70,2^{+}_{2}]}$\\
$1975^{+50}_{-50}$\\
[1ex]
\cline{2-11}

%$|70, ^{2}8, 2, 0,\frac{1}{2}^{+}\rangle 2^{2}S_{\frac{1}{2}^{+}}$   1843   %4
& & & & &  &$N(1520)\pi $  &$N(1535)\pi $ &$N(1650)\pi $ &\\
%\cline{7-9}
$\Delta(1874) \frac{1}{2}^{+}$
&$2.8^{-0.5}_{+0.4}$ 	&$2.5^{-1.8}_{+2.3}$ 	&$0.8^{+0.0}_{-0.1}$  &$2.1^{+0.4}_{-1.1}$ &$6.3^{+2.2}_{-1.9}$ &$19.2^{+18.0}_{-10.9}$ 	 &$269.2^{+80.4}_{-79.8}$ 	 &$28.9^{+23.7}_{-17.6}$ 	 &$331.8^{+122.4}_{-108.7}$  &Ours\\
$\underline{^{2}10[70,0^{+}_{2}]}$\\
$1874^{+50}_{-50}$\\
[1ex]
\cline{2-11}

%$|70, ^{2}10, 2, 2,\frac{3}{2}^{+}\rangle 1^{2}D_{\frac{3}{2}^{+}}$  &1919      %6
& & & & &  &$N(1520)\pi $  &$N(1680)\pi $ &$\Delta(1700)\pi $ &\\
%\cline{7-9}
$\Delta(1973) \frac{3}{2}^{+}$
&$4.8^{+0.0}_{-0.1}$  	&$24.3^{+7.1}_{-6.0}$ 	&$1.8^{+0.1}_{-0.2}$  &$2.1^{+1.0}_{-0.8}$	&$31.9^{+1.3}_{-2.7}$ &$56.4^{+5.2}_{-7.3}$ 	 &$10.5^{+8.1}_{-5.7}$ &$127.4^{+54.7}_{-53.7}$	 &$259.2^{+77.5}_{-76.5}$ &Ours\\
$\underline{^{2}10[70,2^{+}_{2}]}$\\
$1973^{+50}_{-50}$\\
[1ex]
\hline\hline

\end{tabular}

\end{center}

\end{table*}

\section*{Acknowledgement}

This work is supported by the National Natural Science Foundation of China (Grants No.12175065, No.12235018, No.12105203, No.11775078, and No.U1832173). Q.Z. is also supported in part, by the DFG and NSFC funds to the Sino-German CRC 110 ``Symmetries and the Emergence of
Structure in QCD'' (NSFC Grant No. 12070131001, DFG Project-ID 196253076), National Key Basic Research
Program of China under Contract No. 2020YFA0406300, and Strategic Priority Research Program of Chinese
Academy of Sciences (Grant No. XDB34030302).

\bibliographystyle{unsrt}

\end{document}